# Notes from the Physics Teaching Lab: Rubidium Atomic Spectroscopy

Kenneth G. Libbrecht[1]
Department of Physics, California Institute of Technology

**Abstract. We describe a series of rubidium spectroscopy experiments that can be done using the Teachspin Diode Laser Spectroscopy instrument, which is commercially available and is already being used in physics teaching labs at over 150 universities. Our goal here is to provide a detailed examination of the capabilities of this instrument, including numerous examples of measurements and data analysis, presented as a supplement to the manufacturer's user manual. Our hope is that instructors using this product or similar diode-laser-based Rb spectroscopy systems will find the experiments described here useful for designing and implementing the curricula in their own physics teaching labs.**

## Introduction

Atomic spectroscopy is a cornerstone of modern quantum physics, with the electronic structure of the hydrogen atom being a core component of all undergraduate physics curricula. Because atomic hydrogen is an experimentally difficult case study, alkali metal vapors have become quite popular in physics teaching labs for exploring atomic spectroscopy and related phenomena. Because alkali atoms have a single valence electron, each exhibits a hydrogen-like electronic structure that connects well to theory courses and is far simpler to describe than multi-valence-electron atoms.

Looking at the available alkali choices, lithium and sodium would be quite desirable for the teaching lab because they both have electronic transitions in the visible range, making them advantageous regarding eye-safety issues. Unfortunately, lithium vapor cells typically require temperatures above 300 C [1995Lib], which is technically challenging, and there are no inexpensive diode-laser options for observing sodium spectroscopy near 589 nm.

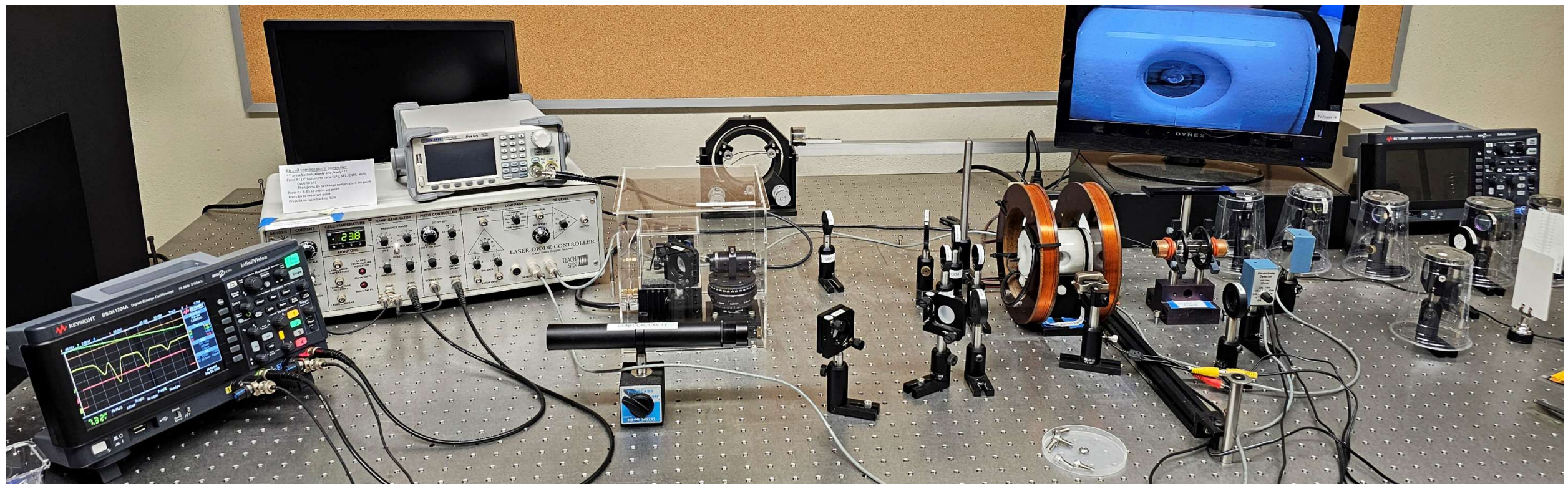

*Figure 1. This photo shows Teachspin's Diode Laser Spectroscopy apparatus and associated hardware on an optical table. This equipment was used for all the experiments described in this paper.*

[1] kgl@caltech.edu

Of all the alkali elements, rubidium has emerged as the easiest to use overall, because the vapor pressure is quite high even at room temperature and the transitions at 780 and 795 nm are accessible using inexpensive diode-laser systems. For this reason, rubidium atomic spectroscopy has long been used in undergraduate teaching labs [1992Mac], and the needed optical technology has steadily (albeit slowly) decreased in price. There has been a steady stream of research papers describing Rb spectroscopy using tunable diode lasers in undergraduate teaching labs [1996Baa, 1996Pre, 1998Rao, 2006Ols, 2009Jac, 2020Der, 2021Buc, 2025Bus] … and this is by no means an exhaustive list of relevant references.

Perusing the literature in this educational area, one quickly perceives a zeal for developing new hardware, new techniques, and new teaching-lab experiments that at least try to keep up with the rapidly advancing frontiers of Atomic/Molecular/Optical (AMO) physics. One result from these efforts, however, is that some of these new experiments are quite sophisticated and require a fairly deep understanding of AMO physics. Although these advanced teaching-lab experiments are fascinating in their own right, they can be quite challenging for students that have only recently had their first exposure to quantum mechanics and atomic structures. The present paper aims to address this knowledge gap (to some degree) by examining the basics of rubidium spectroscopy using relatively simple teaching-lab experiments.

Another problem we hope to address here is that teaching-lab handouts often cover the underlying theory describing an experiment in considerable depth while glossing over many experimental details. In our experience, students can handle the theory but often struggle with even basic laboratory practices, because their courses have focused so little on practical considerations one encounters in the lab. In a typical quantum mechanics course, for example, all atoms are perfectly at rest, in a perfect vacuum, exposed to plane-wave EM fields, etc. While this level of mathematical abstraction is appropriate in theory courses, it does little to prepare students for the harsh realities of real experimental physics – atomic ensembles, imperfect optics and electronics, alignment issues, and the need to juggle so many intertwined physical concepts that are present in even seemingly simple experiments.

With these goals in mind, the present paper focuses on relatively simple Rb spectroscopy experiments that describe basic physical concepts. Of course, avoiding all complexity is impossible in almost any advanced teaching-lab experiment, and laser-based spectroscopy is certainly no exception. But we can at least attempt to develop serviceable physical models that convey a sense of what is needed to fully explain laboratory measurements in real-world settings. At some fundamental level, this is largely why we need teaching-lab courses – to give students some exposure to the skills and the mind-set needed to perform and model practical physical measurements.

We begin by choosing a commercially available experimental system, so instructors do not have to independently develop a great deal of complex hardware from scratch. For Rb spectroscopy, the *Diode Laser Spectroscopy* apparatus offered by *Teachspin* is an excellent starting point. First introduced in 2004, this apparatus remains a popular choice for bringing laser-based atomic spectroscopy into the teaching lab at a reasonable cost. As described on the Teachspin website, this system has been purchased by over 150 universities to date, so it is already widely used around the world.

Of course, the *Diode Laser Spectroscopy* system comes with a user manual, but the manual only scratches the surface regarding the breath of experiments that are possible using this hardware. Moreover, the manual does not include an extensive critical examination of actual experimental data and subsequent analyses that one might incorporate into a teaching-lab course. Teachspin provides a well-crafted experimental platform, but it is left to the teaching-lab community to fully explore how the hardware can be used in an educational setting. A great deal of relevant material can be found in teaching-lab handouts that are scattered about on the internet, but these are increasingly being hidden behind university paywalls using Canvas and other software platforms. Fortunately, arXiv provides an ideal forum for advancing materials used in physics teaching labs, as new insights are immediately available to all with no burdensome subscription costs.

## Rubidium atomic structure

There already exists a vast scientific literature pertaining to the Rb atomic structure, as this atom has been much studied for many decades. We provide just a brief summary here, limiting our discussion to the basic material needed to use this apparatus. Rubidium has two naturally occurring isotopes: $^{85}$Rb, with 72 percent natural abundance and a nuclear spin quantum number $I = 5/2$, and $^{87}$Rb, with 28 percent natural abundance and $I = 3/2$. Although $^{87}$Rb is technically not stable against radioactive decay, it has a half-life of $4.8 \times 10^{10}$ years, which is more than three times the age of the universe. The ground-state electronic configuration of rubidium consists of closed shells plus a single 5S valence electron, giving a hydrogen-like level structure. For the first excited state, the 5S electron is promoted to 5P.

The different energy levels are labeled by the usual Russell-Saunders notation $^{2S+1}L_J$, where S represents the total spin angular momentum ($S = 1/2$ for a single electron), L specifies the total orbital angular momentum (designated S, P, D, etc. for $L = 0, 1, 2$ respectively … alas, this S is different from the total-spin S), and J refers to the total angular momentum.

For the ground state of rubidium, $S = 1/2$ (because only the single valence electron contributes) and $L = 0$, giving $J = 1/2$ and the ground state $^2S_{1/2}$. For the first excited levels we have $S = 1/2$ and $L = 1$, giving $J = 1/2$ or $J = 3/2$. Thus there are two excited states labeled $^2P_{1/2}$ and $^2P_{3/2}$, and these two energy levels are split by spin-orbit coupling. Transitions from the $S_{1/2}$ ground state to the $P_{1/2}$ excited state are called "D1" transitions, or D1 "lines" in spectral jargon, and D2 refers to excitations to the $P_{3/2}$ state. The D1 lines at 795nm are used in Rb optical pumping experiments [2026Lib], but we will focus only on the D2 lines at 780 nm here.

Adding the nuclear spin yields hyperfine splittings to all these electronic states, and Figure 2 shows the $^{85}$Rb and $^{87}$Rb levels that are important for the present discussion. We will postpone our discussion of Zeeman sublevels for now. In general, multi-electron atomic energy levels are difficult to calculate because the various interactions yield a complex many-body problem that makes quantitative calculations quite challenging. On the experimental side, however, modern laser spectroscopy techniques in atomic-state metrology have determined many transition frequencies to remarkable precision. For example, the current best-measured values of the ground-state hyperfine splittings for $^{85}$Rb and $^{87}$Rb are 3.0357324390(60) GHz and 6.834682610904290(90) GHz, respectively [1999Biz, 2025Ste].

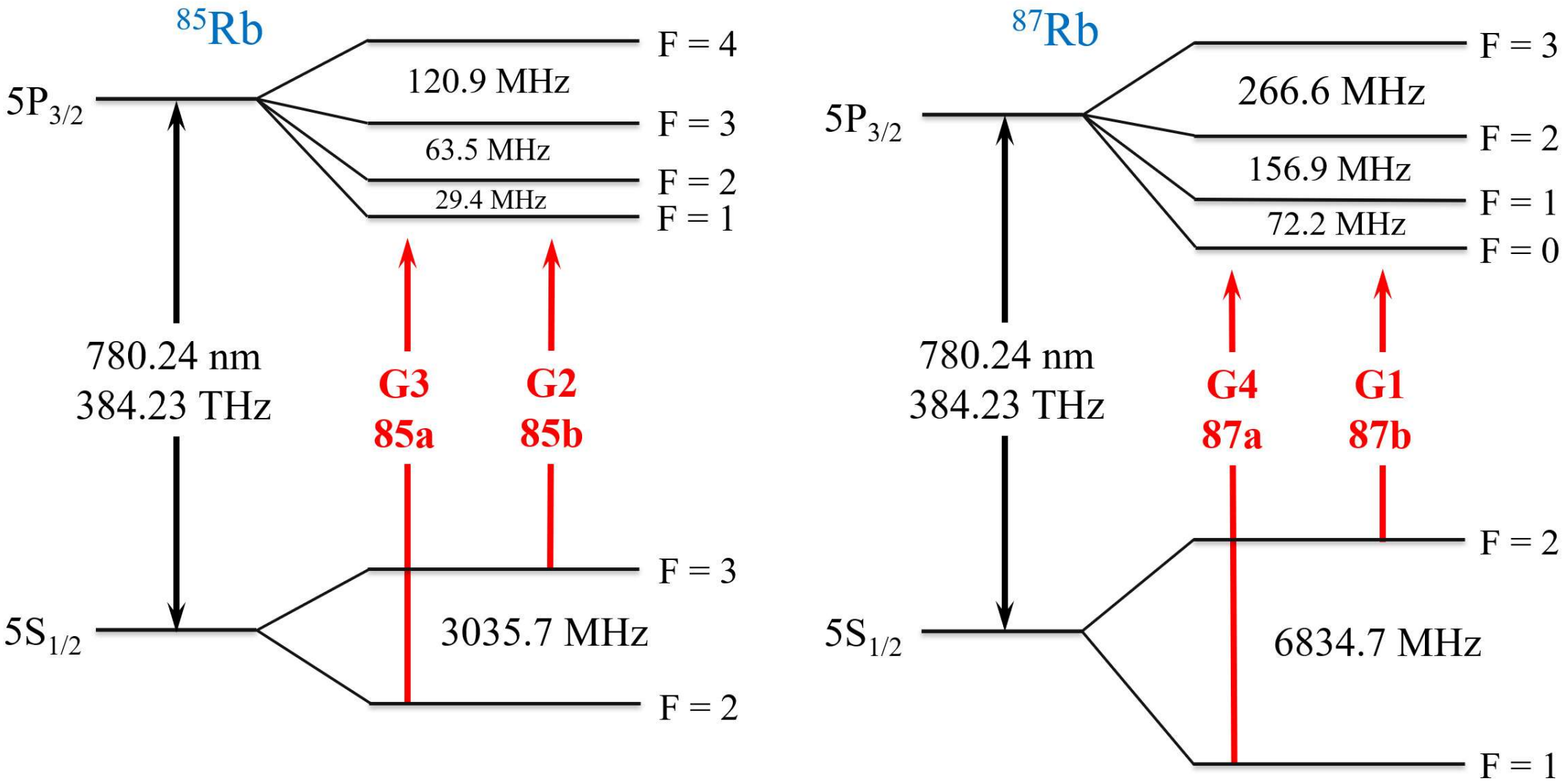

*Figure 2. This sketch shows a simplified level diagrams for Rb87 (left) and Rb85 (right), including the measured hyperfine splittings of the ground and excited states (using numbers from [2025Ste]). The level separations are most decidedly not to scale! The red arrows give the primary D2 spectral features observed in a Rb vapor cell, each coupling one $5S_{1/2}$ ground state (labeled G1-G4) to a manifold of $5P_{3/2}$ excited states. In a Doppler-broadened absorption spectrum like that shown in Figure 3, the four ground states are all resolved while the upper excited states are not resolved.*

In the Rb teaching lab, our main focus is on laser-based spectroscopy using a simple vapor cell, and Figure 3 illustrates a typical laser absorption spectrum. The Doppler width of the gas atoms is somewhat higher than the $P_{3/2}$ hyperfine splittings, so none of these individual levels can be resolved in the Doppler-broadened spectrum. On the other hand, the $S_{1/2}$ splittings are nicely resolved, and the four transmission dips in Figure 3 are each associated with a single ground-state level, as shown in Figure 2. We have labeled these G1-G4 in the figures, and these are simply ordered from left to right as they appear in the absorption spectrum.

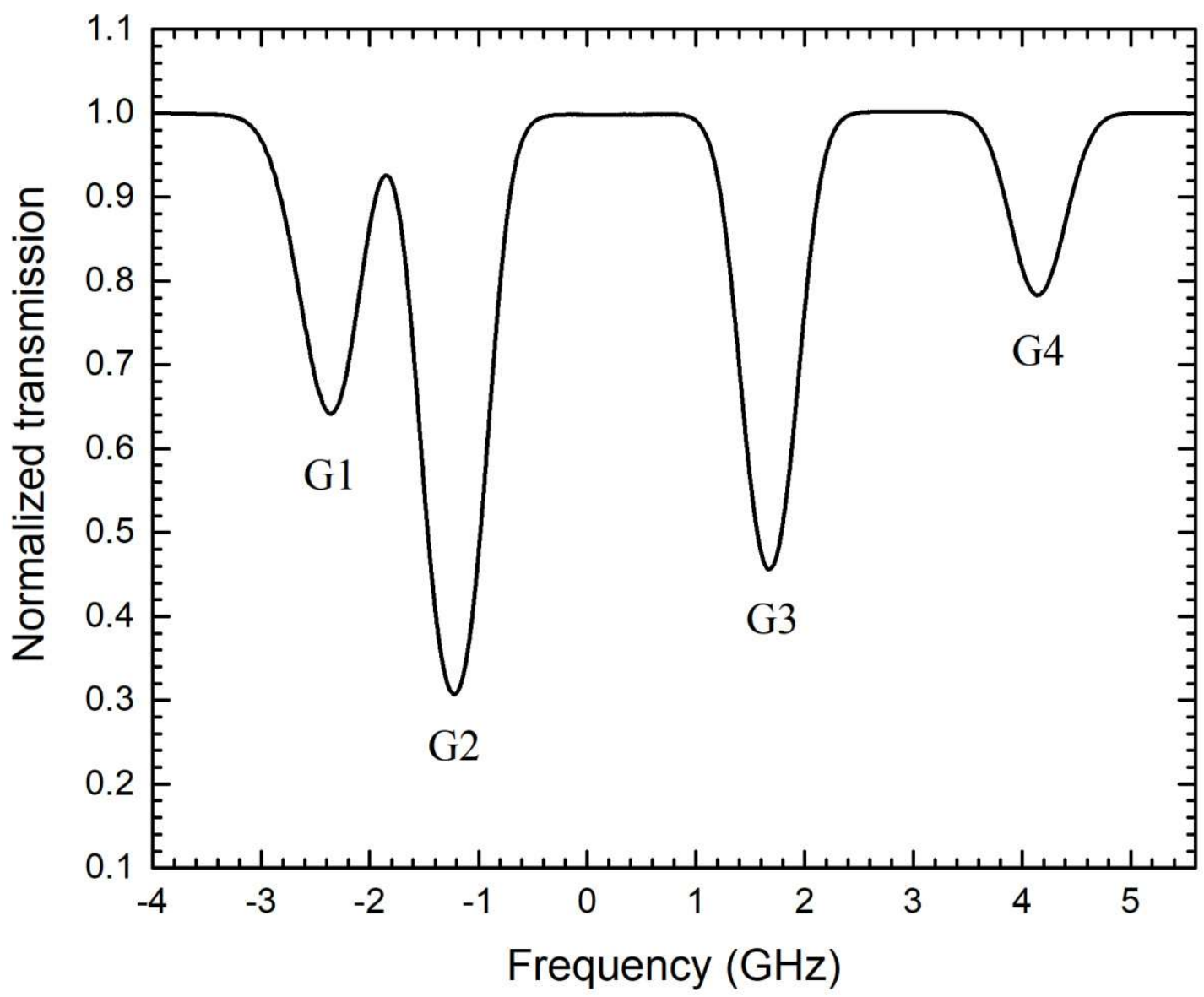

*Figure 3. This plot shows a measured Rb transmission spectrum through a rubidium vapor cell that we obtained using our Teachspin apparatus. The four transmission dips correspond to transitions from the four Rb ground states shown in Figure 2.*

## Basic Rb absorption spectrum

Moving into the lab, we can use the optical setup shown in Figure 4 to reproduce the basic Rb absorption spectrum. The parts shown here are all included in the *Teachspin* package, except for the optical isolator and the camera. We added the optical isolator (Thorlabs IO-5-780-HP) because it greatly reduces the intensity of any back-reflected laser beams that might otherwise go back into the diode laser. These beams will not damage the laser, but they do often degrade the laser's frequency stability. Including the isolator generally makes all the experiments work a bit better. However, this optical element is both expensive and fragile, so we pre-align it with the laser and enclose the assembly in a protective acrylic case, as shown in Figure 1. (The case includes a small hole for the laser beam to pass through, because the optical quality of the acrylic sheet is poor.) The OD2 filter is two filters in series (Thorlabs NE10B+NE20B), as we discuss further below.

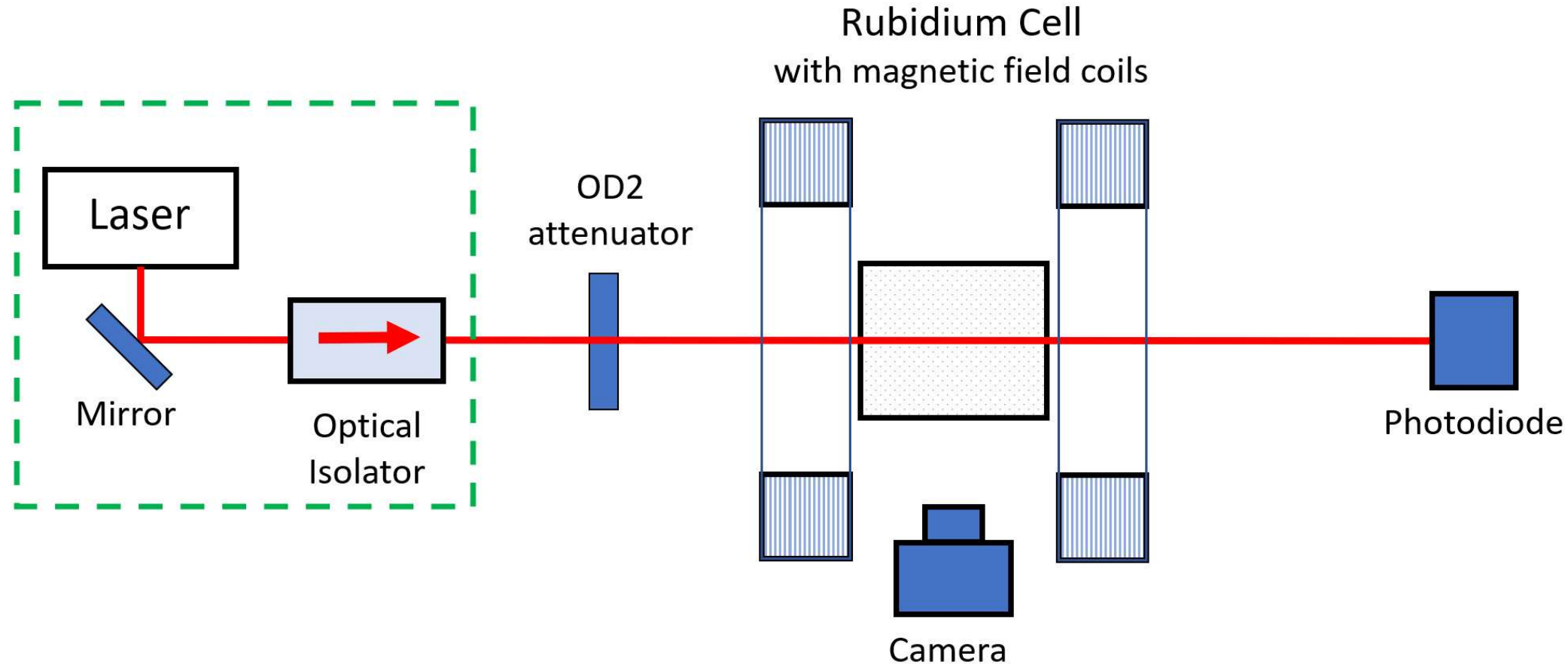

*Figure 4. This sketch shows the optical layout for making a first observation of the rubidium absorption spectrum. The OD2 attenuator should be removed for the initial alignment.*

The camera in Figure 4 is a small automotive "back-up" camera mounted on an optical post, with the image displayed on the TV monitor in Figure 1. The camera can be used as an alignment aid, supplementing the IR card included in the *Teachspin* system, plus it reveals a beautiful flashing laser beam passing through the Rb cell when the laser is scanning through the absorption dips. Viewing the scattered laser light is quite enjoyable if one has not worked with infrared lasers before, making this a worthwhile addition to a teaching lab.

Figure 5 shows the electronics layout needed to observe the Rb absorption spectrum, and you can see that the Teachspin controller and the included twin photodetectors do most of the heavy lifting. Many of the experiments described here can be done using a 2-channel oscilloscope, and the Keysight EDUX1052A has become a workhorse in our teaching lab. For Rb spectroscopy, however, it is quite convenient to have an additional channel or two, and we chose the Keysight DSOX1204A because it integrated well with our other 2-channel 'scopes. The Rigol DS1054Z is a more economical choice, as it provides four channels with 12-bit resolution at a remarkably low price point. Digital oscilloscope technology has been improving rapidly in recent years, and it appears likely that 4-channel 12-bit 'scopes will soon be the norm even in base-level models. Figure 6 shows a typical oscilloscope screen using this optical and electronic layout after setting up the laser sweep as described in the Teachspin manual.

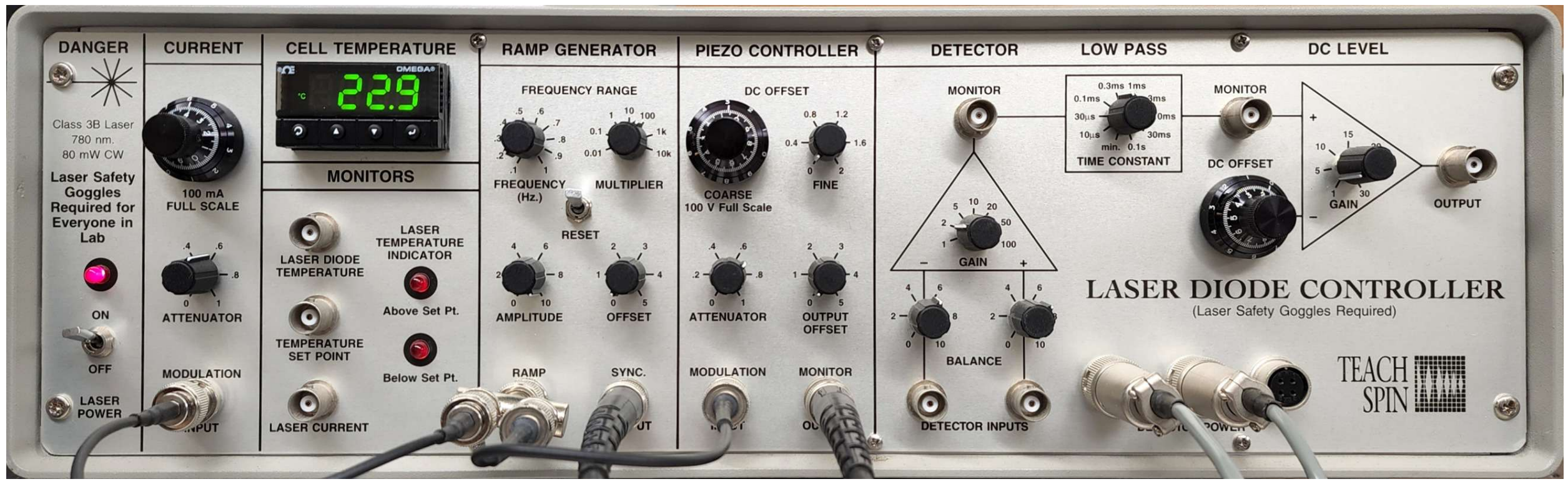

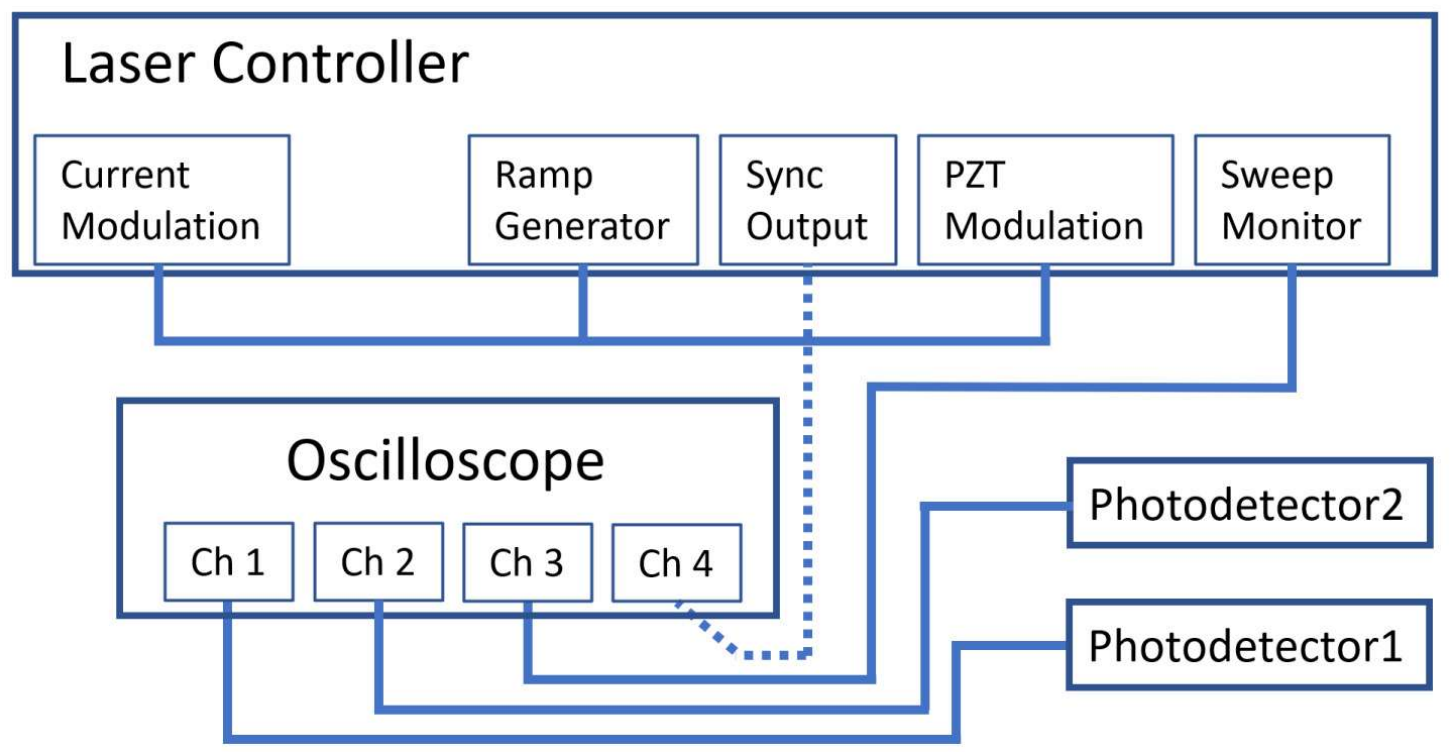

*Figure 5. The photo above shows the Teachspin laser controller, and the electronics layout (left) shows the connections used to scan the laser frequency and observe the light transmitted through the Rb vapor cell. The connections shown are all BNC cables, and we used a Keysight DSOX1204A oscilloscope to collect most of the data described in this paper.*

**Maintaining a stable laser frequency scan.** One notable feature in Figure 5 is that the internal Ramp Generator signal is connected to both the PZT Modulation input and the laser Current Modulation input. This is necessary to obtain a continuous laser sweep that covers all four Rb absorption dips. If the current modulation is not included, then the laser typically experiences “mode hops” over several-GHz frequency scans [2025Lib]. (This reference also includes a primer describing the physics of diode laser spectroscopy, including the use of an external grating/cavity for stabilizing the laser frequency, which is the core technology used in the Teachspin instrument.) An undesirable result from the synchronous PZT/current modulation technique, however, is that the laser intensity changes along with the frequency, as seen in Figure 6. This feature, alas, is one of the prices you pay for having a relatively inexpensive tunable-laser system.

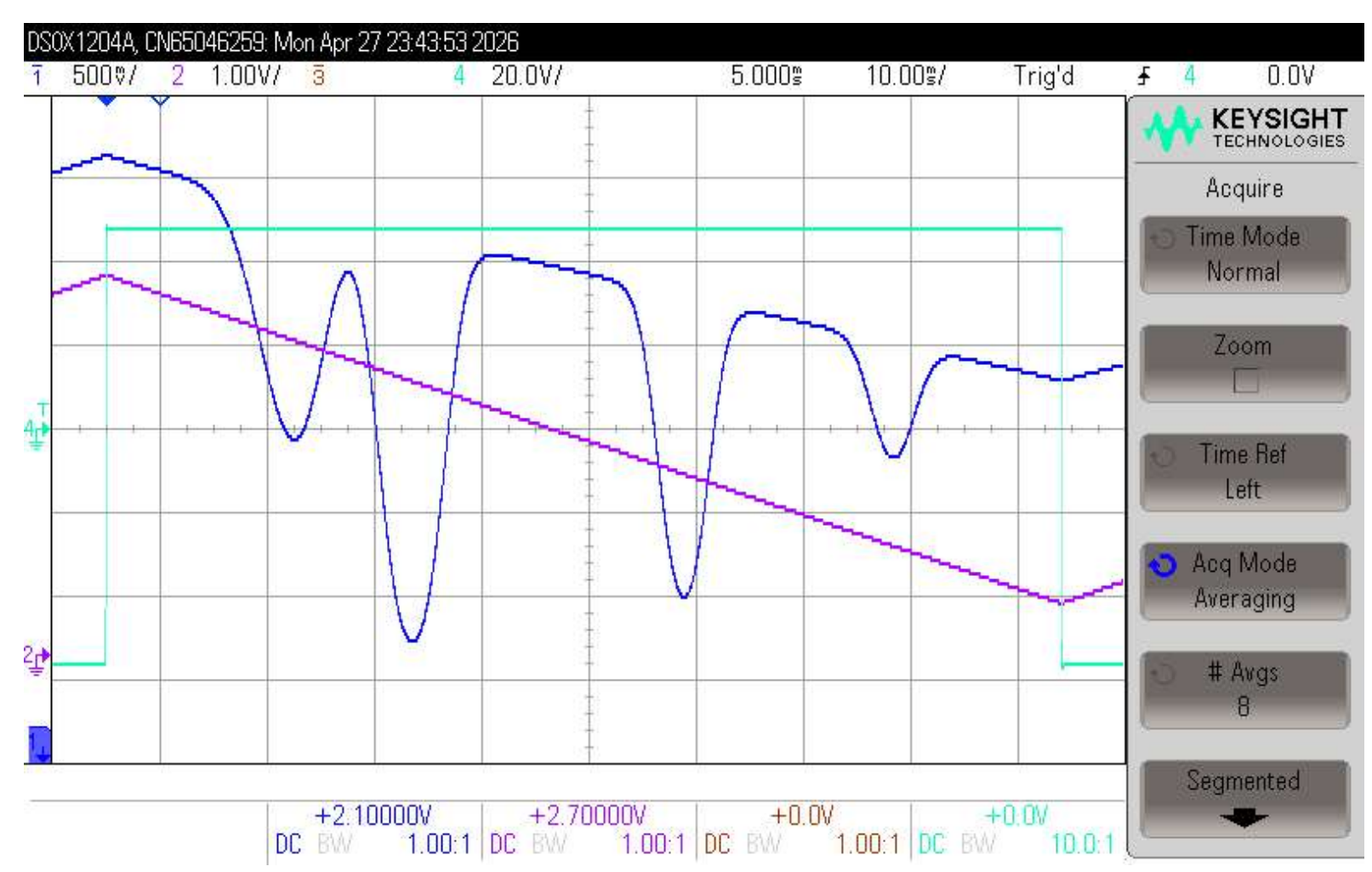

*Figure 6. A typical first oscilloscope screenshot (negative image) showing the four Rb absorption features, also shown in Figure 3. In addition to the photodetector output, the image shows the Sweep Monitor and the Sync Output from the laser controller.*

Another undesirable drawback of this laser system is that drifts in the laser temperature and/or current can distort the frequency scan in sometimes subtle ways. The problem can be readily observed by adjusting the laser current up and down a small amount (by turning the current dial) while observing the absorption spectrum on the ‘scope. If all is working well, then this small current modulation will simply move the spectrum back and forth without changing any of the absorption dips. With a large enough current modulation, however, mode hops will appear that badly distort the observed Rb spectrum. One tricky element of using this laser system is finding a “sweet spot” in the applied current and PZT DC offset that produces a stable absorption spectrum. Even trickier is that sometimes the spectral distortion is relatively minor but will still impact your data. A key operating tip, therefore, is to modulate the laser current up and down from time to time to make sure the spectrum is stable over the entire range of the frequency sweep.

In our experience, it is often difficult to convey to students that this occasional checking of the spectrum and tweaking the parameters is necessary to obtain quality experimental data. This is where a bit of instructor oversight can make a big difference. And it leads to an important teaching moment – while these kinds of quirky problems are largely absent in most consumer electronics and optics, they are not uncommon in cutting-edge research labs … and even in cutting-edge physics teaching labs.

## Modeling Rb absorption

An important part of the teaching-lab experience is to go beyond passively observing a physical phenomenon and take the next step of trying to form a mathematical model of quantitative measurements. The Rb absorption spectrum is a good place to practice and develop some of these basic analytical skills. Beginning with a .csv file containing the spectrum in Figure 6, a normalized version can be created using

$$I_{norm}(t) = I_{in}(t)/(A_0 + A_1 t + A_2 t^2) \tag{1}$$

where $I_{in}(t)$ is the measured spectrum (in volts), $t$ is the sweep time, and the constants $A_i$ are from a fit to the background level away from the absorption dips.

We encourage students to use a “chi-by-eye” fit to the data (plotting the data together with a fit line and then adjusting the parameters manually until the fit looks good by eye), because it is difficult to tell a fitting algorithm to ignore the large absorption dips. Using the chi-by-eye method also moves students away from the notion that there must always be some button to press when analyzing experimental data. Note that the normalized spectrum $I_{norm}(t)$ is dimensionless and ideally runs from 0 to 1. A reasonable conversion from the frequency sweep time $t$ to the actual laser frequency change $\Delta f$ can be obtained by assuming that $\Delta f$ is proportional to $t$ and that the spacing between the G1 and G4 dips is 6.5 GHz (which is *not* equal to the ground-state hyperfine splitting of 6.8 GHz, as one might expect … more on that below).

Note that we have no way to measure the absolute frequency of the laser, so the zero point of $\Delta f$ is somewhat arbitrary in the observed absorption spectrum. It is customary to set $\Delta f = 0$ when the laser frequency is equal to the bare $S_{1/2} \rightarrow P_{3/2}$ transition in the absence of any hyperfine splitting. This bare transition does not exist in nature, however, so the traditional $\Delta f = 0$ point is set by measurements of real transitions combined with theoretical modeling of the hyperfine splittings.

To model the absorption dips, we note that the photodetector signal is expected to follow

$$\begin{aligned} I_{norm}(t) &= e^{-\alpha z} \\ &= e^{-\tau} \end{aligned} \tag{2}$$

to good approximation, where $z$ is the propagation distance through the Rb vapor cell of length $L$, $\alpha$ is the *absorption coefficient*, and $\tau = \alpha L$ is the *optical depth* of the cell. We further simplify the problem by treating each absorption dip as if it were a single atomic Doppler-broadened transition (more on this below), writing the frequency-dependent absorption as

$$\tau(\Delta f) = \tau_0 \exp[-(\Delta f/\sigma)^2] \tag{3}$$

for each absorption dip, where $\sigma$ is the $1/e$ half-width of the Doppler profile and $\tau_0$ is the on-resonance optical depth. At a typical cell temperature of 40C, statistical mechanics gives $\sigma \approx 315$ MHz, with a negligible isotope shift of about 3 MHz.

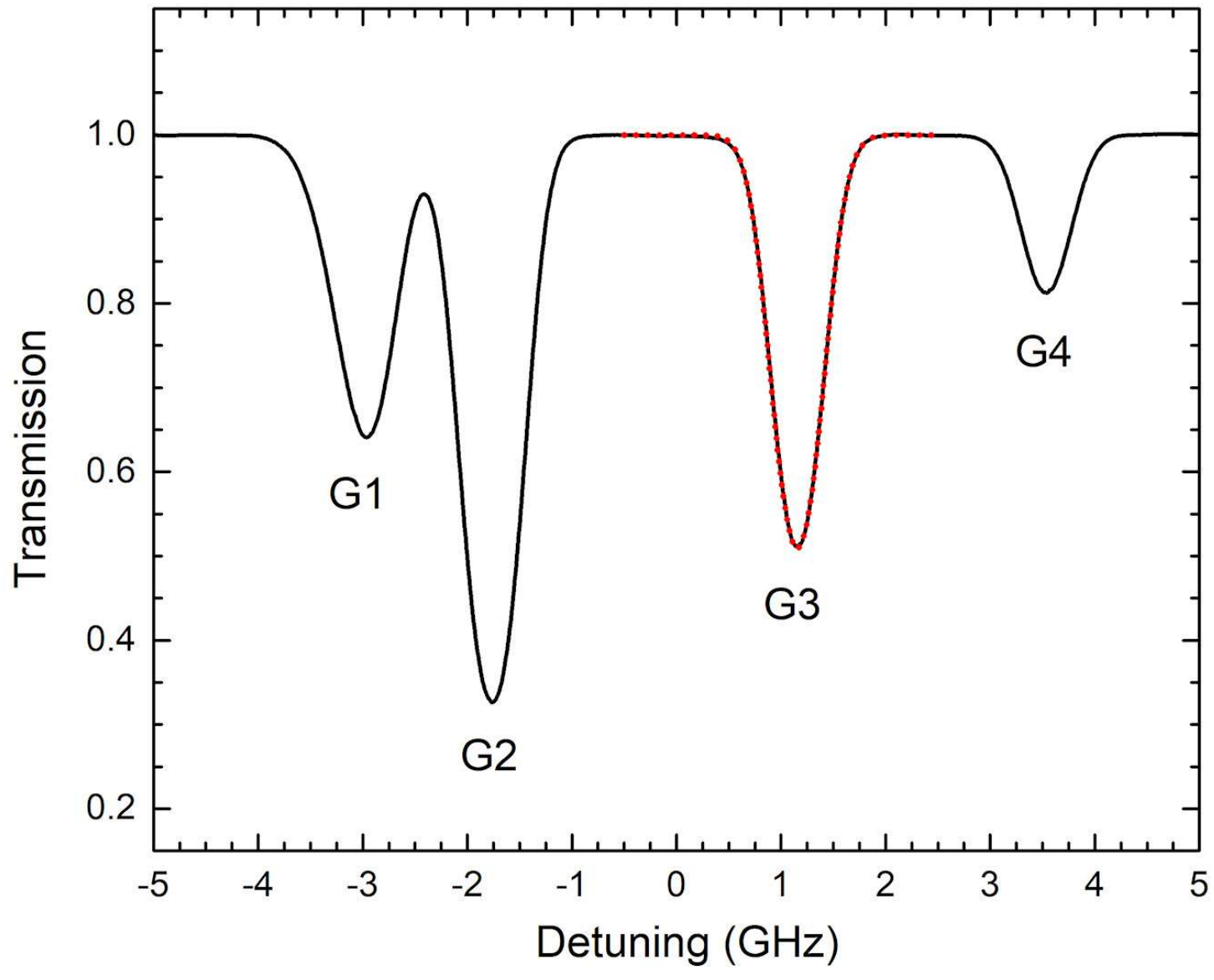

*Figure 7. This graph shows digital oscilloscope data converted to a normalized Rb transmission spectrum. The red curve is from a chi-by-eye fit to the G3 dip, equal to exp[-0.675*exp[-(Δf-1.155)^2/0.31^2]]). Note that the zero point of the laser frequency detuning Δf is essentially an arbitrary experimental parameter in this and other spectra data plots.*

The absorption profile for a single dip is then the nested exponential function

$$\frac{I(z)}{I_{in}} = \exp\left\{-\tau_0 \exp\left[-\frac{\Delta f^2}{\sigma^2}\right]\right\} \tag{4}$$

and Figure 7 shows some sample data overlaid with a fit curve. Here we see that a bit of trace averaging on the oscilloscope yields a spectrum with very good signal-to-noise, and the model in Equation (4) reproduces the G3 profile with quite high fidelity.

## Temperature-dependent absorption

With this basic Rb spectrum in hand, the next logical step is to investigate how the absorption changes with the temperature of the vapor cell. The overall setup is the same as shown in Figure 4, although for quantitative data it is now important to consider the background photodetector signal. We augmented the Teachspin hardware by installing narrow-band optical filters (Thorlabs FBH05780-10) over both photodetectors, as these filters greatly reduce unwanted signals from ambient room lights. And we installed an LED floor lamp in the room, driven by DC current, to avoid flickering from the overhead lights. We have found that students are often inclined to turn off all the lights and work in total darkness, but doing so tends to reduce productivity substantially.

Even with good stray-light management, the background photodetector (PD) signal is not always zero, as illustrated in Figure 8. This background is easily measured by blocking the laser light and by adding a constant term to the model in Equation (4), For these spectra and those at other cell temperatures, we fit the G4 dip to determine $\tau_0(T)$, as this dip gives a good range of $\tau_0$ measurements and avoids especially high $\tau_0$ values (that are difficult to measure accurately). As mentioned above, it is important to begin the series of measured spectra by adjusting the laser current back and forth to make sure that the laser is free of mode hops and other spectral distortions that can corrupt the $\tau_0$ data. With reasonable experimental care, it is possible to obtain a clean set of measured $\tau_0(T)$ values for the G4 dip, as seen in Figure 9.

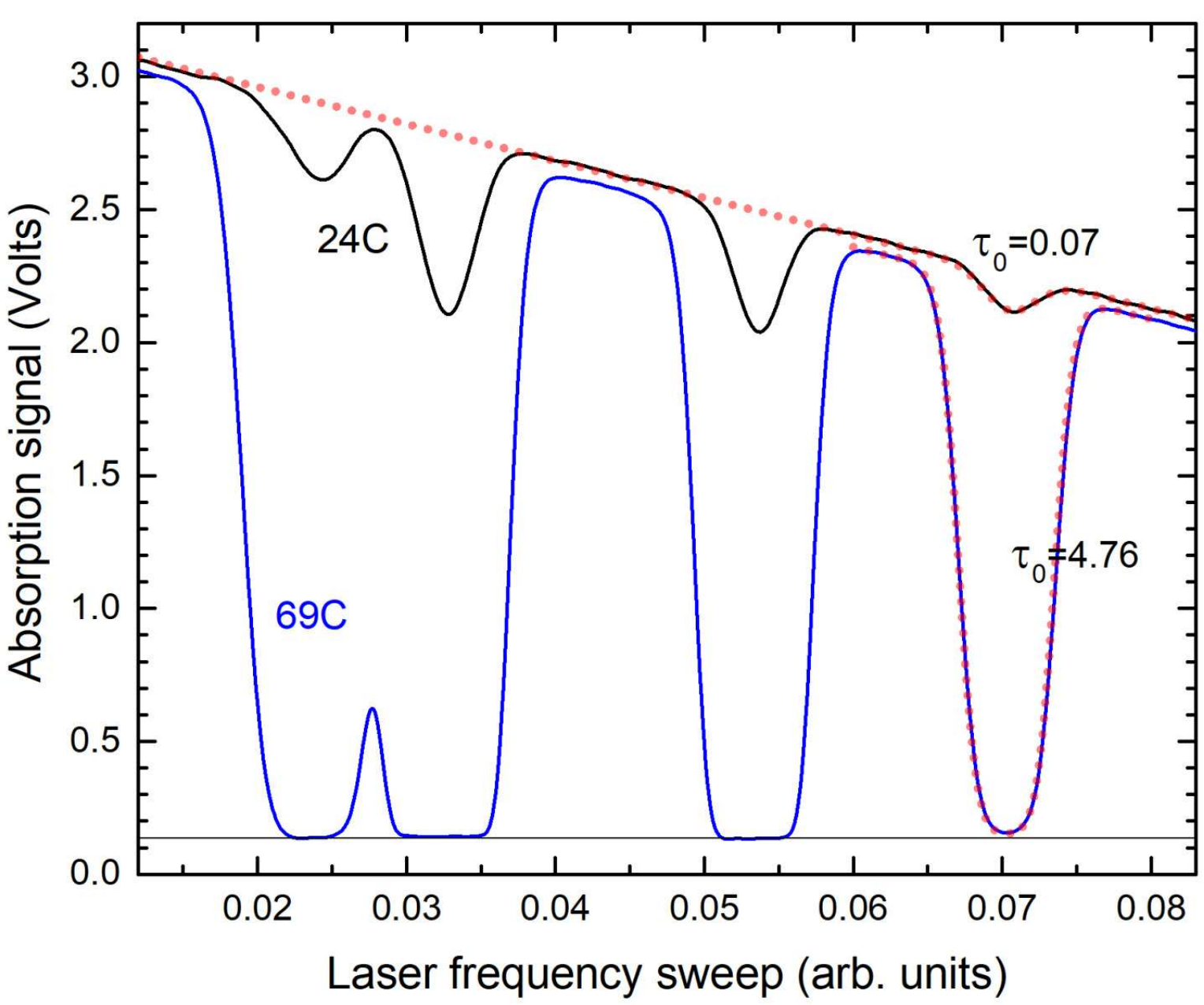

*Figure 8. This graph shows two measured Rb absorption spectra at different vapor-cell temperatures, along with fits to the G4 dip. Note that the background PD signal (nonzero flat line) strongly affects the 69C data and fits.*

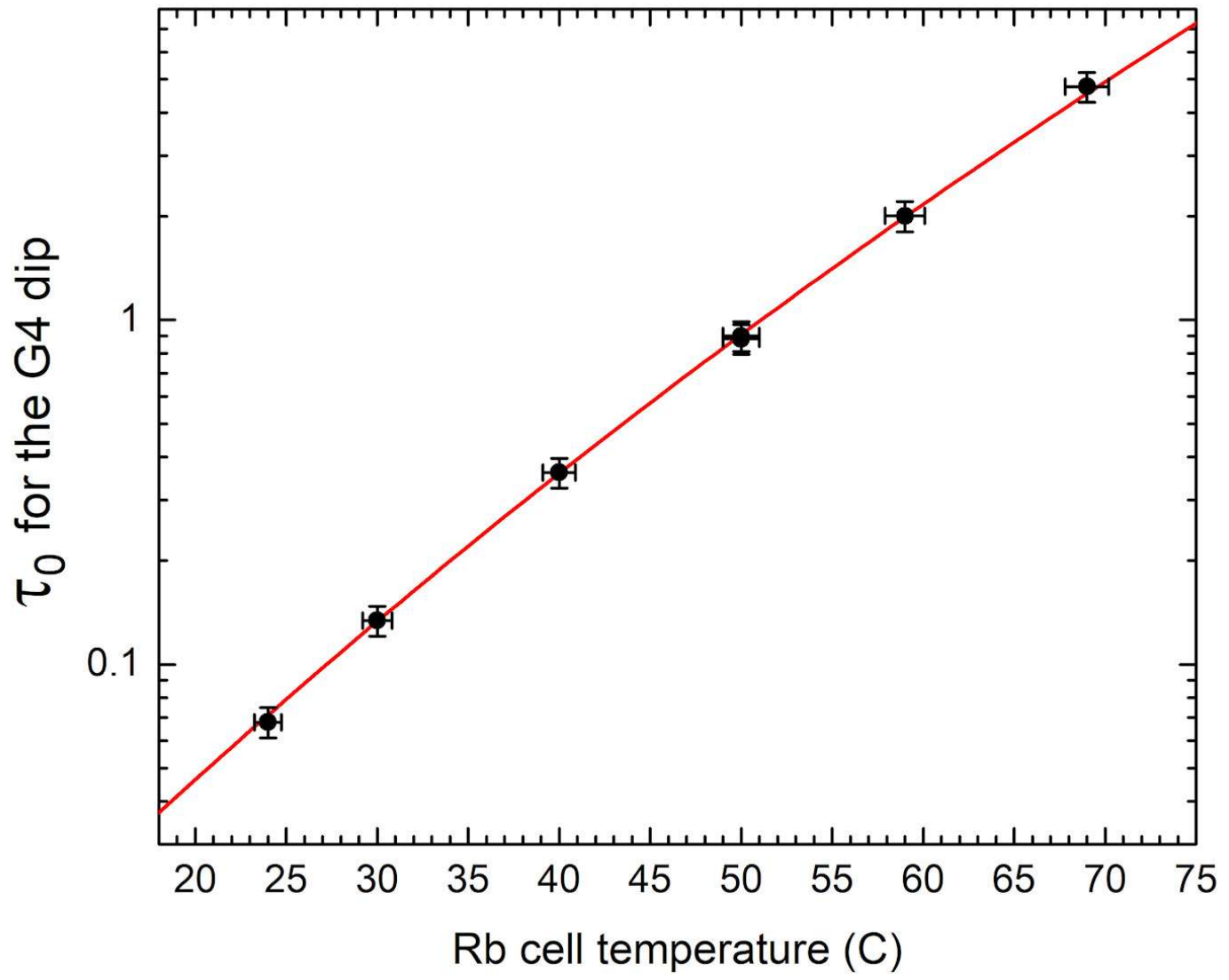

*Figure 9. The points in this plot show measured values of $\tau_0(T)$ for the G4 dip as a function of the temperature of the Rb vapor cell, with error estimates for both $\tau_0$ and the cell temperature. The red curve shows the Arrhenius model described in the text.*

## Modeling temperature-dependent absorption

To model the data in Figure 9, we begin by noting that the Rb vapor in the cell is in equilibrium with solid Rb metal on the cell walls, and we assume a uniform temperature at all points in the cell body. Then the usual Arrhenius model gives a vapor number density

$$n(T) \propto \exp\left(-\frac{E_a}{kT}\right) \tag{5}$$

where $E_a$ is an activation energy specific to the material. This is often written in terms of the vapor pressure

$$\log_{10} P(T) = A_1 - \frac{B_a}{T} \tag{6}$$

where $T$ is in Kelvin, $A_1$ is a constant, and $B_a$ is called the *Arrhenius temperature* of the material. For rubidium, published measurements give $B_a \approx 4215 \pm 200$ K, and any isotopic difference is dwarfed by the overall measurement uncertainties [2025Ste]. Because we expect $\tau_0(T) \propto n(T)$, we rewrite the above expression as

$$\log_{10}[\tau_0(T)] = A_2 - \frac{B_a}{T} - \log_{10}(T) \tag{7}$$

which is nearly a straight line when $\log_{10}[\tau_0(T)]$ is plotted as a function of $1/T$. Figure 10 shows that (with some attention to experimental details) this apparatus can produce measurements of $\tau_0(T)$ that match published measurements to good accuracy. We estimate a measurement uncertainty of roughly $\pm 400$ K in our determination of the Arrhenius temperature, limited mainly by potential systematic errors in the experiment (the Rb cell temperature is not precisely controlled, and effects of saturated absorption are not completely negligible). The G3 dip is also well isolated in the Rb absorption spectrum and could be used to measure $\tau_0(T)$ over a higher range. We leave that as an additional exercise for the interested student.

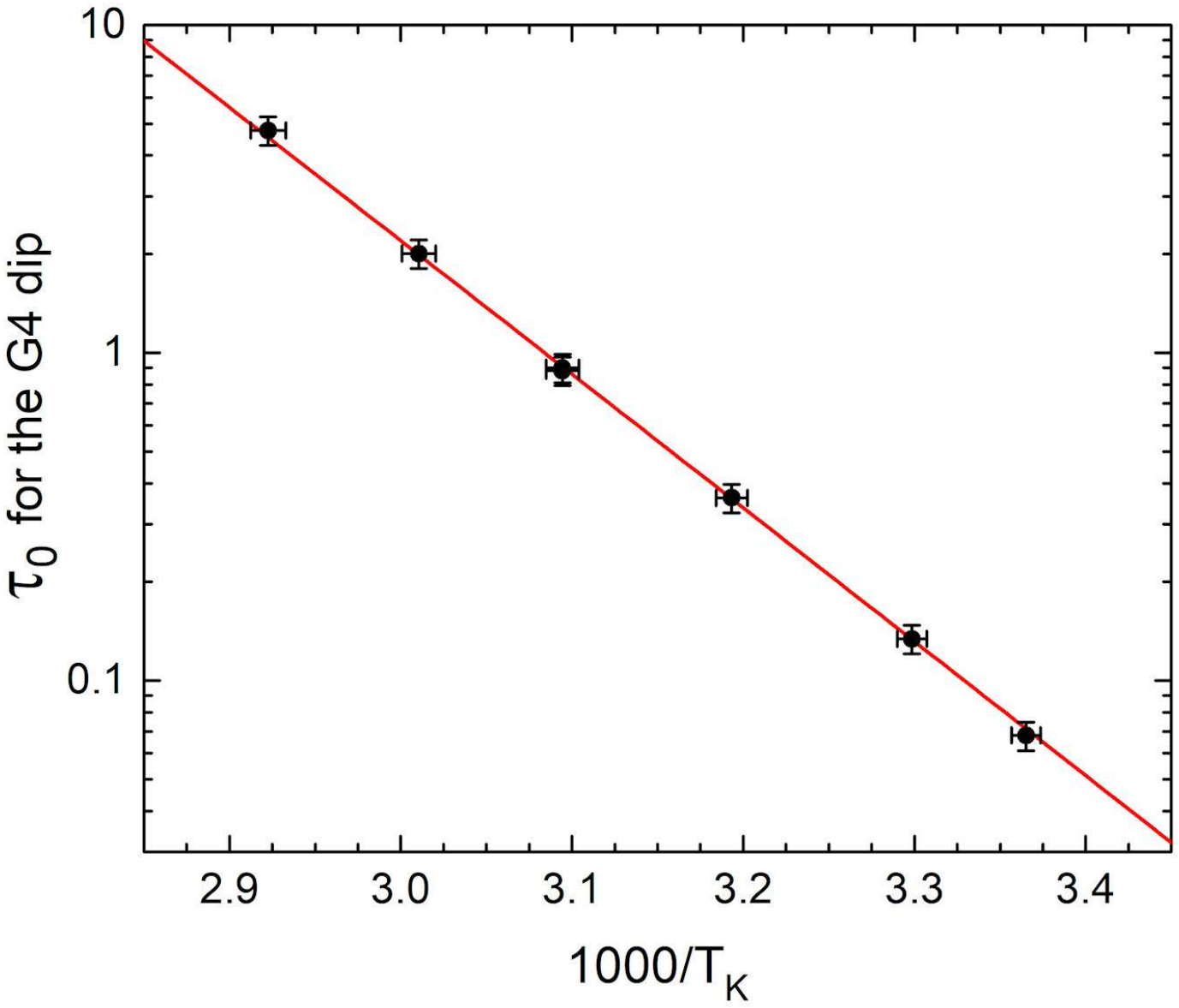

*Figure 10. These data and the included model are the same as in Figure 9, just displayed in a log-log plot. The model is from Equation (7) using $A_2 = 15.51$ and $B_a = 4215$ K, in good agreement with published measurements.*

## Saturated absorption

Another fundamental phenomenon one can observe with the basic setup in Figure 4 is how the Rb absorption spectrum changes with laser intensity. The theory can be understood by considering a collection of two-level atoms undergoing laser-driven transitions between the ground and excited state. For the case on on-resonance radiation, we can write the (simplified) rate equations as

$$\frac{dP_1}{dt} = \Gamma P_2 - gI(P_1 - P_2)$$

$$\frac{dP_2}{dt} = -\Gamma P_2 + gI(P_1 - P_2) \tag{8}$$

where $P_1$ and $P_2$ are the relative populations of the ground and excited states (with a total population $P_1 + P_2 = 1$), $I$ is the incident laser intensity, $g$ is a transition-probability constant, and $\Gamma$ defines the spontaneous decay rate [1999Met]. The first terms in both these expressions come from spontaneous emission, and the second terms are from stimulated emission. Note that stimulated emission is proportional to the intensity $I$, while spontaneous emission does not depend on $I$. Note also that stimulated emission is a purely quantum phenomenon, so these rate equations have no direct analog in classical physics.

In steady state we have $dP_1/dt = dP_2/dt = 0$, and plugging this in gives

$$P_2 = \frac{gI/\Gamma}{1+2gI/\Gamma} \tag{9}$$

This result is usually written in the form

$$P_2 = \frac{s/2}{1+s} \tag{10}$$

for on-resonant radiation, where $s = I/I_{sat}$ and for dipole radiation the value of $I_{sat}$ is given by [1999Met]

$$I_{sat} = \frac{2\pi^2 hc\Gamma}{3\lambda^3} \tag{11}$$

For the case of rubidium, $\Gamma/2\pi \approx 6.07$ MHz and $I_{sat} \approx 16.7$ W/m$^2$ [2025Ste]. Atomic physicists call $I_{sat}$ the *saturation intensity*, because the transition begins to "saturate" ($P_2$ becomes comparable to $P_1$) when $I$ is increased substantially above $I_{sat}$. However, the transition is only fully saturated with $P_2 = P_1 = 1/2$ at infinite intensity.

Assuming that each atom sees a constant laser intensity $I$ (not a especially good assumption, as we will soon see), we can then write an effective optical depth

$$\begin{aligned} \tau_{0,eff} &\approx \tau_0(P_1 - P_2) \\ &\approx \tau_0 \frac{1}{1+s} \end{aligned} \tag{12}$$

where $\tau_0$ is the optical depth at line center in the limit of zero laser intensity (so all atoms are in the ground state).

## Observing saturated absorption

Quantitative measurements of saturated absorption in the lab can be obtained quite easily with the optical setup in Figure 4; one needs only to replace the OD2 filter with a series of optical filters and thereby record spectra at a range of laser intensities. It is important to make sure that the filter transmissions are accurately known at a wavelength of 780 nm, which requires some care in the filter selection and characterization. We used the four filters Figure 11, placing them in the optical path in different combinations to produce a roughly logarithmic series of transmission values ranging of 1 (no filters in place) to ~$10^{-4}$. The OD2 filter we use for many of the measurements in this paper was made by combining the F2 and F3 filters in Figure 11. Note that all these filters were manufactured for the visible range, but measurements of the transmission as a function of wavelength are conveniently available at the Thorlabs website.

| Filter number | Thorlabs part | Transmission at 780 nm | OD at 780 nm |
|---|---|---|---|
| F1 | NE03B | 0.510 | 0.29 |
| F2 | NE10B | 0.168 | 0.77 |
| F3 | NE20B | 0.0459 | 1.34 |
| F4 | NE40B | 0.00183 | 2.74 |

*Figure 11. We used these four optical filters to measure saturated absorption. The transmission values of these filters at 780nm (last column) were estimated from the Thorlabs website.*

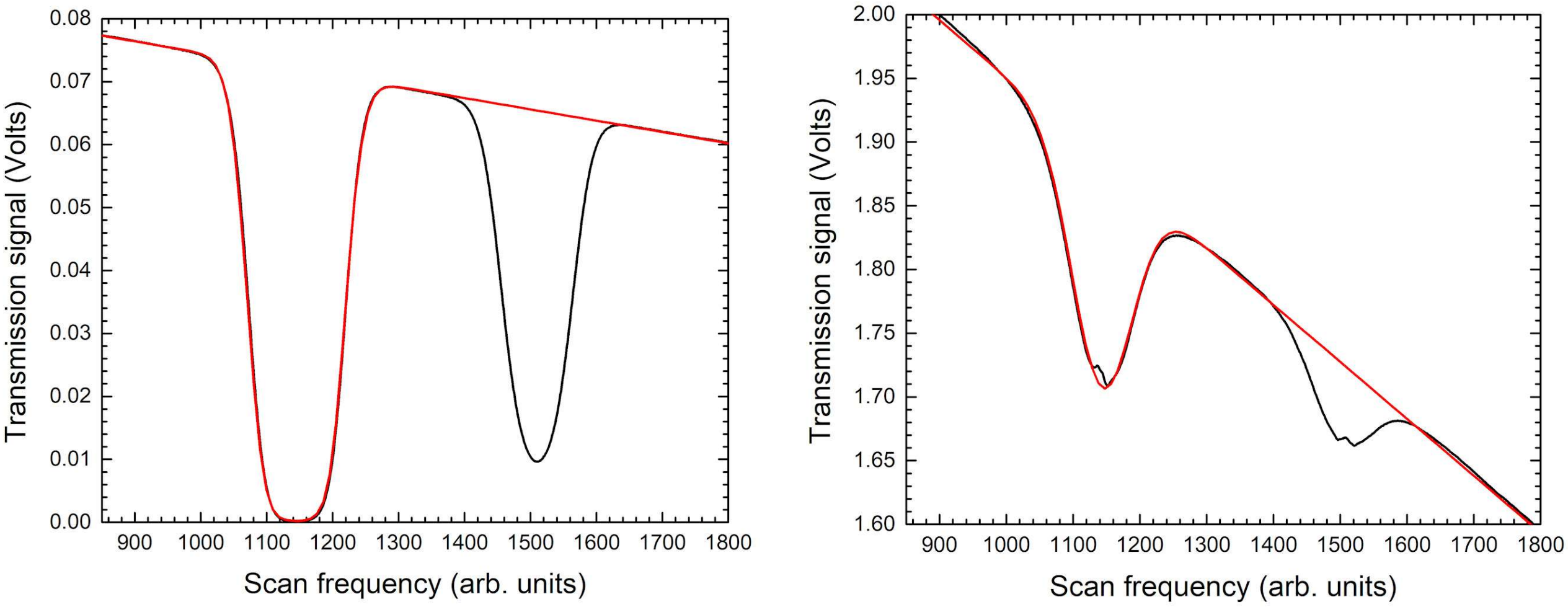

*Figure 12. These sample spectra show the G3 and G4 absorption dips at a Rb vapor cell temperature of 60C, with the F3+F4 filters (left) and no filters (right) placed in the beam. Red lines show Note the quite different vertical axes on the graphs. The black lines are measured spectra, while the red lines are fits to the G3 dip. The small wiggles at the dip centers in the high-intensity spectrum are caused by laser reflections from the cell windows, explored further in the Doppler-free saturated absorption spectroscopy section below.*

Figure 12 shows Rb absorption spectra taken at the maximum and minimum laser intensities we used in this series, revealing how a high-intensity laser beam "punches a hole" through the Rb vapor at the G3 and G4 resonance frequencies, greatly increasing the transmitted power. Extracting measurements of $\tau_{0,eff}(I)$ from the spectra yields the results in Figure 13, and we see that the basic theory curve in Equation (12) provides an excellent fit to the data if we assume a maximum saturation parameter $s_{max} = I_0/I_{sat} = 60$.

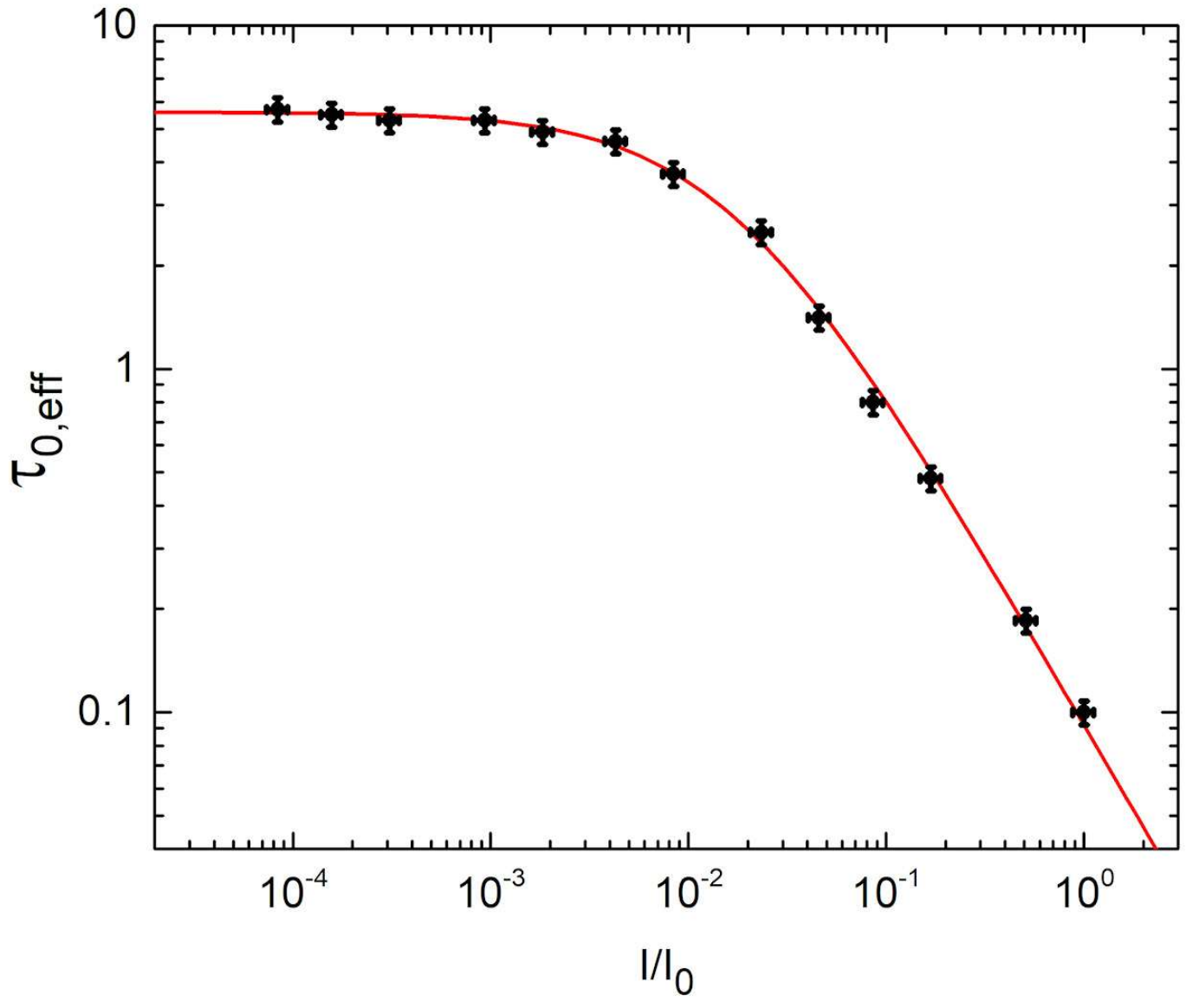

*Figure 13. The data points in this graph give the measured effective optical depth $\tau_{0,eff}$ of the G3 absorption feature as a function of the laser attenuation factor (equal to the optical filter transmission). The red line is a fit to the data using Equation (12) with $s_{max} = I_0/I_{sat} = 60$.*

## Modeling saturated absorption

It is tempting at this point to stop with the results shown in Figure 13, satisfied that the theory curve in Equation (12) nicely passes through the measurements. A closer inpection, however, reveals that this basic theory is not telling the whole story. An important clue that this is the case comes from considering an independent estimate of the model parameter $s_{max} = I_0/I_{sat}$.

Figure 14 shows our laser beam projected onto a paper card printed with various scale lines, revealing the diode laser's elliptical beam shape. From this image we estimate a beam waist size of $w_x \approx 0.6$ mm and $w_y \approx 1.0$ mm, and combining these with the measured laser power of 6.4 mW (minus 10% for reflection losses by the entrance window of the Rb cell) gives an intensity of $I_0 \approx 6100$ W/m$^2$ at the center of the beam. Using the known value of $I_{sat} \approx 16.7$ W/m$^2$ then gives $s_{max} \approx 365$, which is substatially higher than the fit value of in Figure 13. Put another way, accepting $s_{max} = 60$ from Figure 13 would require an overall beam height of $2w_y \approx 5$ mm, which is clearly incompatible with Figure 14. This discrepancy suggests that our simple picture of saturated absorption may indeed be too simple, and we could do better with a more careful theoretical model of the measurements.

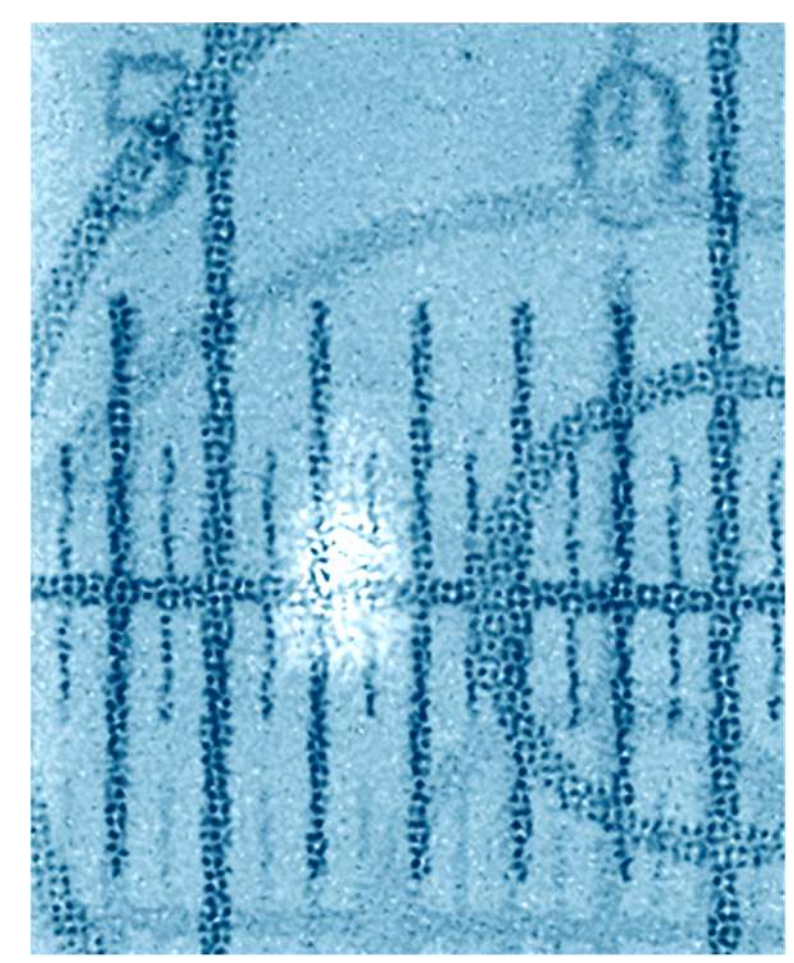

*Figure 14. This photo shows our laser beam projected onto a paper card with 1-mm scale markings (and smaller 0.5mm divisions).*

A bit of thought reveals two immediate shortcomings of the simple theory presented above. First, the laser beam intensity was assumed to be constant throughout the Rb vapor cell, and this is certainly not true if the cell has a high optical depth (which it does in this experiment). And second, the simple theory assumes a plane-wave electromagnetic field, while the laser beam has an elliptical Gaussian profile.

We can address the first shortcoming by dividing the Rb vapor cell into small $dz$ slices along the optical axis and then calculating the change in laser intensity $dI$ within each slice. The simple saturated-absorption theory applies within each slice, and the overall absorption can be obtained by numerically integrating the combined effect of all the slices.

The second shortcoming of the basic theory can be eliminated by simply assuming a collection of laser beams of varying intensity, following a distribution function that reproduces that of an elliptical Gaussian beam, again summing all the individual contributions numerically. Although producing these numerical models presents a nontrivial coding task, your favorite AI tool can do nearly all the heavy lifting for you, and we used ChatGPT to produce the models shown in Figure 15.

Walking through this figure, Model A is nothing more than the basic saturated-absorption theory described by Equation (12), setting $\tau_0 = 5.6$ to match the data at low laser intensities. Including attenuation in the Rb cell gives Model B in Figure 15, using the measured $\tau_0 = 5.6$ once again as a model input. Here we see that Model B agrees with Model A when $I \to 0$ and when $I \to \infty$. The low-$I$ agreement makes sense because there are no saturation effects in this limit. so all the models converge to $\tau_0 = 5.6$ when $I \to 0$. In the opposite limit, the two models agree when $I \to \infty$ because there is essentially no attenuation of the laser beam intensity in that limit. The detailed shape of Model B between these limits depends on $\tau_0$.

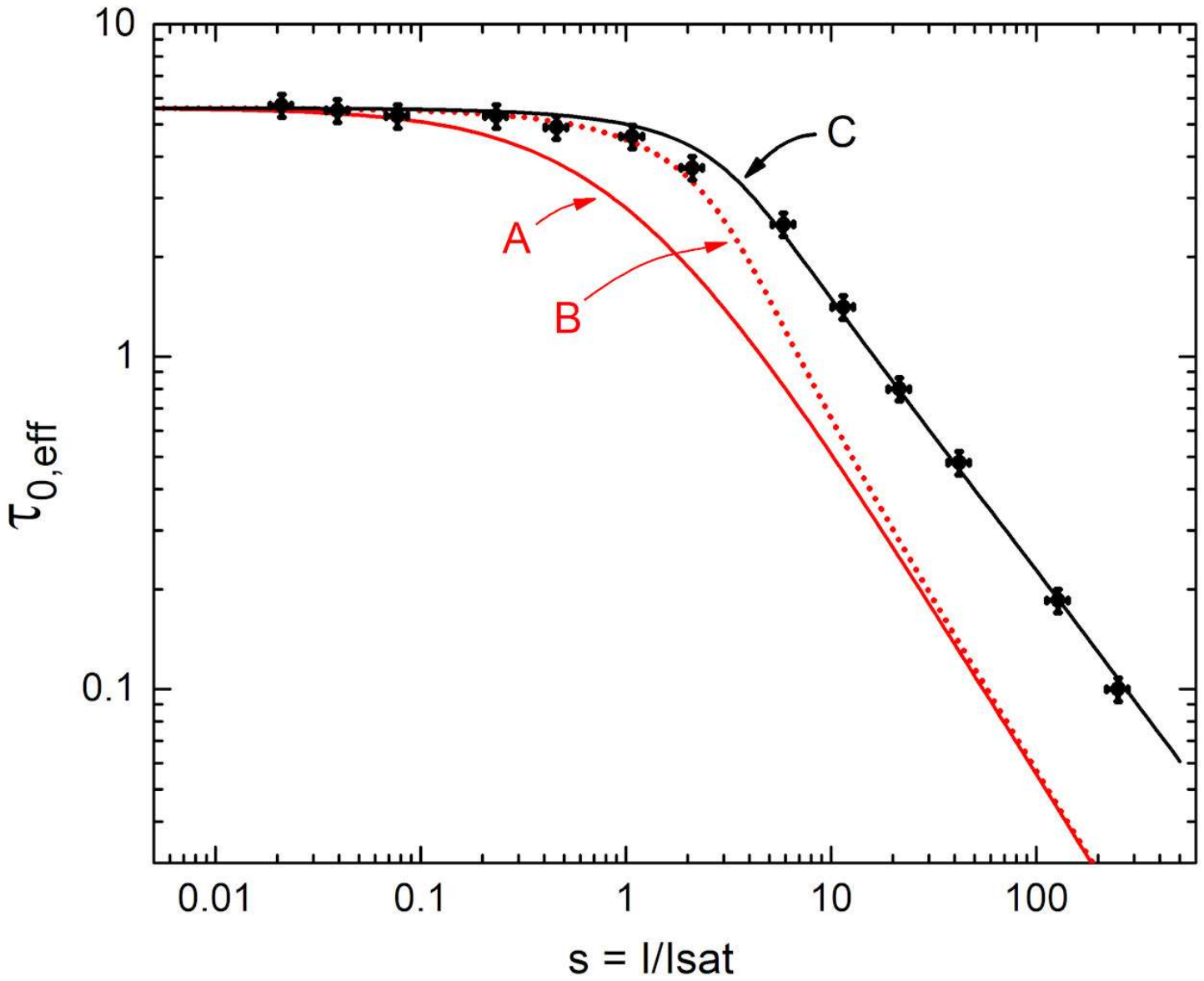

*Figure 15. This plot shows the same data points as in Figure 13 but with different theoretical models. Model A is the basic theory described by Equation (12), Model B includes beam attenuation within the Rb cell, and Model C includes both attenuation and a Gaussian laser beam profile. The experimental data were renormalized to fit Model C by setting* $s_{max} = 250$.

Model C starts with Model B and adds the Gaussian shape of the laser beam. The resulting curve depends only on $s_{max}$ at the beam center and is independent of the size or ellipticity of the beam profile. Having calculated these three models using $\tau_0 = 5.6$ as the only input parameter, we then adjusted the data points by converting the experimentally known $I/I_0$ to the theoretical parameter $s = I/I_{sat}$ by adjusting $s_{max}$ by fitting the data points to Model C. In a roundbout fashion, this data analysis procedure yielded an experimental measurement of $s_{max} \approx 250$ based on our theoretical Model C of the saturated absorption. This measured value of $s_{max}$ is still less than our independent measurement of $s_{max} \approx 365$ from Figure 14, but there is considerable measurement uncertainty in the latter number. There is much one could do to improve this experiment (beginning with the purchase of a proper laser beam "profiler" to measure $s_{max}$ directly with good precision), but taking such steps starts to go beyond the teaching lab and into serious quantitative experimental physics.

There are many potential teaching moments in this discussion. It is worthwhile to reiterate, for example, that stimulated emission is a purely quantum-mechanical effect with no classical analog. Thus this relatively basic lab experiment provides a clear demonstration of a fundamental aspect of how electromagnetic radiation interacts with atomic transitions. The phenomenon also makes many connections to modern AMO physics.

If one wants to focus mainly on the physical phenomenon of saturated absorption and its significance, then stopping at the result in Figure 13 might be desirable from a pedagogical perspective, even if the quantitative outcome is a bit deceptive. Sometimes an ounce of deception is worth a pound of explanation. Taking the next steps to further develop the model and better explain the data is laudable, but this path is also somewhat confusing and may distract students from the core physical principles in the lesson.

On the other hand, the teaching lab is about more than the theory that underlies the experiments. Other lessons may include: 1) real experiments in the lab are never as neat and tidy as theory courses suggest, 2) the simplest theory might not tell the whole story behind a complex phenomenon, and 3) experimental physics is often about model-building – developing a bridge that connects abstract theoretical concepts and real-world experimental measurements. If one's teaching-lab goals include an introduction to the substance and subtleties of experimental physics, then a walk through the details of this seemingly simple experiment might well be a worthwhile learning experience.

## Calibrated Rb absorption spectrum

In our first observations of the Rb absorption spectrum described above, we glossed over numerous important details related to the experiment itself as well as the Rb atomic structure. So let us now return to this basic measurement and examine it with more care. Our first step is to add an optical resonator (a.k.a. optical cavity) to measure the laser sweep, following the optical layout in Figure 16. The 780 nm resonator (also a Teachspin product) consists of a matched pair of curved, ~99% reflecting mirrors in a "confocal" arrangement, where the mirror spacing is equal to the radius of curvature of both concave mirrors. The optical physics describing the cavity transmission is outside the scope of this paper, so we state without explanation that the cavity transmission consists of a series of sharp peaks separated in frequency by the *free-spectral-range*

$$\Delta f_{FSR} = \frac{c}{4L} \tag{13}$$

where $c$ is the speed of light and $L$ is the mirror separation. With the Teachspin cavity $L \approx 20$ cm and $\Delta f_{FSR} \approx 375$ MHz.

Unlike the signal going to PD1, the laser beam going into the cavity must be carefully aligned to obtain sharp transmission peaks. The beamsplitter (Thorlabs BSS11) and folding mirror (Thorlabs BB1-E02) must align the beam so it hits the center of the first cavity mirror (2 degrees of freedom) and enters along the cavity axis (another 2 degrees of freedom). Also the cavity length must be carefully set to produce a confocal cavity.

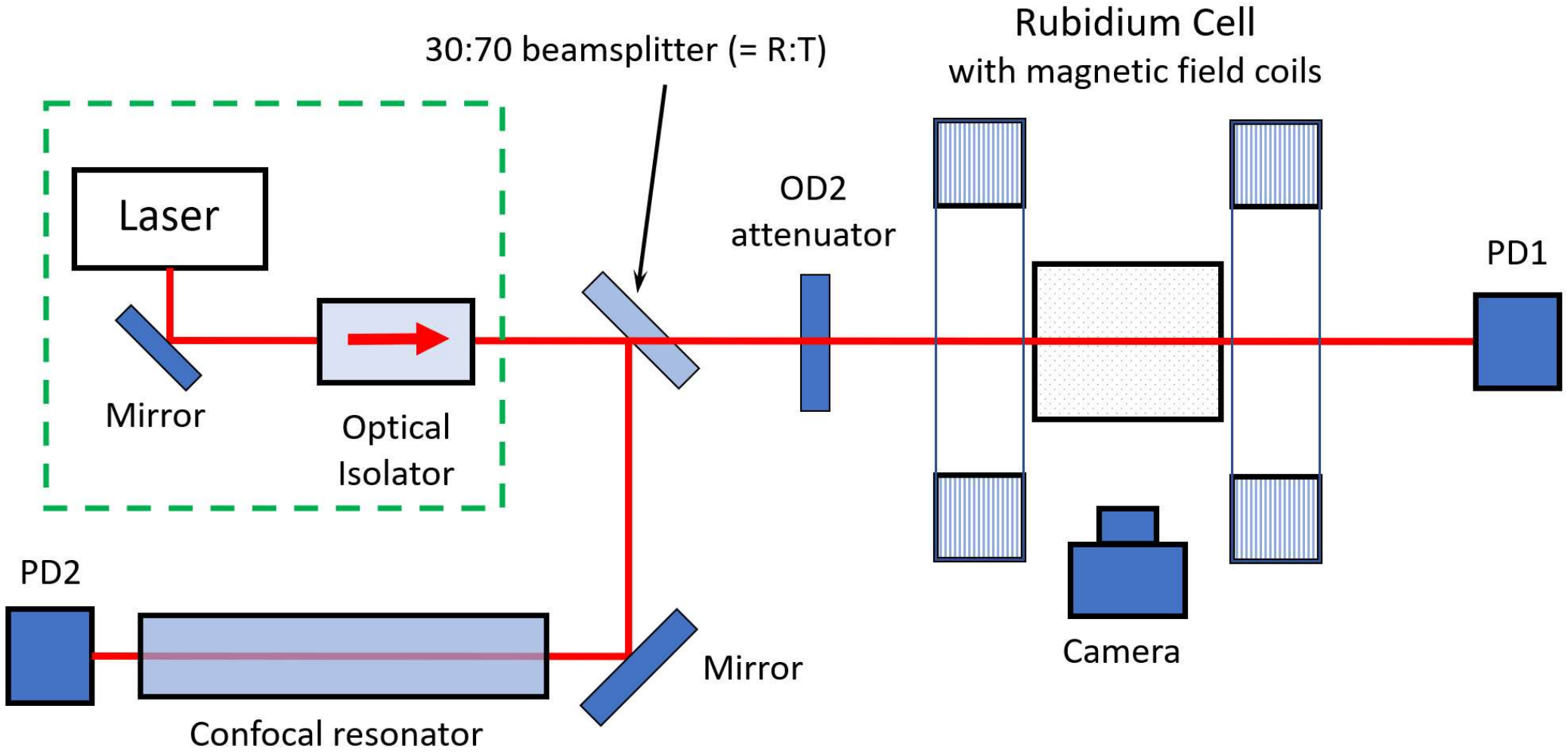

*Figure 16. We used this optical layout to produce a Rb absorption spectrum with a calibrated frequency scan.*

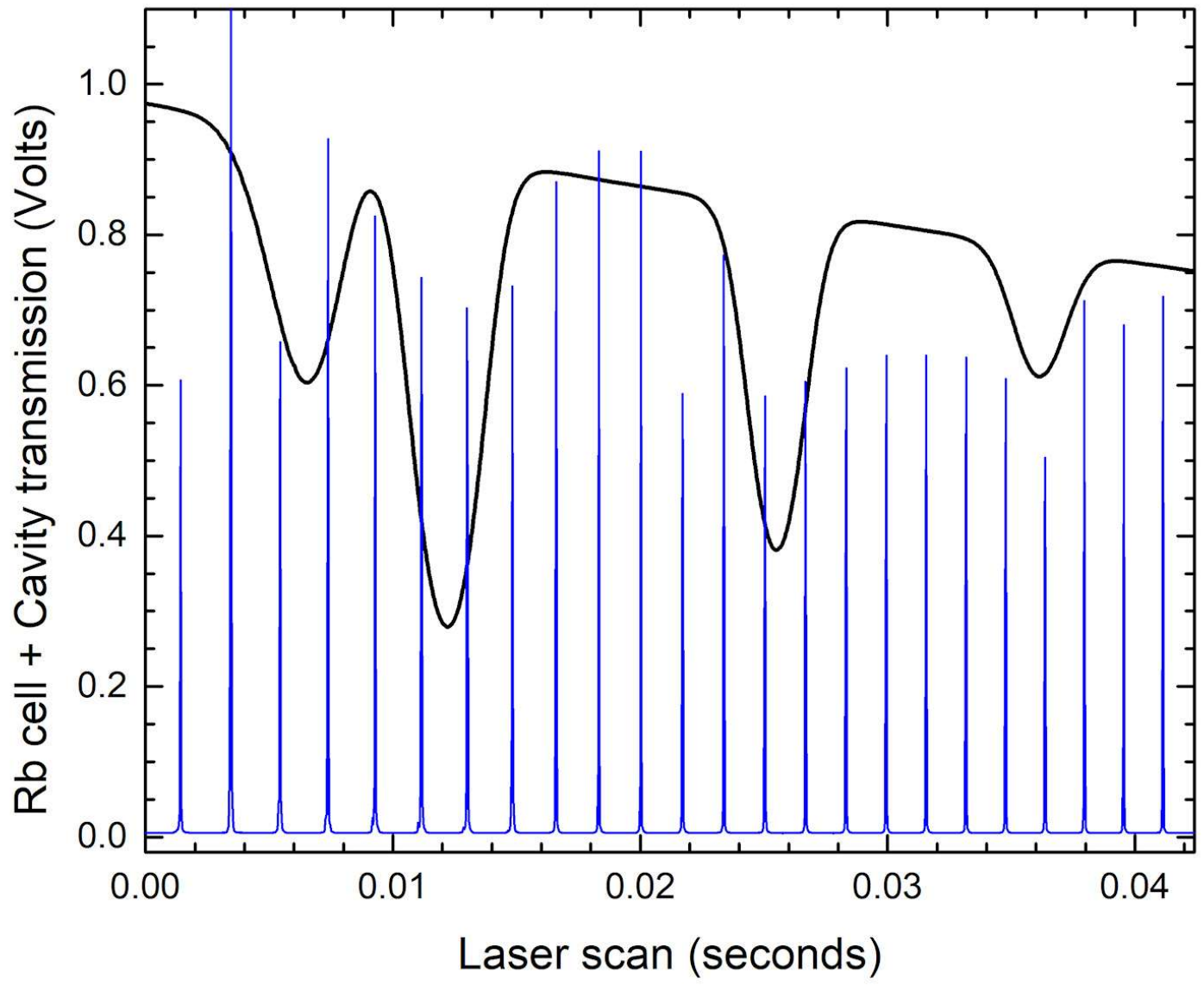

*Figure 17. The black line in this graph shows the Rb absorption spectrum recorded by PD1 in Figure 16, and the blue line shows the cavity transmission spectrum recorded by PD2. The series of sharp cavity peaks provides a set of frequency markers separated by $\Delta f_{FSR}$ that allows us to calibrate the laser frequency scan.*

With proper alignment, Figure 17 shows digital oscilloscope traces taken with this optical layout, and Figure 18 shows the measured positions of all the cavity peaks. In both plots one can see that the spacing of the cavity peaks is not constant across the scan, reflecting a nonlinearity in $f_{laser}(V)$ – the laser frequency as a function of the voltage that simultaneously drives the PZT and laser current (see Figure 5 and the accompanying discussion). Diode laser design is generally somewhat complex, making it difficult to produce a perfectly linear frequency scan. In contrast, the optical cavity is simple and stable, so the cavity peaks provide an accurate "ruler" for measuring the laser frequency scan. By using this ruler, we can convert the Rb spectrum $I(t)$ (as a function of the scan time measured by the oscilloscope) to a proper spectrum as a function of laser frequency $I(\Delta f)$. Note that the cavity does not provide a measurement of the absolute laser frequency, but $I(\Delta f)$ will let us measure the spacing between absorption dips.

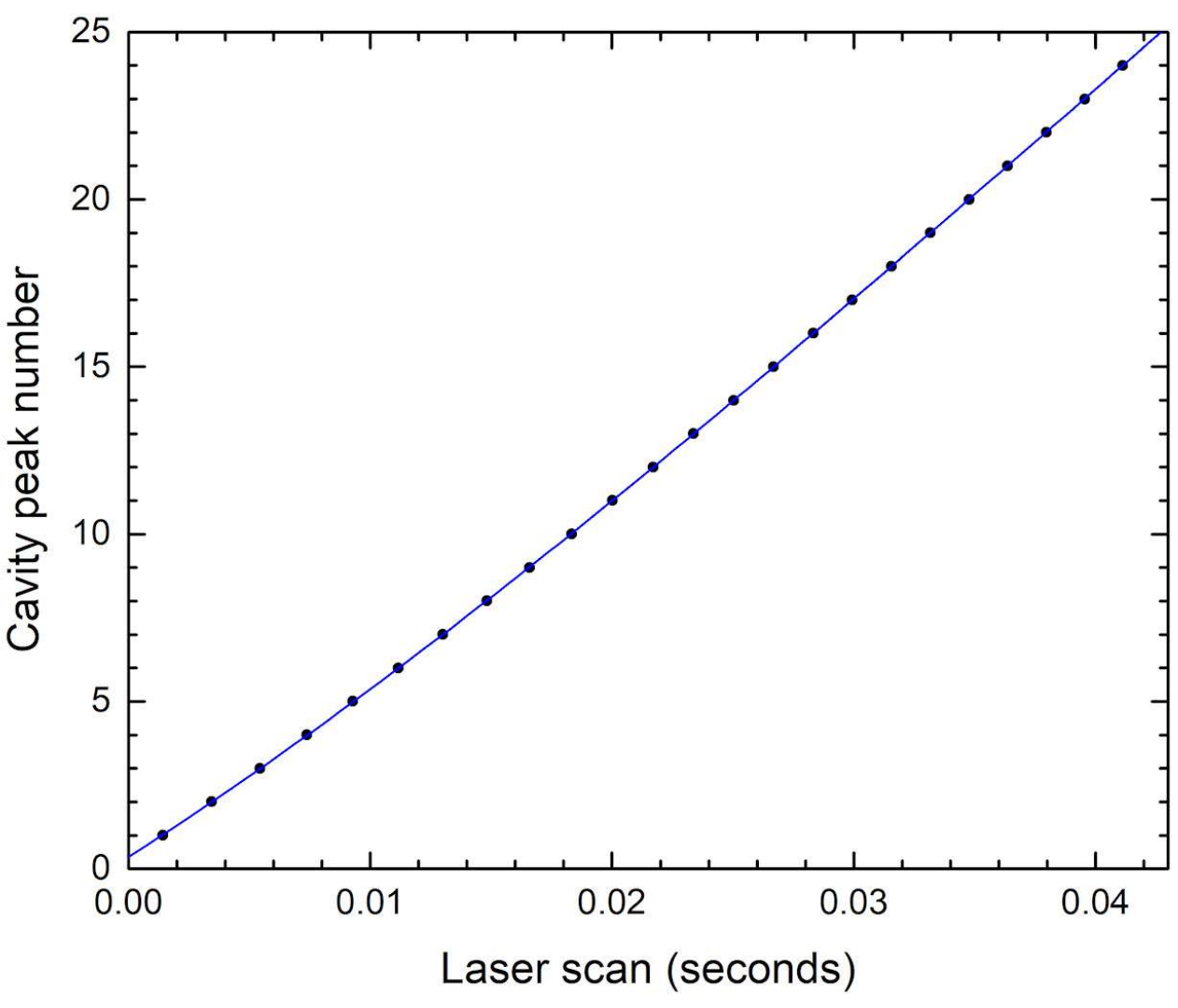

*Figure 18. The data points on the right show measurements of the positions of each of the cavity peaks in Figure 17, assigning each cavity peak and index number $m$. The laser frequency is then given by $f_{laser} = f_0 + m\Delta f_{FSR}$, where $f_0$ is an unknown constant. Using a fit line $m(t)$, we can accurately convert the oscilloscope scan time to laser frequency.*

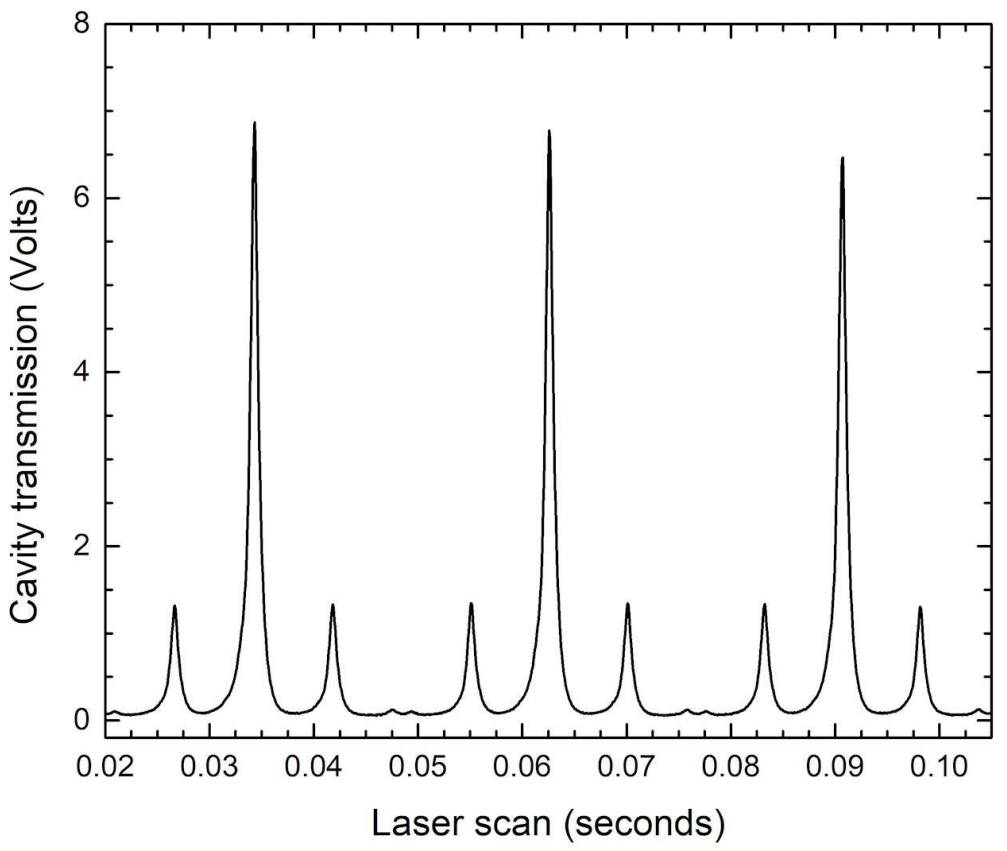

*Figure 20. For this measured cavity scan, we used a 100 MHz sine-wave signal to modulate the laser frequency, thus adding 100-MHz sidebands to the cavity peaks. By measuring the positions of these sidebands relative to the main cavity peaks, we can measure $\Delta f_{FSR}$ relative to the sine-wave frequency, ultimately giving $\Delta f_{FSR} = 375.45 \pm 0.5$ MHz.*

To complete this diversion into laser frequency metrology, we note that the cavity length $L$ is difficult to measure directly with curved mirrors, and this introduces a significant uncertainty in $\Delta f_{FSR}$. To reduce this uncertainty, we modulate the laser frequency to add sidebands to the cavity peaks, as illustrated in Figure 20. The sidebands thus provide a "ruler within the ruler" to further calibrate the spacing between the cavity peaks. The sidebands were created by modulating the laser current using a signal generator going into a supplied modulation port in the Teachspin system. Because even an inexpensive signal generator has an absolute frequency accuracy of a few parts per million, the sidebands provide our ultimate frequency reference. Our analysis is mainly limited by our ability to measure the peak positions from the oscilloscope data, and a close look at the various measurement residuals indicates that our determination of the frequency axis is accurate to <0.05 GHz across the spectrum.

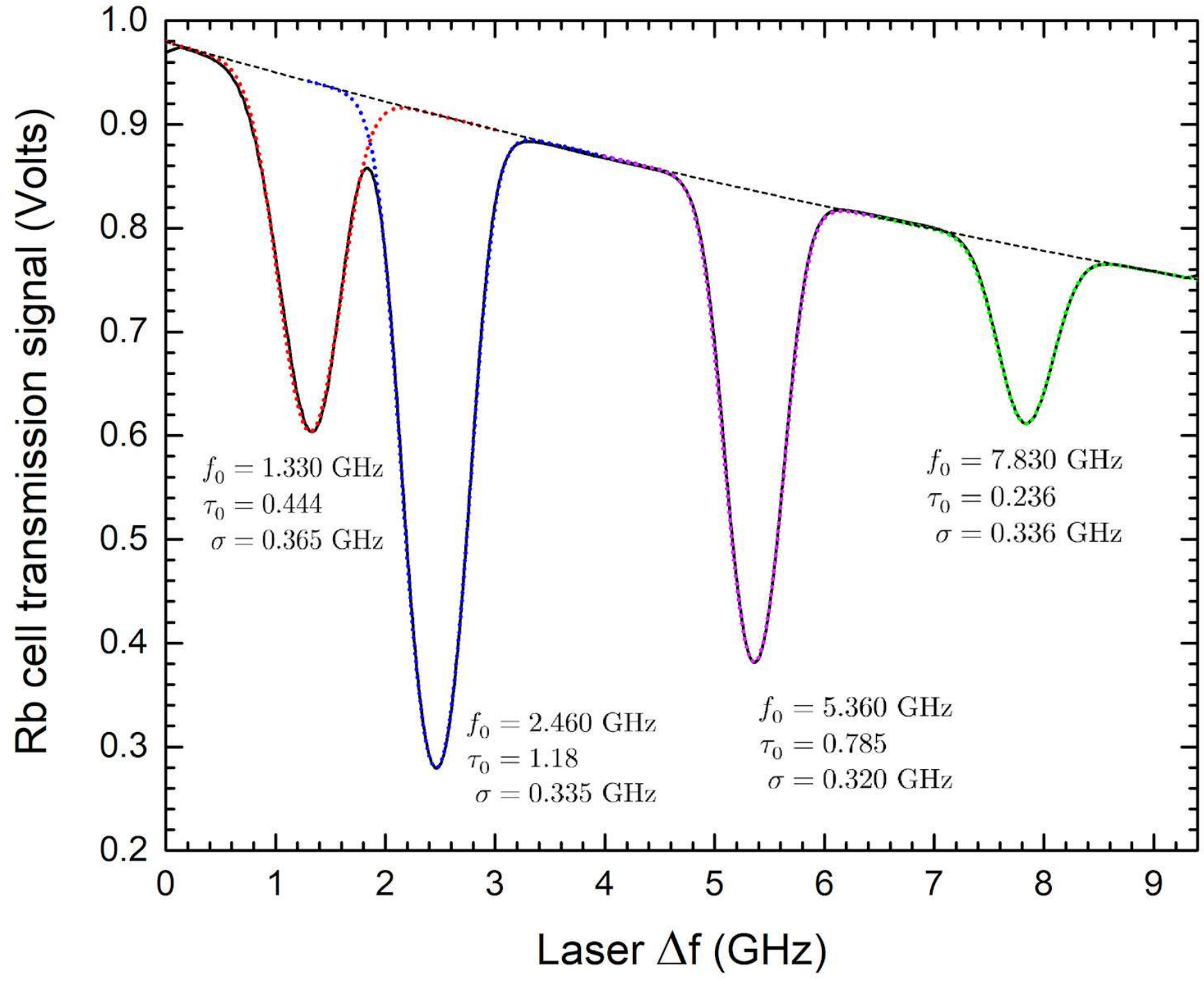

*Figure 19. This plot shows the same Rb absorption spectrum shown in Figure 17, along with a series of chi-by-eye fits to the individual dips. We estimate frequency uncertainties of about ±0.02 GHz for all the fits, although the G1 dip has a noticeable asymmetry compared to the assumed Gaussian profile. From these fits, we estimate a spacing of 2.90±0.05 GHz between the Rb85 dips and 6.50±0.05 GHz for Rb87. These numbers are not equal to the ground-state hyperfine splittings because of different transition probabilities to the manifold of excited states.*

Figure 19 shows the Rb absorption spectrum as a function of the laser frequency after this calibration, along with fits to the four absorption dips. Uncertainties in the fit frequencies for each of the four dips were estimated by adjusting this fit parameter and zooming in to each dip to see when the frequency was clearly too low or too high. From this qualitative assessment of the fit parameters, we measured a spacing of 2.90±0.05 GHz between the Rb85 dips and 6.50±0.05 GHz for Rb87, in general agreement with numbers that have been reported in the literature [2008Sid]. The frequency of G1 dip was likely influenced by small systematic errors, as the G1 dip profile shows a significant asymmetry in relation to the assumed Gaussian Doppler profile.

We also examined the use of nonlinear fitting algorithms in this analysis, first normalizing the transmission spectrum and then assuming a global fit with the functional form

$$T(f) = \prod_{X=1}^{4} \exp\left[-\tau_X \exp\left(-\left(\frac{f-f_X}{\sigma_X}\right)^2\right)\right] \tag{14}$$

which includes 12 fit parameters for the four absorption dips. Because the dips are so cleanly separated, the fit always converged quickly, and Figure 21 shows one example. [Pro tip: if you provide a screen grab of the above equation along with your .csv data file, the AI should have no trouble interpreting the image and performing the nonlinear fit.] In our experience, nonlinear fitting algorithms tend to return wildly optimistic error estimates when applied to large numbers of data points, and our spectroscopic data were no exception. In this example, the algorithm returned an uncertainty of ±0.0004 GHz for the G1 dip and even smaller uncertainties for the other dips.

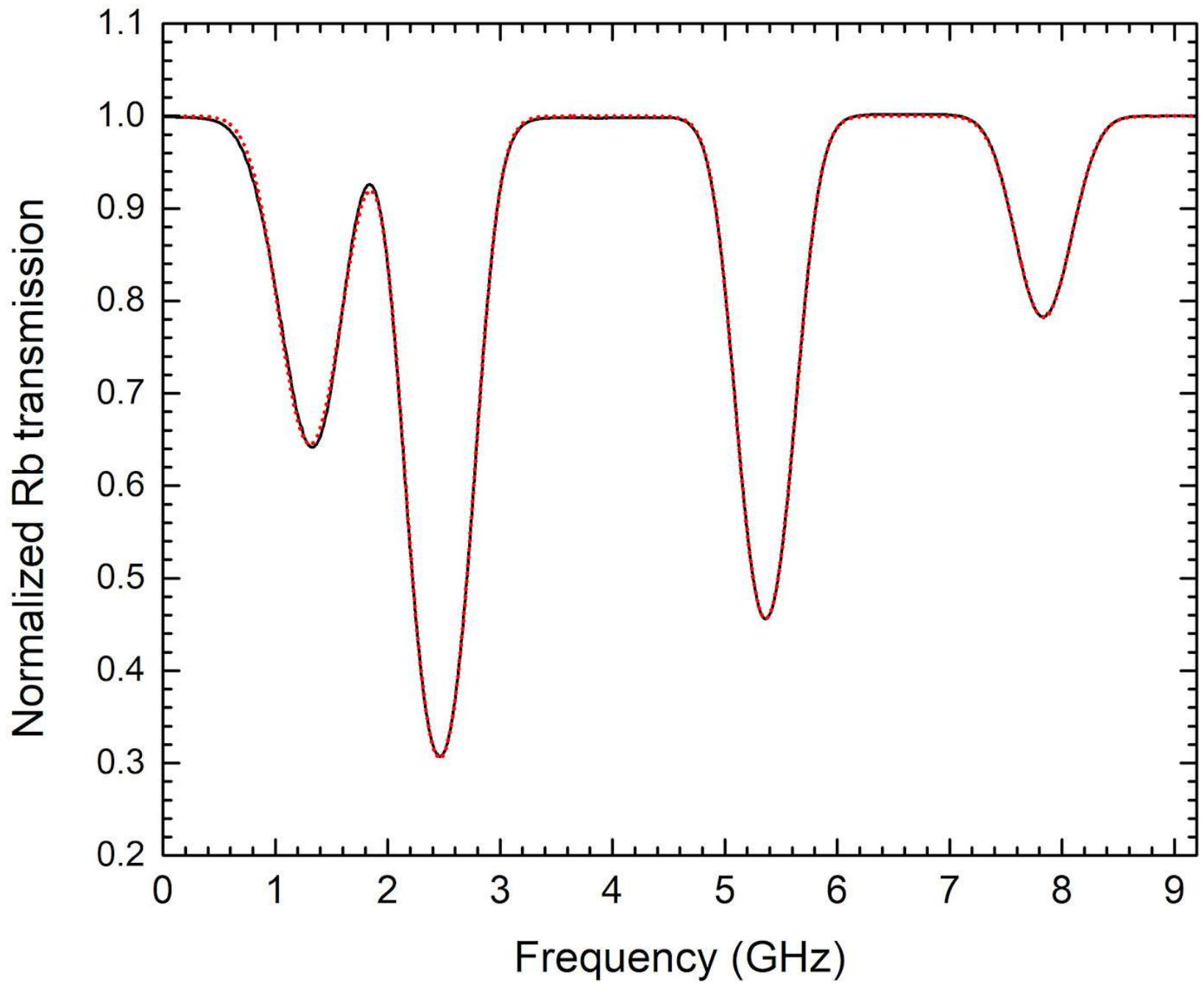

*Figure 21. This plot shows a normalized transmission spectrum derived from the same data as in Figure 19 (black line) together with a global fit to the data using ChatGPT and Equation (14) (red line). Again, the G1 dip is a bit of an outlier, showing a greater dip asymmetry than the others. The fit parameters here are comparable to those in Figure 19, but the nonlinear fitting algorithm reported frequency uncertainties that were about 30x smaller than seemed reasonable.*

Although students generally balk at precision measurements and metrology, this exercise has some educational merits. First, ChatGPT produced an excellent nonlinear fit to these data with almost no user effort, and it clearly incorporated a sophisticated fitting algorithm that is quite expensive if purchased separately. Although these AI tools are still new in 2026, it is already obvious that they have tremendous potential for application in physics teaching labs. And second, this exercise makes it clear that nonlinear fitting algorithms are no substitute for common sense. While the tiny fit uncertainties no doubt have a solid basis in the underlying mathematics, they do not produce sensible measurement uncertainties. If one simply takes two spectra separated in time by a short period, the nonlinear fitting algorithms will yield frequency measurements that differ by many sigma.

## Spectral modeling

Although making a full spectral model of the Doppler-broadened Rb absorption spectrum is certainly beyond the scope of this paper, the results are informative. As described in [2008Sid, 2024Bal], the modeling process involves calculating the line strengths of all the allowed electric dipole transitions and combining them to form an overall absorption profile for a Doppler-broadened vapor. Figure 22 shows one model in the limit of low laser intensity (thus avoiding saturated-absorption effects and optical pumping of the hyperfine ground states), showing how each absorption dip represents a sum of the different contributing transitions. The precise spectrum one obtains in the lab will likely change slightly with the laser intensity, laser polarization, the Rb vapor density, and other details. In day-to-day operation, we usually specify that the G4-G1 spacing is 6.50±0.05, explain why this is less than the ground-state splitting, and then quickly move on to the next experiment.

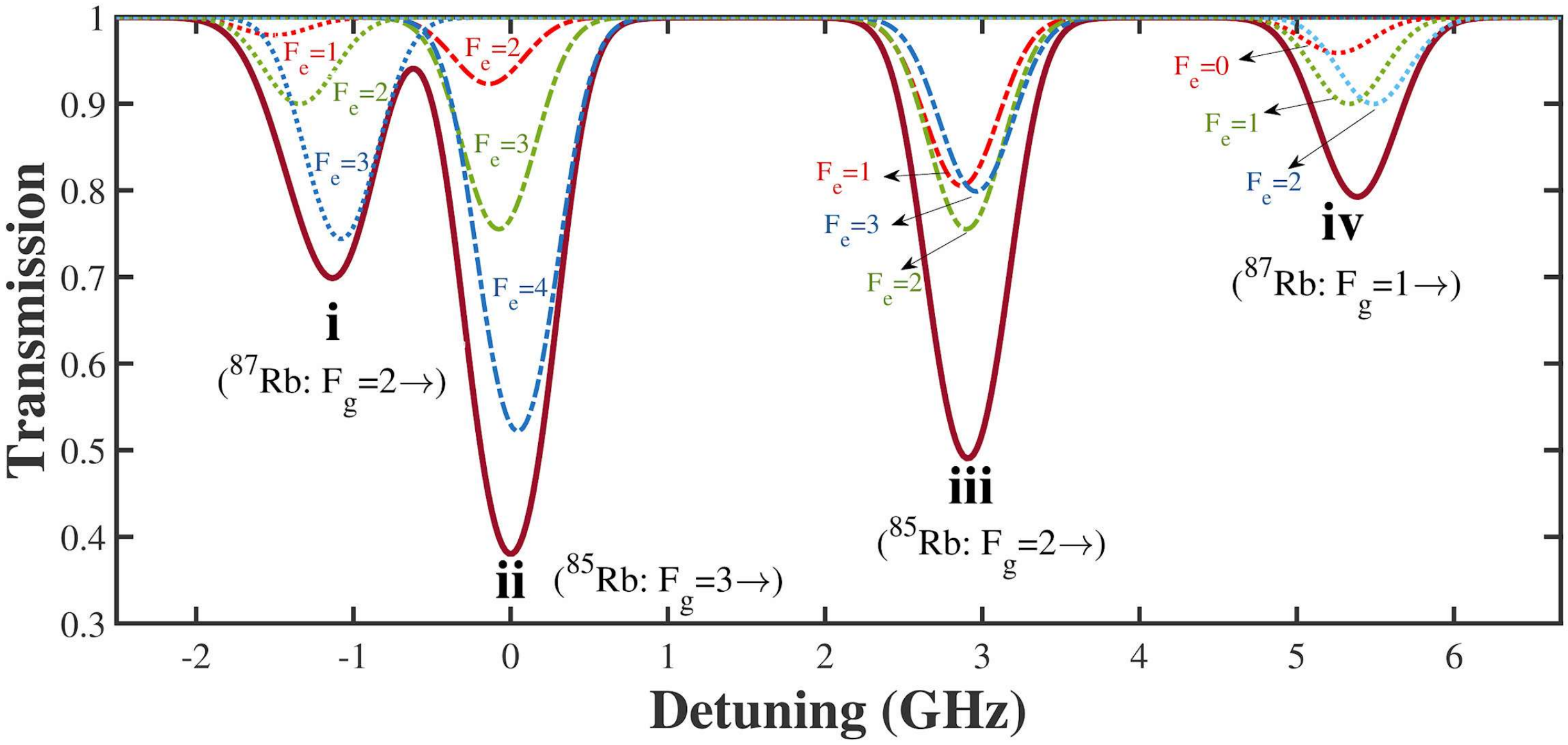

*Figure 22. This plot from [2024Bal] shows a theoretical model of the Rb absorption spectrum at 780 nm, including the individual contributions of all the allowed transitions. Here we see that the G1 dip is dominated by the highest energy F=2 to F=3 transition, while the G4 dip is dominated by lower energy transitions. Thus the G4-G1 dip separation of 6.50 GHz is lower than the Rb87 ground-state splitting of 6.83 GHz. This plot also explains why the G1 dip is especially asymmetrical in shape, while the G3 dip looks more like a clean two-level transition.*

## Doppler-free saturated-absorption spectroscopy

The technique of *Saturated-Absorption Spectroscopy* (SAS) is a fascinating and frequently used method for observing narrow atomic spectral features with a frequency resolution limited by the natural linewidth of the transition ($\Gamma \approx 6$ MHz for the rubidium D lines) even in the presence of much larger Doppler broadening ($\sigma \approx 315$ MHz in our Rb vapor cell). As can be seen in Figure 2, the enhanced spectral resolution obtained using SAS can allow us to resolve all the $P_{3/2}$ hyperfine levels, even though these levels were completely unresolved in the basic absorption spectra presented above.

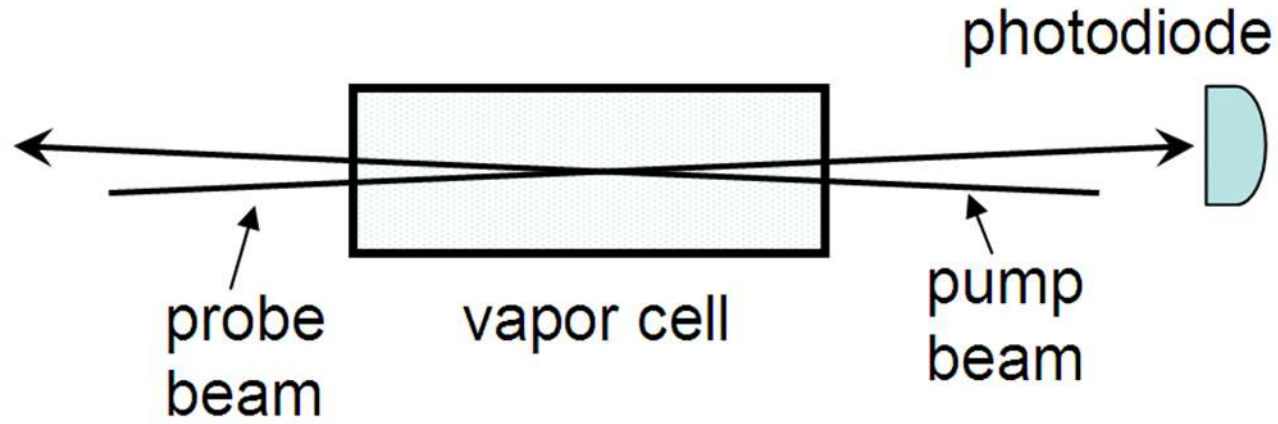

*Figure 23. The basic saturated absorption spectroscopy setup. The two laser beams are tilted here for clarity; in practice the beams are collinear and fully overlap.*

To see how SAS works, consider the experimental setup sketched in Figure 23. Two lasers are sent through an atomic vapor cell from opposite directions; one beam, called the "probe" beam, is very weak, while the other, called the "pump" beam, is strong. But both beams are derived from the same laser and therefore have the same frequency. As the laser frequency is scanned, the probe beam intensity is measured by a photodetector. If one has 2-level atoms in the vapor cell, this setup would yield spectra like those shown in Figure 24. The left sketch gives the probe beam absorption without the pump beam, showing usual Doppler-broadened absorption.

When the pump beam is turned on, the spectrum remains unchanged except when the laser frequency is tuned to the atomic resonance for zero-velocity atoms. At any other laser frequency, the probe laser excites atoms with some non-zero axial $v_z$, while the pump excites atoms moving at $-v_z$. Because those velocities represent different atoms, the probe absorption is unchanged by the pump beam whenever $v_z \neq 0$. When the laser is tuned to the center of the Doppler broadened dip, however, then the pump and probe beams both excite atoms with $v_z = 0$. The stronger pump beam saturates the transition, and this reduces the probe absorption in accordance with the saturated-absorption mechanism described above. Thus a narrow peak appears at the center of the absorption dip, and the width of the peak is roughly the natural linewidth $\Gamma \approx 6$ MHz.

The SAS picture becomes somewhat richer if we consider atoms with a single ground state and two excited states (typically an electronic level split by the hyperfine interaction), and this case is illustrated in Figure 25. Importantly, the separation of the excited states is less than the Doppler width in this sketch, which is true for rubidium. The peaks on the left and right in Figure 25 are the normal saturated absorption peaks at $f_1$ and $f_2$, where these are the two atomic resonance frequencies. If the two resonances have different strengths, then the center of the overall Doppler dip may be closer to one resonance than the other.

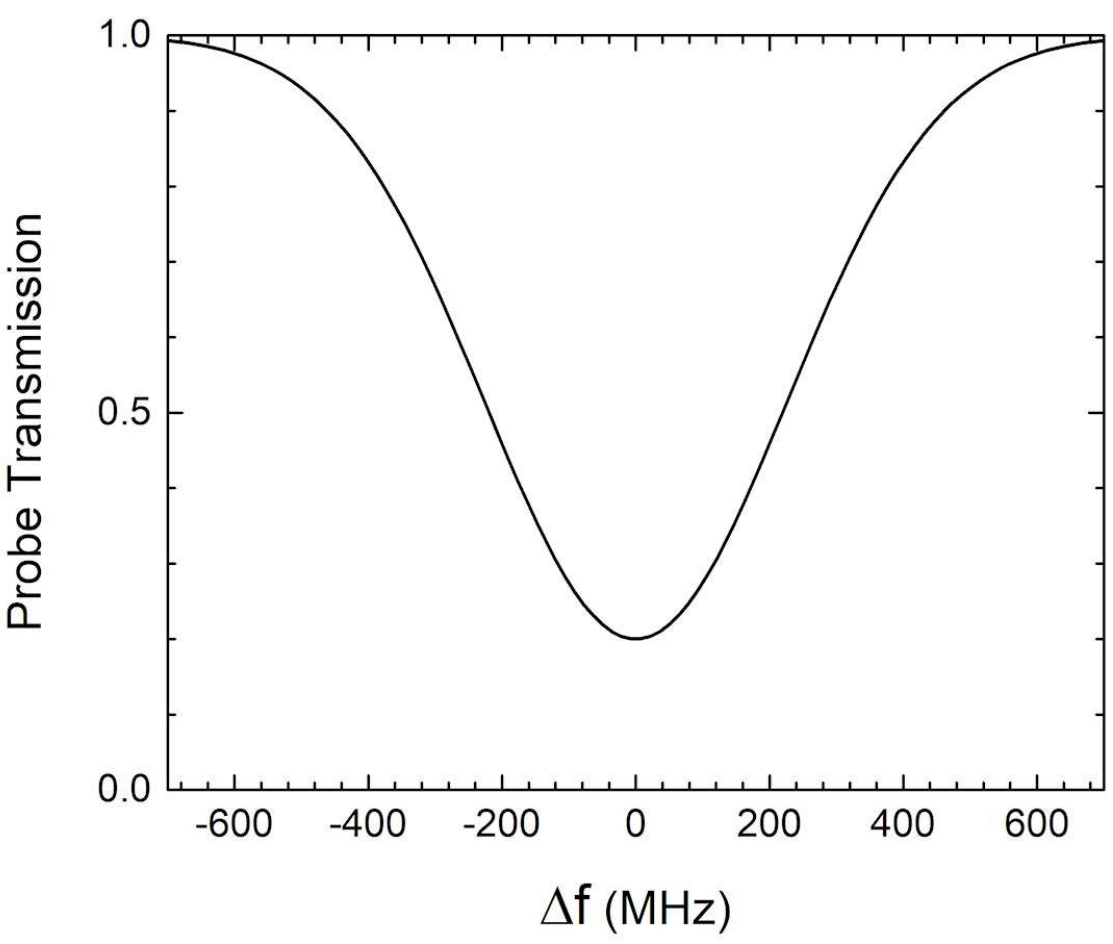

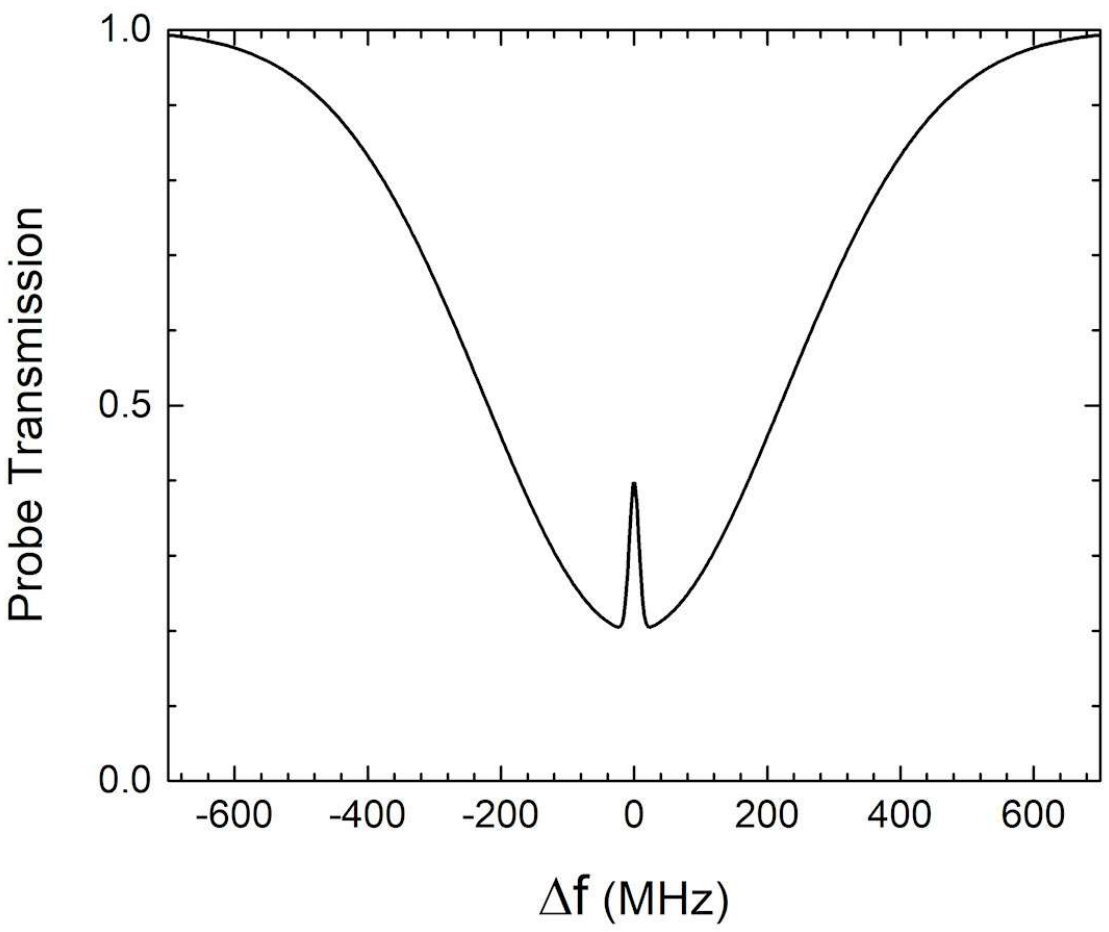

*Figure 24. These sketches show a qualitative model of saturated-absorption spectroscopy for 2-level atoms, illustrating the probe transmission without (left) and with (right) the pump beam. Because the pump and probe lasers with both excite the 2-level transition for zero-velocity atoms (and not for atoms with nonzero velocities), the probe absorption is diminished by saturated absorption only at the zero-velocity resonance frequency.*

The middle peak in Figure 25 is called a *crossover peak* or *crossover resonance*, and it arises from atoms moving at specific velocities where one laser is redshifted and hits the $f_1$ transition while the other laser is blueshifted and hits the $f_2$ resonance. It is an exercise for the student to show that this happens at velocities

$$v_z = \pm \frac{f_1 - f_2}{f_1 + f_2} c \tag{15}$$

and produces a crossover peak at $f_{crossover} = (f_1 + f_2)/2$. Once you convince yourself (and your students) that the basic SAS concepts make sense, the next step is to observe the phenomenon in the lab.

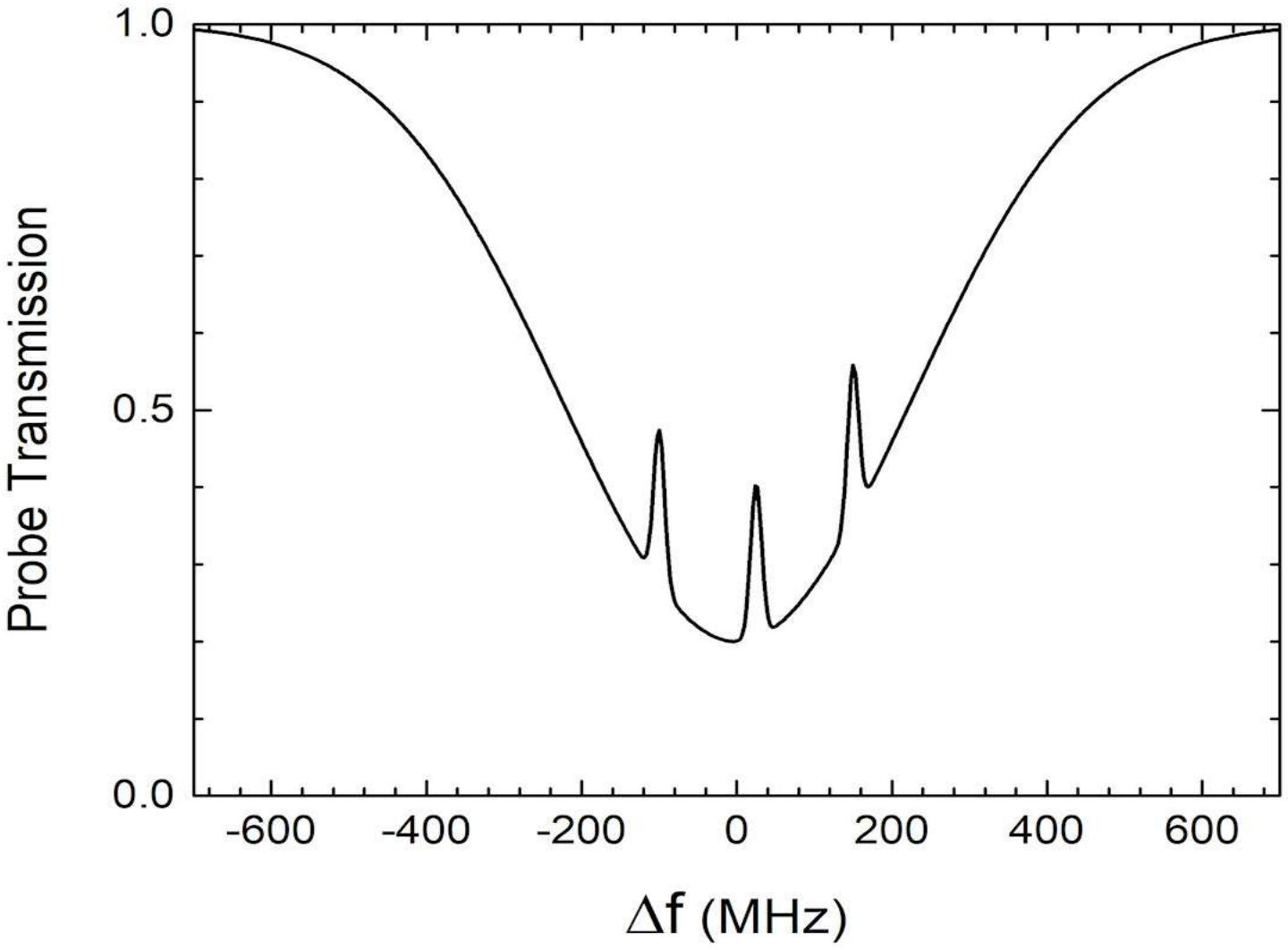

*Figure 25. Applying SAS to atoms with one ground state and two excited states yields two normal SAS peaks and an additional "crossover peak" at a frequency halfway between the two normal peaks. The crossover peak is caused by atoms moving as a velocity where one laser is redshifted and excites the lower-energy transition while the other laser is blue-shifted and excites the higher-energy transition.*

## Laboratory SAS observations

Figure 26 shows an optical layout that works quite well for performing Doppler-free saturated-absorption spectroscopy in the teaching lab. The rotatable polarizer in the pump beam is not essential in this layout, but it is useful for observing how the SAS peaks change with the pump intensity. On technical detail in this layout is that the beamsplitters must have the correct "handedness" to avoid blocking the transmitted beams. In Figure 26 the 30:70 beamsplitters (Thorlabs BSS11) have the same handedness (and are thus interchangeable), while the 50:50 beamsplitter (Thorlabs BSW11) has the opposite handedness. Students are generally unaware of vignetting issues in optics, and with these beamsplitters the correct way is the only way to arrange the components.

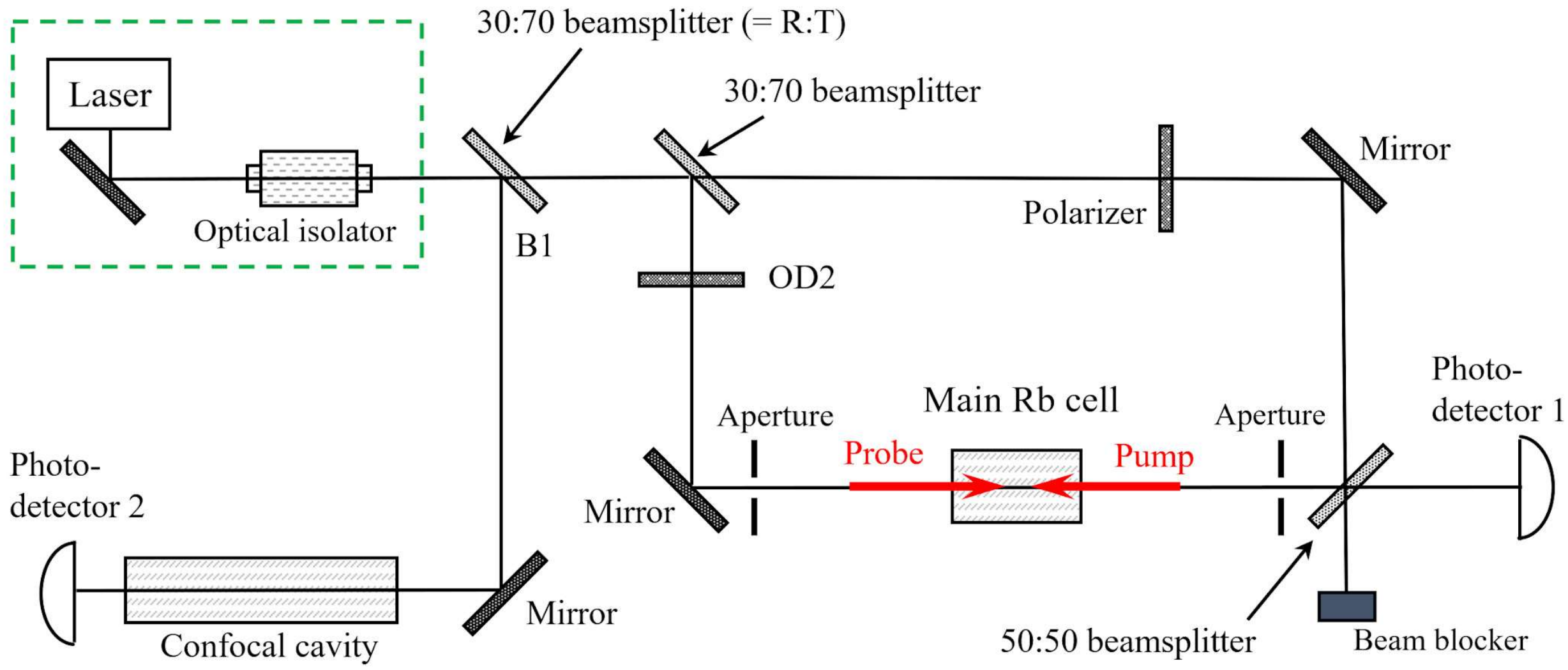

*Figure 26. This optical layout can be used to observe saturated-absorption spectroscopy and to measure the hyperfine splittings of the Rb $P_{3/2}$ levels for both isotopes. The polarizer in the pump beam serves as a variable attenuator, allowing one to observe how the SAS features change with pump intensity.*

We also added a pair of fixed 3-mm apertures because we found that students often struggle with the SAS alignment process. While the previous experiments were rather straightforward to set up, in this setup the probe and pump beams must overlap rather precisely to produce a good signal. If the apertures are not present, the alignment process can be frustrating for all involved. Here the two apertures are fixed to the table using bolts for which the students do not have a corresponding wrench.

The source of these alignment problems stems from the many degrees of freedom (d.o.f.) in the problem. Aligning the probe beam, for example, requires that the beam be positioned to pass through the 1st aperture (2 d.o.f.) with an angle that sends it through the 2nd aperture (an additional 2 d.o.f.). The problem is like that with the optical cavity discussed above, and again two mirror mounts (each with two adjustment knobs) provide the needed degrees of freedom. And the same goes for the pump beam, requiring two additional mirror mounts to provide the necessary d.o.f. Inexperienced students can usually handle 2-3 d.o.f. on their own, but eight is a lot to manage. In our experience, the fixed apertures were quite helpful in guiding students toward the correct alignment strategy. It is also useful to block one beam while the other is being aligned, with the OD2 filter removed during the initial alignment process.

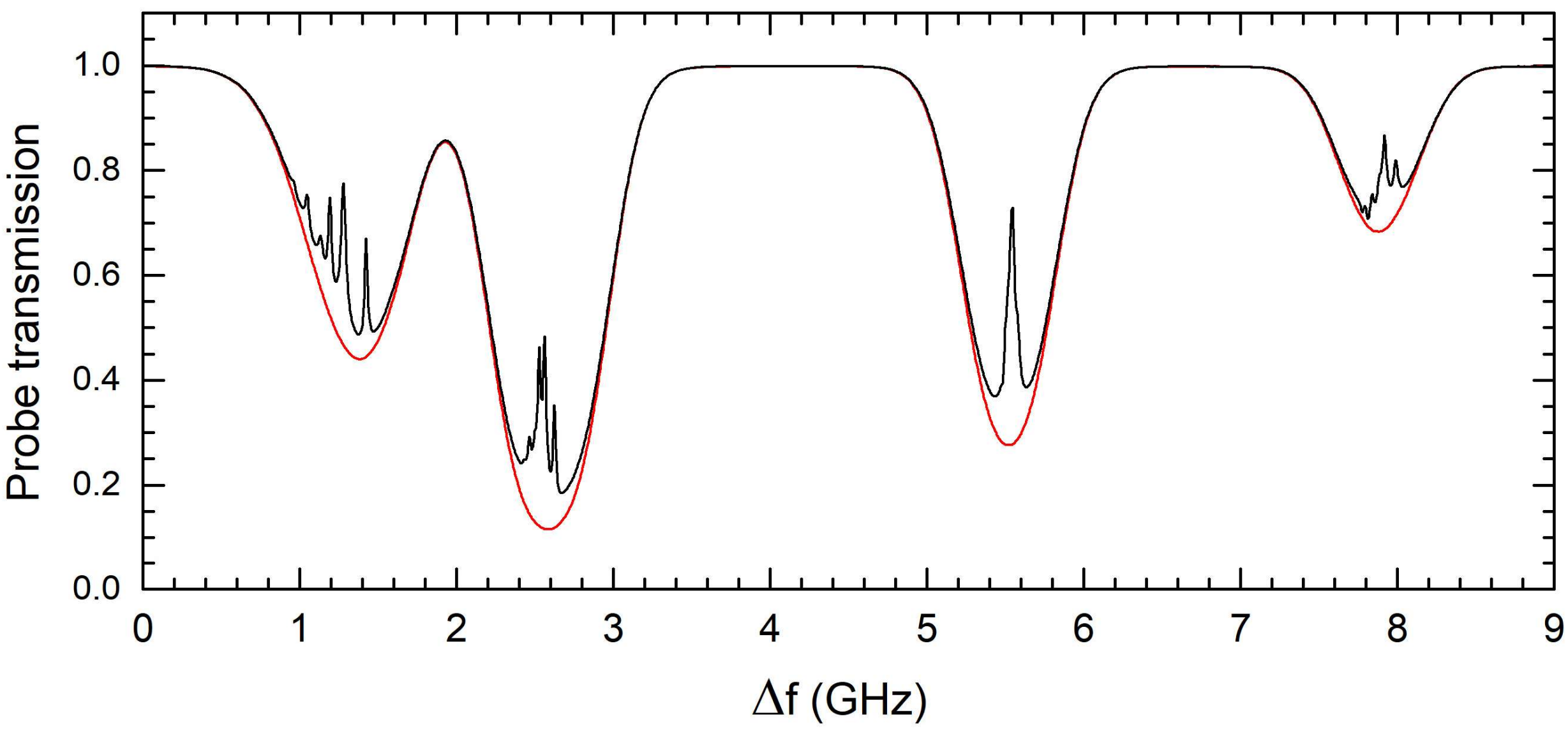

*Figure 27. These SAS data show the normalized probe transmission with the pump beam on (black line) and with the pump beam blocked (red). The Rb cell temperature was 45C, and some care was taken to reduce unwanted scattered light from the (very bright) pump beam entering PD1.*

Once the optics have been roughly aligned, one can then usually observe the SAS peaks and tweak the alignment further to obtain the best signal. Here again, we have found that inexperienced students are often unaware that optical alignment is often an iterative process, requiring a lot of back-and-forth adjustments of the various components to optimize the signal. Many assume that the experiment will provide its best signal immediately upon placing the optics on the table. Such has been their experience with consumer electronics and other products requiring user assembly. The lesson for educators (in our opinion) is that laboratory instruction is becoming a much more important part of the physics curriculum, as it provides some much-needed experience with complex tasks that require careful attention to setup and alignment.

Figure 27 show some saturated-absorption-spectroscopy data taken using the optical layout in Figure 26. Note that the G1 SAS peaks are fully resolved in this spectrum while the G3 peaks are completely unresolved, reflecting the difference in the $\mathrm{P}_{3/2}$ upper-state splittings for the two cases. The G1 dip presents the clearest SAS signal, and Figure 28 shows some close-up data that better illustrates the three normal SAS peaks along with three crossover peaks.

Using the transmission peaks from the confocal cavity to calibrate the laser sweep, it is possible to measure the hyperfine splitting of the $P_{3/2}$ upper states, specifically the 266.6 and 156.9 MHz level separations shown in Figure 2. The measurement accuracy is mainly limited by one's ability to determine the positions of the spectral peaks, which can be done to $\pm 5$ MHz or better if one is careful. If you do have not purchased the confocal cavity, another option is to modulate the laser current to produce sidebands on the spectrum, as illustrated in Figure 29. This technique is a bit

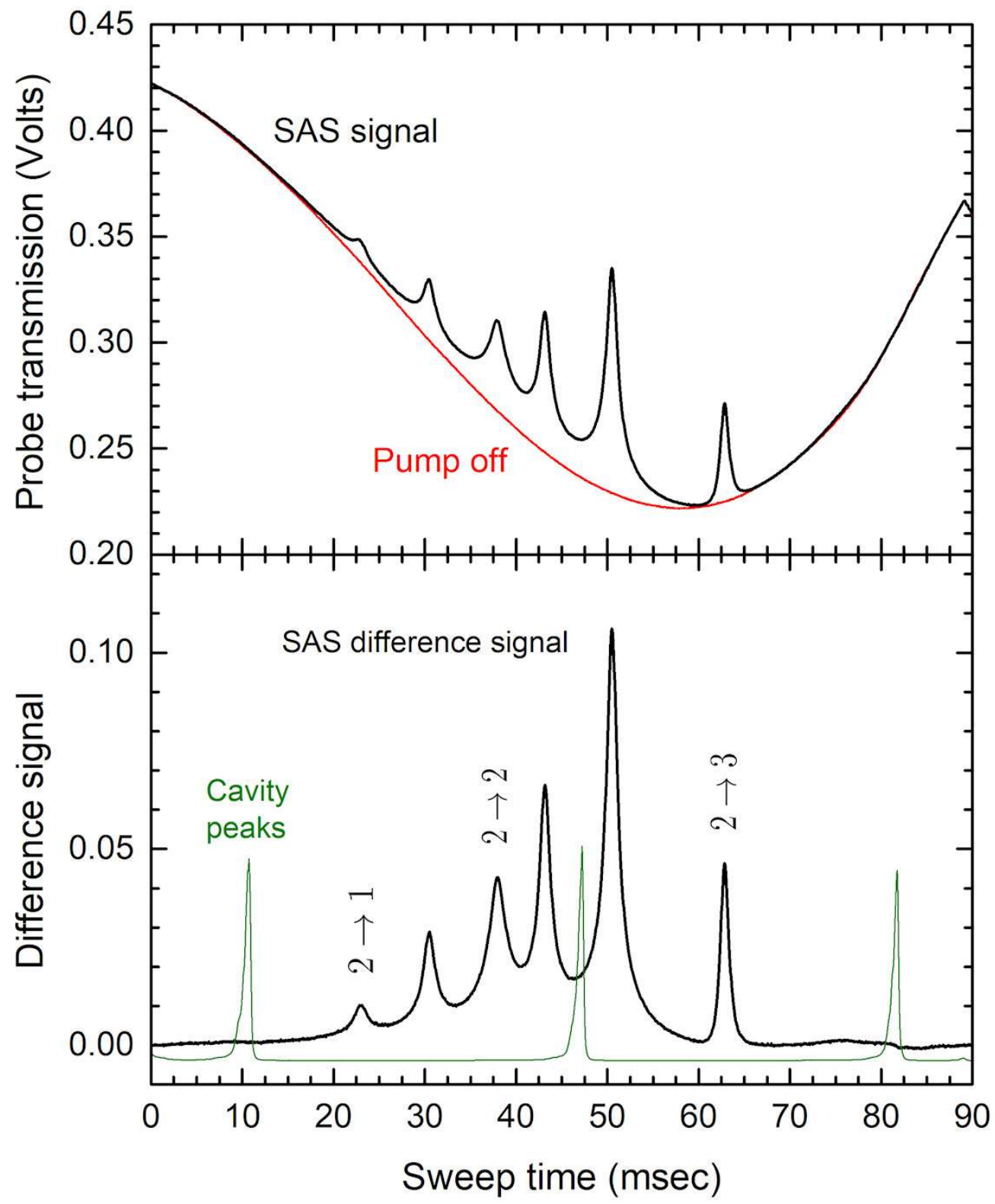

*Figure 28. The top plot on the left shows the normalized SAS probe transmission with the pump beam on (black line) and with the pump beam blocked (red). The Rb cell temperature was 45C, and some care was taken to reduce unwanted scattered light from the (very bright) pump beam entering PD1.*

*The lower plot shows the difference of the pump-on and pump-off signals, together with a spectrum from the confocal cavity (green line). Here the normal SAS peaks are labeled by their corresponding transitions, and the unlabeled peaks are crossover resonances.*

more confusing than using the cavity peaks, but it also bypasses the need to separately calibrate the cavity length, thus allowing a somewhat more direct measurement of the splittings.

To close out this section, Figure 30 shows SAS data focusing on all four Rb absorption dips. Note that observing clear SAS peaks in all cases required quite a lot of trial-and-error adjusting the intensities and even polarizations of both the pump and probe beams. In our experience, students cannot produce such clean spectra in a short period of time, but knowing that such result are possible with this apparatus motivates them to produce substantially better spectra than they would have done in the absence of that information.

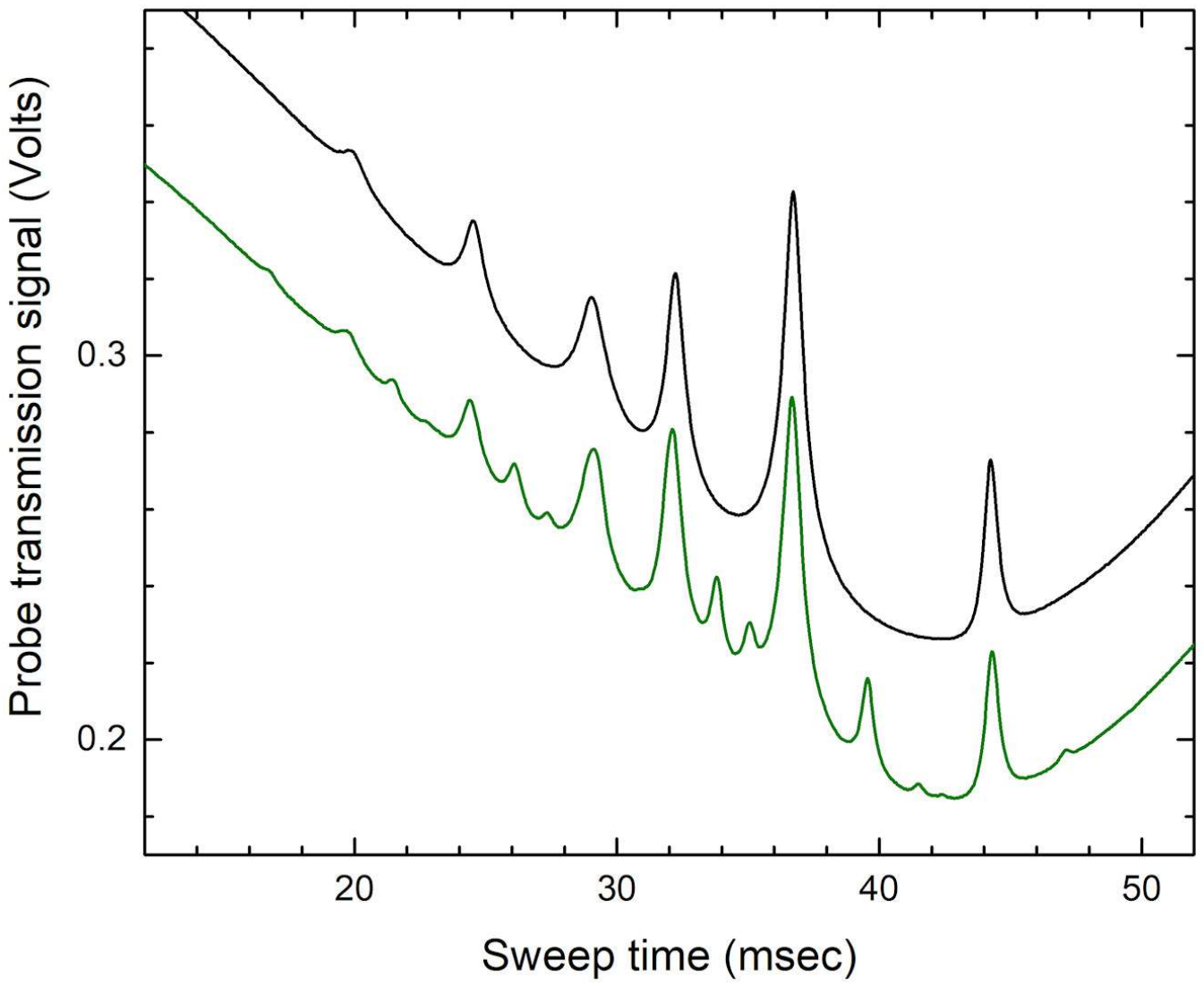

*Figure 29. The top spectrum on this plot shows a measured SAS data of the G1 dip, while the bottom trace shows the same spectrum but with 100 MHz sidebands on all the peaks. Although the spectrum is a bit confusion-limited with overlapping peaks, it is straightforward to disentangle the data sufficiently to measure the upper-state hyperfine splittings with considerable accuracy.*

While the positions of the SAS peaks are set by the Rb hyperfine splitting, the amplitudes of the peaks (including their sign – it is possible to observe SAS dips as well as peaks) are a complex function of the various transition probabilities and level populations, the latter often influenced by optical pumping. An excellent description of how one can observe and model a wide variety of SAS phenomena is given in [1994Sch].

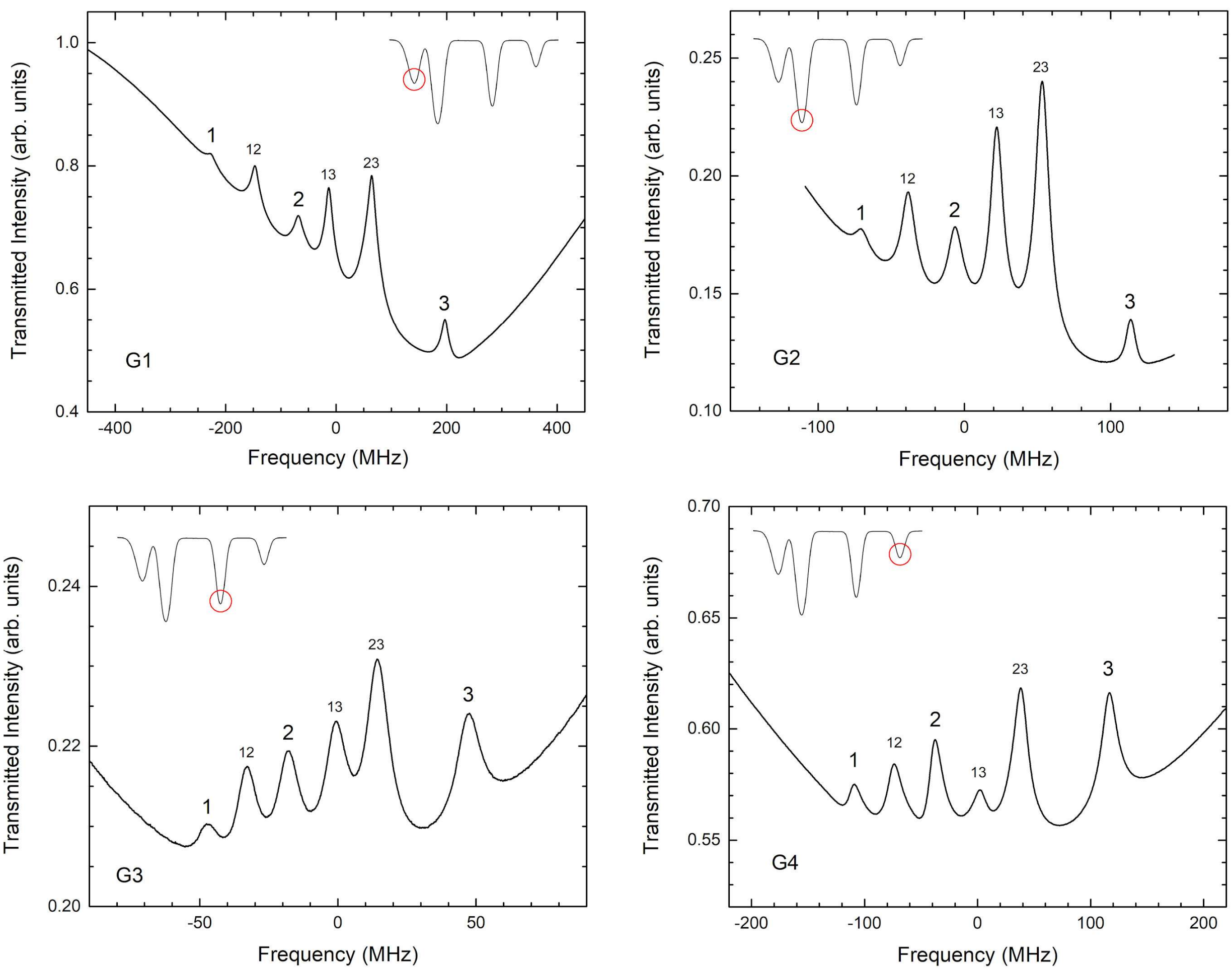

*Figure 30. Examples of saturated-absorption spectra for all four Rb dips, each exhibiting three normal SAS peaks and three crossover peaks.*

# Zeeman spectroscopy

One notable feature of laser-based atomic spectroscopy, and the Teachspin apparatus in particular, is that it can be used to demonstrate a wide variety of fundamental physical concepts using hands-on laboratory experiments. In what follows, we next add a longitudinal magnetic field to the Rb vapor cell (with the B-field aligned along the laser beam) and explore the resulting transmission spectra under different conditions.

Throughout these investigations, we can explain much of the physics using the minimal $F = 0$ to $F = 1$ "toy" model shown in Figure 31. While this model is much simpler than the actual Rb level structure, we will find that it captures many central features in the real laboratory spectra. Of course, it falls short of explaining every detail in the data, but this toy model is quite useful for its pedagogical simplicity, as most students have little experience with polarization-dependent optical phenomena when they enter the teaching lab.

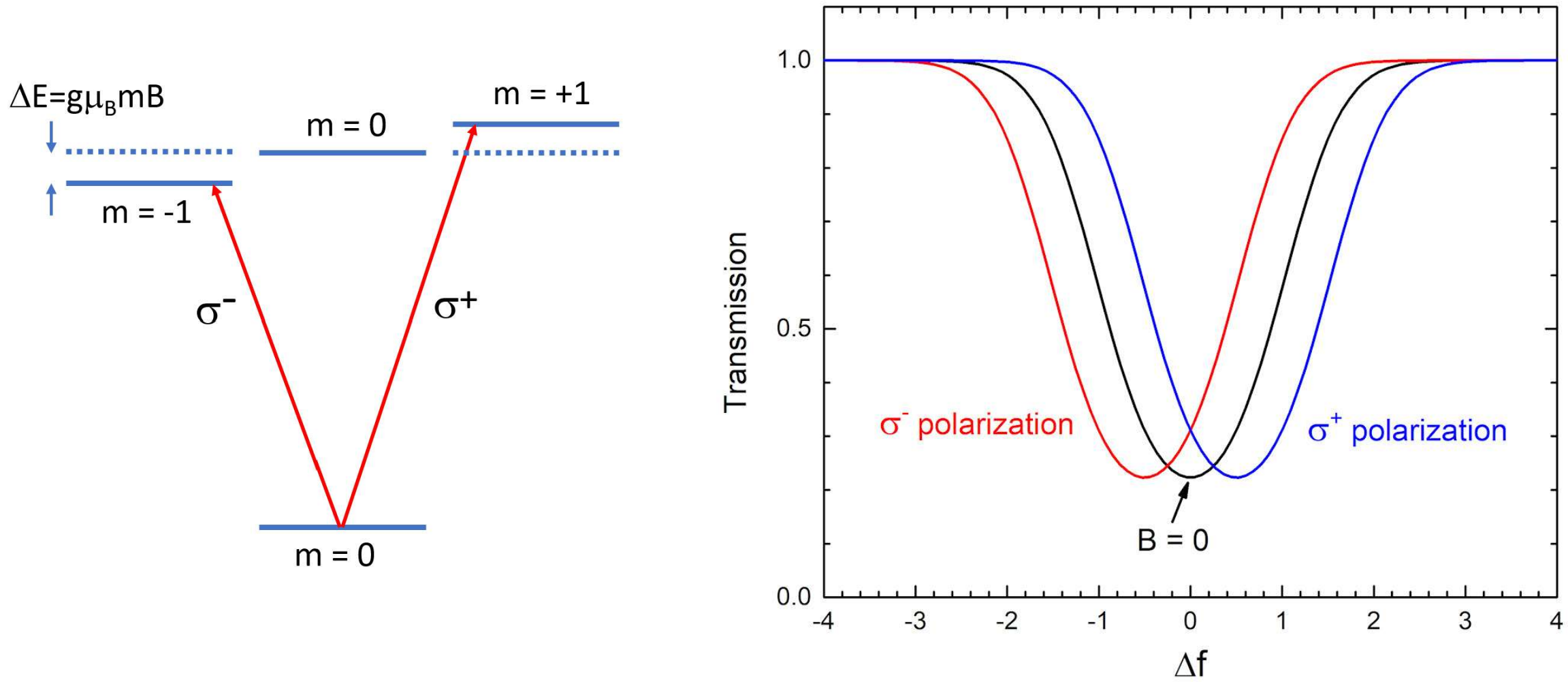

*Figure 31. This "toy" model of the Zeeman effect is sufficient to explain many aspects of the Rb transmission spectra observed in the teaching-lab experiments presented in this paper. The sketch on the left shows an F=0 to F=1 level structure, with the upper levels split by the Zeeman effect. Laser light with $\sigma^-$ circular polarization (see Appendix 1) will excite the m=-1 upper state only, while $\sigma^+$ light will excite the m=+1 state only. The sketch on the right shows the resulting transmission spectra for a Doppler-broadened atomic sample.*

Because the magnetic field is aligned along the axis of the laser light propagation, the Zeeman sublevels shown in Figure 31 are stable eigenstates of the system, and circularly polarized light will drive transitions with $\Delta m = \pm 1$ only. Notably, $\Delta m = 0$ transitions are forbidden whenever the optical electric field vector has no component along the magnetic field.

## Measuring the Zeeman shift

We have found it best to launch quickly into the lab with these experiments, postponing a detailed discussion of the theory until it is needed to understand the observations. Setting up a relatively simple experiment and being unable to understand the resulting oscilloscope trace is a strong motivator for students to then learn the corresponding theory. To this end, Figure 32 shows the optical layout of a first experiment to observe the Zeeman splitting. Walking along the optical path, we see that:

1) The first polarizer (Thorlabs LPNIRE100-B) is present to produce laser light with a vertical polarization. (Light coming out of the optical isolator is polarized at 45 degrees, but vertical polarization is needed to match the axis of the polarizing beamsplitter.)
2) The OD2 filter (Thorlabs NE10B+NE20B) is there to reduce saturated-absorption effects, as was done above.
3) The quarter-wave plate (QWP, Thorlabs WPQ10E-780) and the polarizing beamsplitter (PBS, Thorlabs PBSW-780) together serve to separate the different circular polarizations (see

Appendix 1). If the QWP birefringence axis is set to 45 degrees, then $\sigma^+$ laser light will be sent to PD1 while $\sigma^-$ light will be sent to PD2 (or the opposite if the QWP is set to -45 degrees).

In our experience, most students enter the lab with a good grasp of linear polarization states, but they have had little or no exposure to circular polarization. The Jones matrices are a good way to describe polarization vectors, and we present a brief introduction in Appendix 1. In this case, however, we have found that it is best to gain some lab experience before getting bogged down by the theory. Once students gain some practical experience working with $\sigma^+$ and $\sigma^-$ polarizations as orthogonal basis vectors (like linear polarization states, of which they are already familiar), understanding the mathematics becomes a much simpler task.

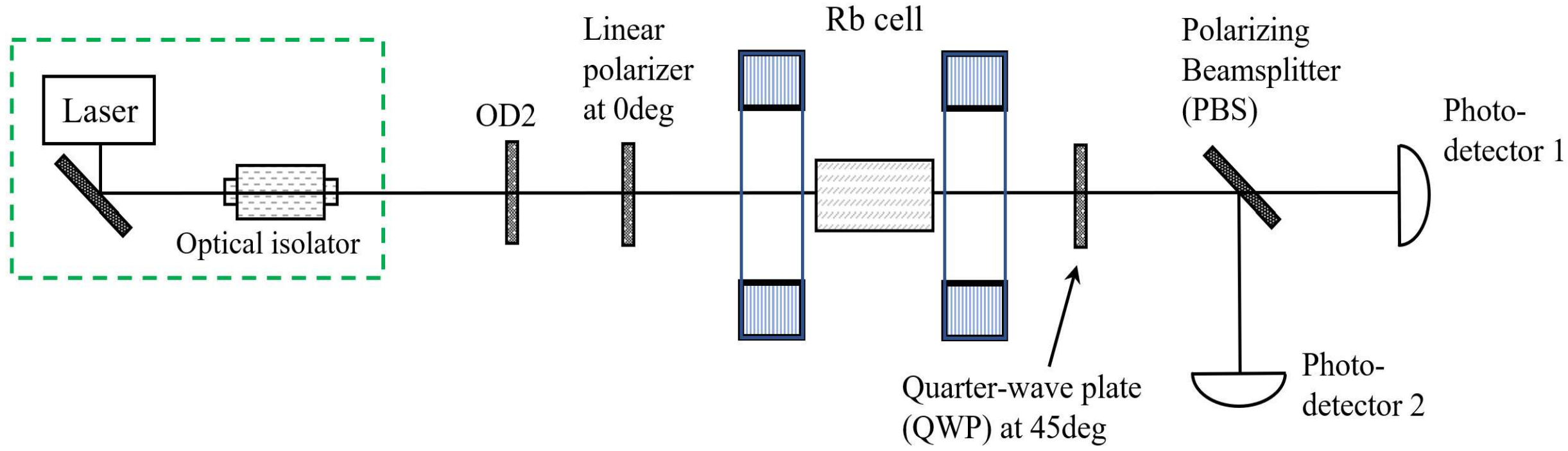

*Figure 32. This sketch shows the optical layout used to measure the Zeeman splitting.*

Pressing forward, there are some important alignment details one must attend to before observing the resulting transmission signals:

1) The PD1 and PD2 gains must be set to the same value.
2) The oscilloscope channels for PD1 and PD2 should also have the same Scale and Offset.
3) PD1 and PD2 should be aligned (by moving them side-to-side and up-down) to maximize both signals.
4) The QWP should be rotated to its optimal position, which maximizes the PD1 signal and simultaneously minimizes the PD2 signal.
5) Usual it is best to iterate Steps 1-4 to correct any errors.

Once these alignment steps have been done, the PD1 and PD2 signals should be essentially overlapping when the magnetic field coils are off, showing nearly identical Rb absorption features.

Starting from this point, one can apply a longitudinal magnetic field using the field coils supplied with the Teachspin apparatus to obtain spectra like those in Figure 33. Measuring the splittings for the G3 dip, we obtained a Zeeman splitting of $\Delta f_{Zeeman} = 280 \pm 10$ MHz at a field strength of 133 Gauss (4A to the field coils), giving a splitting of Zeeman g-factor of $g = 1.50 \pm 0.05$. Similar results were obtained for the other dips. Calculating this $g$ value would involve a detailed analysis of the level structure of Rb its various Zeeman splittings, which is beyond the scope of this paper. With a measured Doppler width of $\sigma \approx 330$ MHz, we obtained normalized Zeeman splitting value equal to $\delta_Z = \Delta f_{Zeeman}/\sigma \approx 0.21$ per amp of coil current.

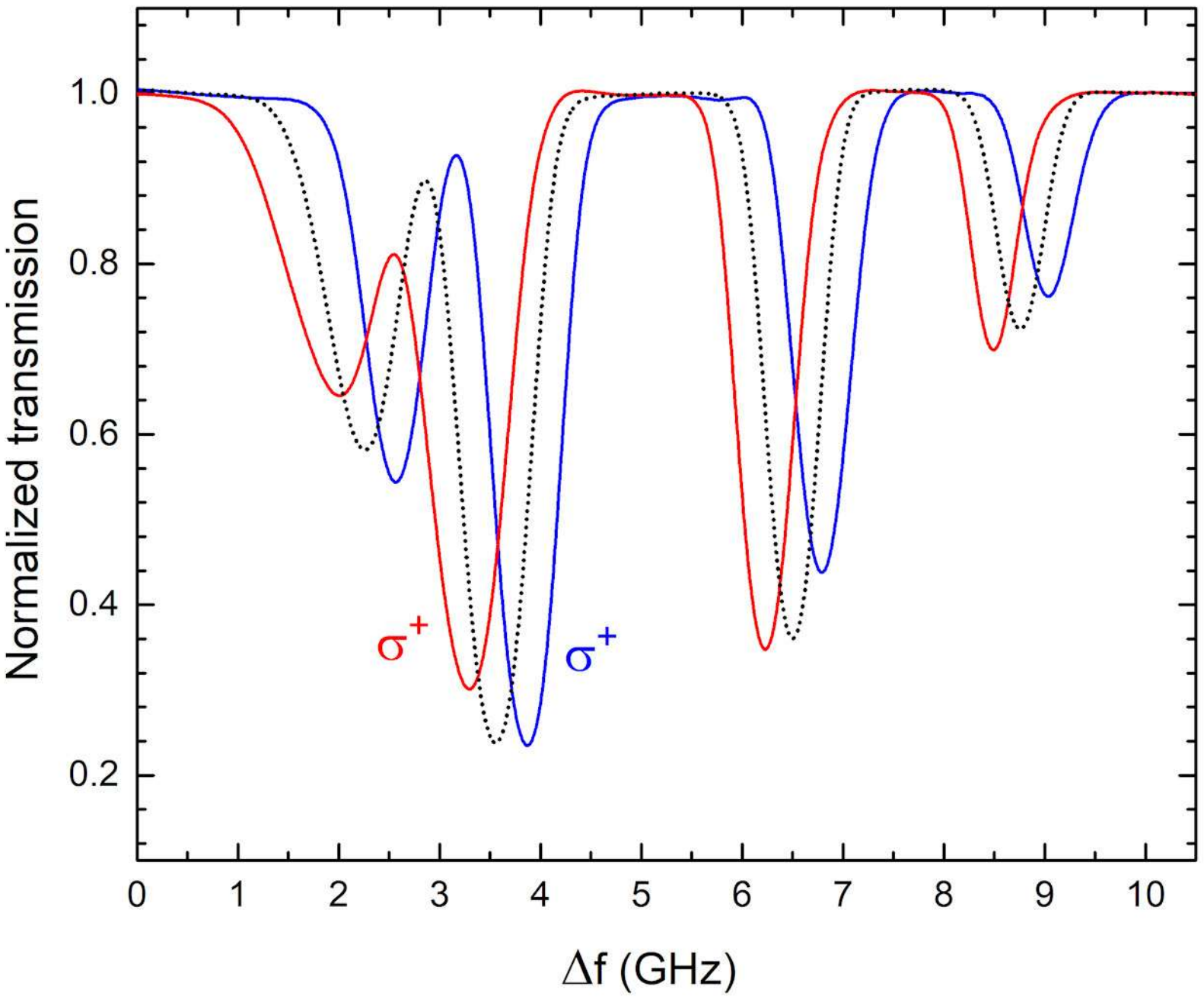

*Figure 33. This plot shows Zeeman shift data using the optical layout in Figure 32, at a Rb cell temperature of 40C and a longitudinal B field of 133 Gauss (for a coil current of 4A). The dotted line gives the zero-B spectrum while the blue and red curves correspond to the toy-model transitions shown in Figure 31. The systematic trends in the dip depths arises from a nontrivial consideration of the full Rb level diagram, which is not included in the toy model.*

While the Zeeman splittings are in general agreement with expectations from our toy model, the systematic differences in the dip depths in Figure 33 are not so easy to explain. For example, we see that the $\sigma^+$ absorption dips are deeper than the $\sigma^-$ dips for G1 and G2, while the opposite is true for G3 and G4. Unfortunately, a full explanation would involve taking a close look at how all the transition probabilities vary with polarization, which is beyond the scope of this paper. The toy model does a nice job of roughly explaining Zeeman shifts for all the dips, and measuring the g-factor is a good student exercise. Going beyond that, however, is a challenging exercise that is best left for another day.

As a final note, one can do this experiment another way by removing the polarizer in Figure 32, removing the PBS, and placing the QWP right after the optical isolator. Because the isolator output is linearly polarized, the QWP can be rotated to produce either $\sigma^+$ or $\sigma^-$ light entering the Rb cell. Thus the two absorption spectra can be observed separately using PD1. This setup is simpler than the one shown in Figure 32 and it requires fewer optical components. But it removes the pleasure of watching the $\sigma^+$ and $\sigma^-$ spectra separate in real time on the oscilloscope as you change the size of the magnetic field (and its sign, if so desired).

## A laser locking system

Another advantage of having both outputs in Figure 32 is that you can difference the two signals in real time using the oscilloscope's math feature, thus producing the output like that shown in Figure 34. Because the difference signal goes through zero at each of the absorption dips, it can be used in conjunction with an electronic feedback system to "lock" the laser at a fixed frequency. And by adding an additional small offset voltage to the difference signal, the frequency lock point can be tuned over about a 1 GHz range in this example.

This method of producing a stable laser frequency near any of the Rb D1 lines was called a *dichroic atomic vapor laser lock* (DAVLL) by its inventors [1998Cor], and it provides a valuable tool in the field of laser cooling and trapping of neutral atoms, in addition to other areas of AMO physics. With a suitably strong and stable signal, the frequency noise and/or drift in lock can be as low as ~1 MHz, which is comparable to the natural linewidth of the Rb transitions [1998Cor].

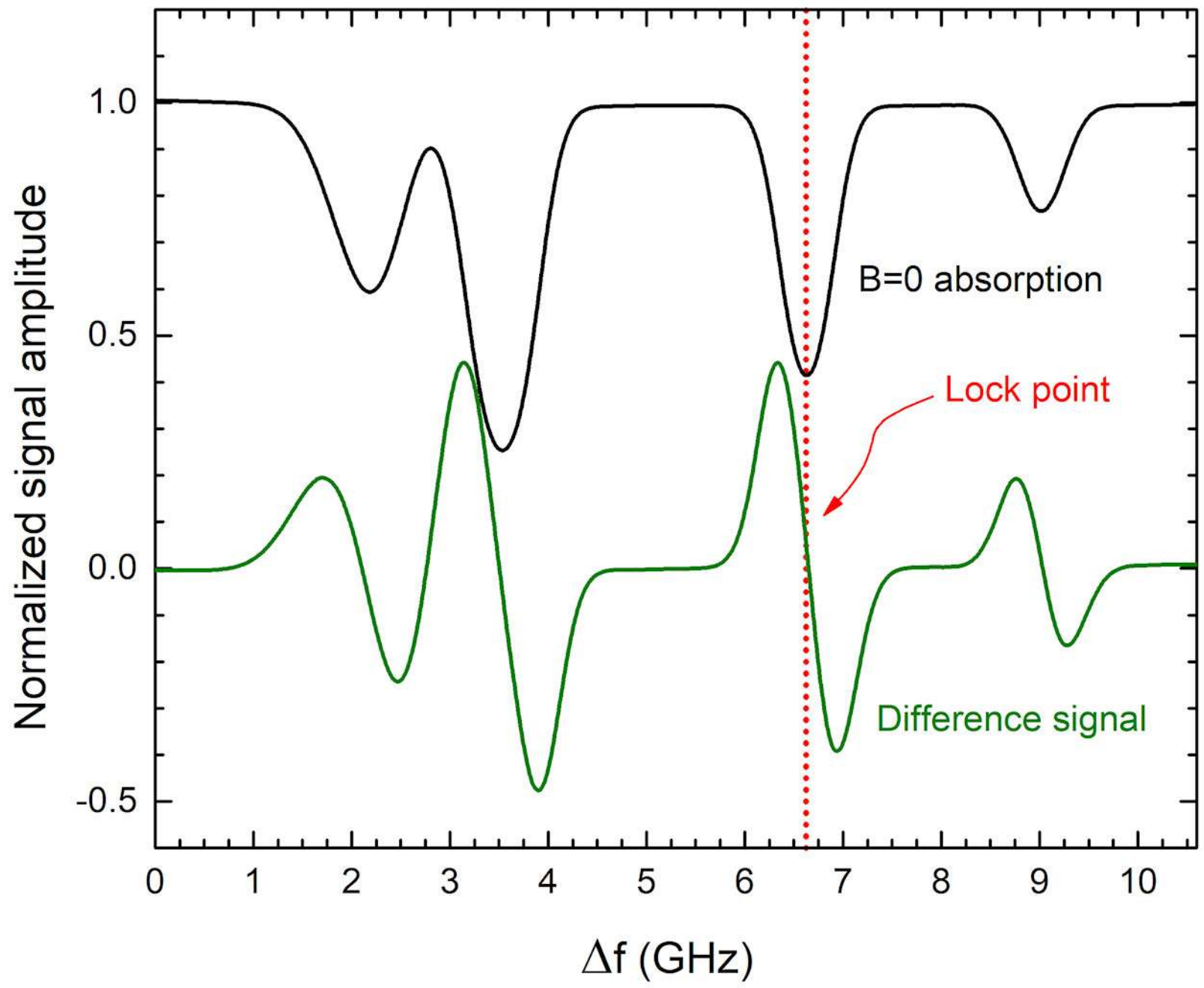

*Figure 34. The top (black) line in this plot shows the normal Rb absorption spectra with no applied magnetic field, and the lower (green) line shows the difference of the $\sigma^+$ and $\sigma^-$ spectra. The lower curve can be used in a servo feedback loop to "lock" the laser at a frequency near any of the zero crossing points. This laser frequency stabilization technique is called a "dichroic atomic vapor laser lock (DAVLL)" [1998Cor].*

# A Magneto-Optical Filter (MOF)

We can take our exploration of Zeeman spectroscopy to the next level if we simply remove the quarter-wave-plate from the optical layout in Figure 32 to give the new setup in Figure 35. With no magnetic field and the polarizer set to zero degrees as shown in the figure, then the polarizing beamsplitter sends all the laser light to PD2. The signal seen by PD1 is essentially just the Rb cell between crossed polarizers, which gives zero signal. When we apply a longitudinal magnetic field, however, a nonzero PD1 signal appears at the Rb transitions. This optical device – a resonant atomic gas between crossed polarizers – is often called a *Magneto-Optical Filter* (or MOF) because it only passes light in a set of narrow frequency bands that are fixed by the atomic optical transitions.

The MOF experiment is remarkably easy to do in the lab, but explaining the observed signal requires an understanding of the electromagnetic *susceptibility* of atomic gases. A brief introduction to this subject is presented in Appendix 2, in which we define our notation and treat the specific case of a Doppler-broadened atomic gas with a single resonant transition. Using the information in both Appendix 1 and 2, we are then able to analyze the optical layout in Figure 32.

We begin by writing the electric field of the linearly polarized input laser using the Jones matrix notation (see Appendix 1)

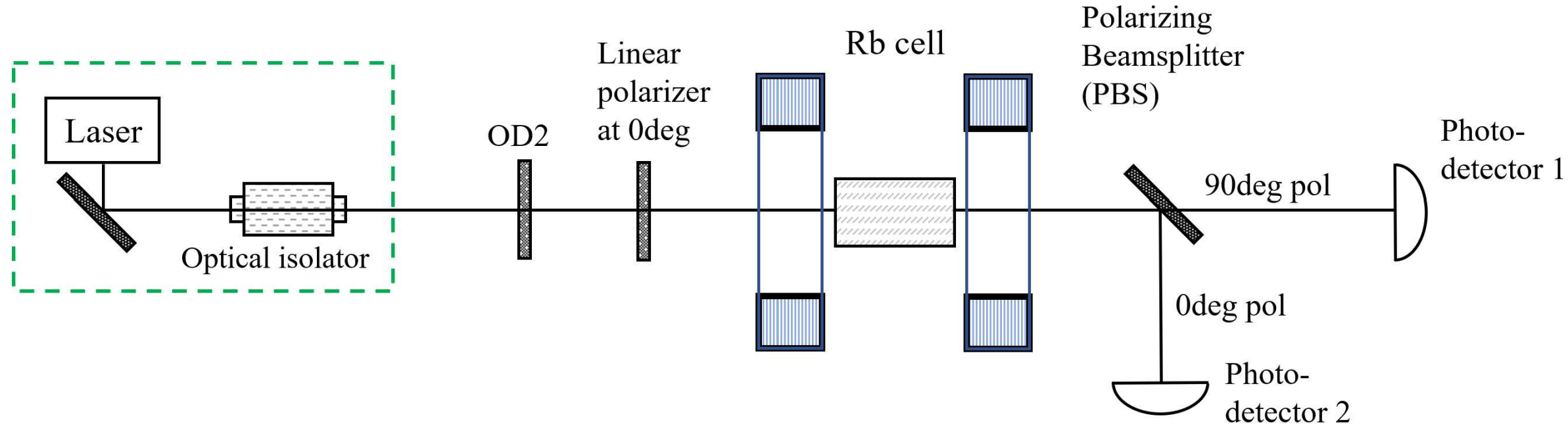

*Figure 35. This sketch shows the optical layout for the MOF experiment. In the absence of the Rb cell, the PBS sends all the laser light to PD2. What PD1 sees is essentially the Rb cell between crossed polarizers.*

$$\begin{aligned}\vec{E}_{in} &= \begin{bmatrix}1\\0\end{bmatrix} \\ &= \frac{1}{2}\left\{\begin{bmatrix}1\\i\end{bmatrix} + \begin{bmatrix}1\\-i\end{bmatrix}\right\}\end{aligned} \tag{16}$$

which rewrites the linear polarization as a sum of two circular polarizations.

After passing through the Rb cell, the electric field vector is modified by the atomic susceptibility, which includes contributions from both the absorption and the index of refraction of the atomic gas near resonance. In keeping with our "toy model" approach, we will develop the theory assuming a single atomic transition with a simple Zeeman level structure, as this model is sufficient to describe the observed signals at a level that is suitable for the undergraduate teaching lab. Using the theoretical results in Appendix 2, the electric field emerges from the Rb cell as

$$\vec{E}_1 = \frac{1}{2}\left\{\begin{bmatrix}1\\i\end{bmatrix} e^{i\chi^+} + \begin{bmatrix}1\\-i\end{bmatrix} e^{i\chi^-}\right\} \tag{17}$$

where

$$\begin{aligned}\chi^{\pm}(\delta) &= \frac{\tau_0 \chi_{cell}}{2} \\ &= \frac{\tau_0}{\sqrt{\pi}} D_+(\delta \pm \delta_Z) + i\frac{\tau_0}{2} e^{-(\delta\pm\delta_Z)^2}\end{aligned} \tag{18}$$

and $\tau_0$ is the optical depth on resonance, $\delta = \Delta f/\sigma$ is the normalized detuning, $\delta_Z = \Delta f_{Zeeman}/\sigma$ is the normalized Zeeman shift, and $D_+$ is Dawson's integral (see Appendix 2).

After $\vec{E}_1$ passes through the PBS, the laser beam sent to PD1 is

$$\begin{aligned}\vec{E}_{\text{out1}} &= \begin{bmatrix}0 & 0\\0 & 1\end{bmatrix} \frac{1}{2}\left\{\begin{bmatrix}1\\i\end{bmatrix} e^{i\chi^+} + \begin{bmatrix}1\\-i\end{bmatrix} e^{i\chi^-}\right\} \\ &= A\frac{1}{2}\left(e^{i\chi^+} - e^{i\chi^-}\right)\begin{bmatrix}0\\1\end{bmatrix}\end{aligned} \tag{19}$$

where $A$ is an overall phase factor (which is unimportant in the end because PD1 records only the beam intensity). Note that if $B = 0$, then $e^{i\chi^+} = e^{i\chi^-}$ and $\vec{E}_{\text{out1}} = 0$. This makes sense; an empty cell between crossed polarizers will produce zero signal at PD1. The final (normalized) MOF transmission signal recorded by PD1 is then

$$T(\delta) = \frac{I(\delta)}{I_0} = \frac{1}{4}\left|e^{i\chi^+} - e^{i\chi^-}\right|^2 \tag{20}$$

and this expression must be evaluated numerically to produce a model of the experimental results. Fortunately, AI tools can do all the calculations with remarkable ease, and doing the exercise provides an interesting educational experience. For example, we fed *Claude* (a popular AI tool in 2026) small screen captures of Equations (18) and (20) along with the prompt:

*Use these equations to plot (I/I0) as a function of delta over the range (-5,5). Include sliders for tau0 (0-8) and deltaZ (0-3). Note that D+ is Dawson's integral.*

which gave the result in Figure 36. Claude understood the request, easily interpreted the screen captures, knew how to set up the numerical analysis, and could handle evaluating $D_+(\delta)$ with no more difficulty than $\cos(\theta)$. (The fact that AI could do any of this was relatively new in 2026.)

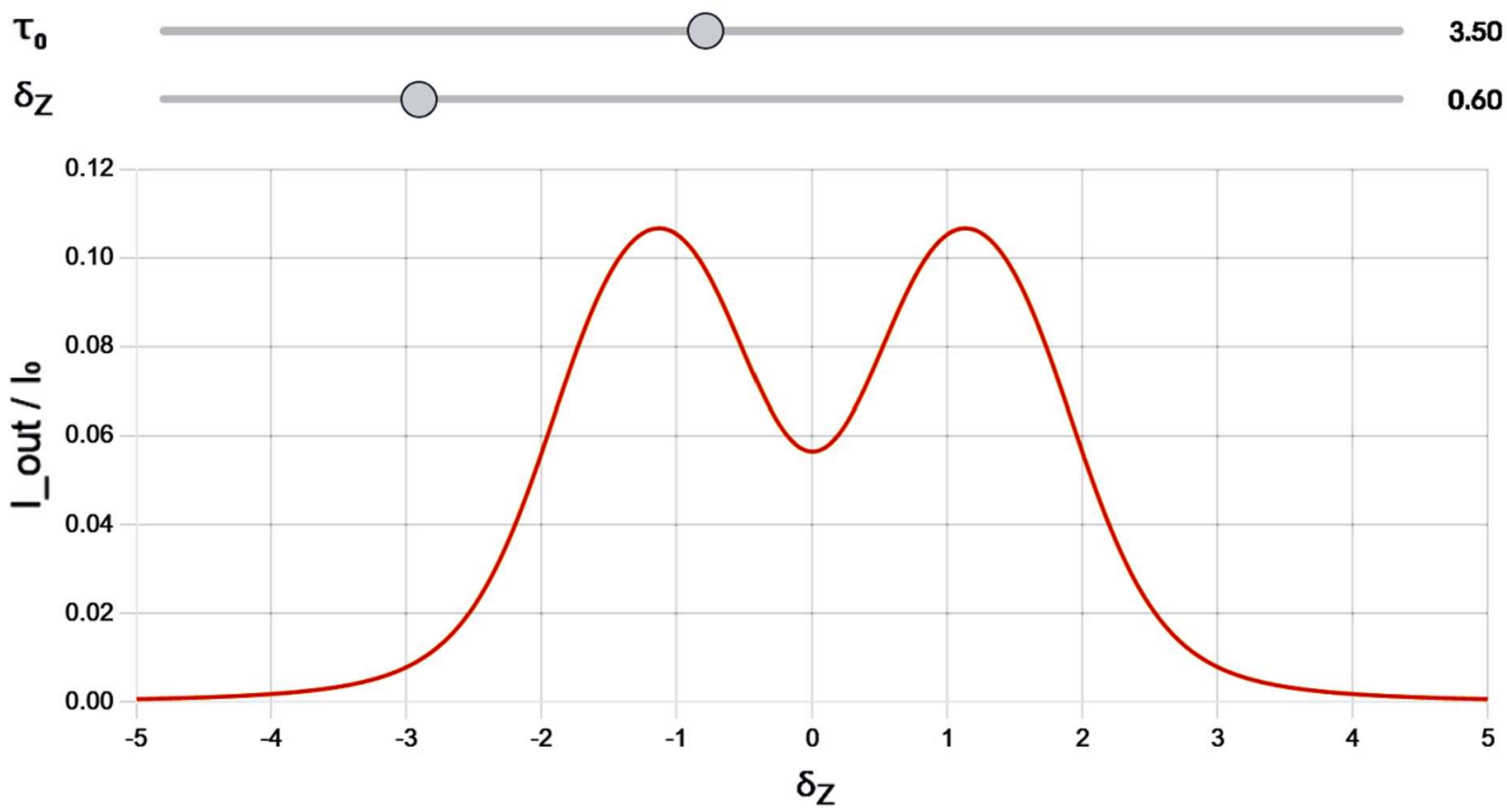

*Figure 36. This graph (generated by the AI tool Claude) shows a theoretical model of the expected MOT signal described by Equations (18) and (20). Within Claude this is an interactive graph: the theory line responds in real time as the user changes $\tau_0$ and $\delta_Z$.*

Of course, a theory plot like the one in Figure 36 may be of little use if students do not understand the underlying physics and the meaning of the input equations. But having an interactive visual representation of the theory is a fantastic learning tool, and it certainly facilitates easier comparisons between theory and measurements. In the past, instructors might have provided such interactive computation tools for their students on a teaching-lab computer. Now that we have these AI tools, instructors can focus on explaining the theory (a task for which AI can also be quite helpful) while students can use AI on their own computers to generate models of their experimental data. It is not clear in 2026 how these new AI tools will develop, but they clearly have much potential in physics teaching labs.

To compare this theory with experimental data, note that the input parameters for this toy model can all be obtained in the lab. For example, if one turns off the longitudinal B-field, then PD2 gives

$$I(\delta) = I_0 \exp\left(-\tau_0 e^{-\delta^2}\right) \tag{21}$$

(for the case of a single isolated transition like G3) and this signal will provide measurements of $I_0$, $\tau_0$, and the Doppler width $\sigma$. And our previous measurement of the Zeeman splitting $\Delta f_{Zeeman}$ gives us $\delta_Z$ as a function of the coil current.

Of course, putting all these pieces together – the experimental setup, data acquisition, model building, and data presentation – presents a formidable challenge for students, so we human instructors still have numerous indispensable roles to play in the process. If all parties are up for the challenge, the final results should look something like that in Figure 37. Here we see that the toy model assuming a single Doppler-broadened transition matches the data quite well for the G3 and G4 dips, even though the model has no free parameters. The G1 and G2 dips overlap to some extent, producing a significant distortion of the two signals that is not included in the models.

While the full Rb spectrum shown in Figure 37 is a bit much for a student exercise, the task becomes substantially less tedius if one focuses on the G3 dip. The experiment itself is quite easy to set up, the data acquisition can all be done using the oscilloscope, and the model calculations are straightforward using an AI tool. The crux of the problem is understanding the underlying physics described by the atomic susceptibility, which is a challenging but certainly worthwhile endeavor. Plus it is not necessary to obtain a 100% understanding of susceptibility theory during the process. Much of the undergraduate educational experience in physics is learning and accepting the fundamental concepts and equations without any real proof that they are correct (e.g., quantum mechanics). Because the MOF experiment is itself fairly easy, it provides a good first exposure to the underlying physics.

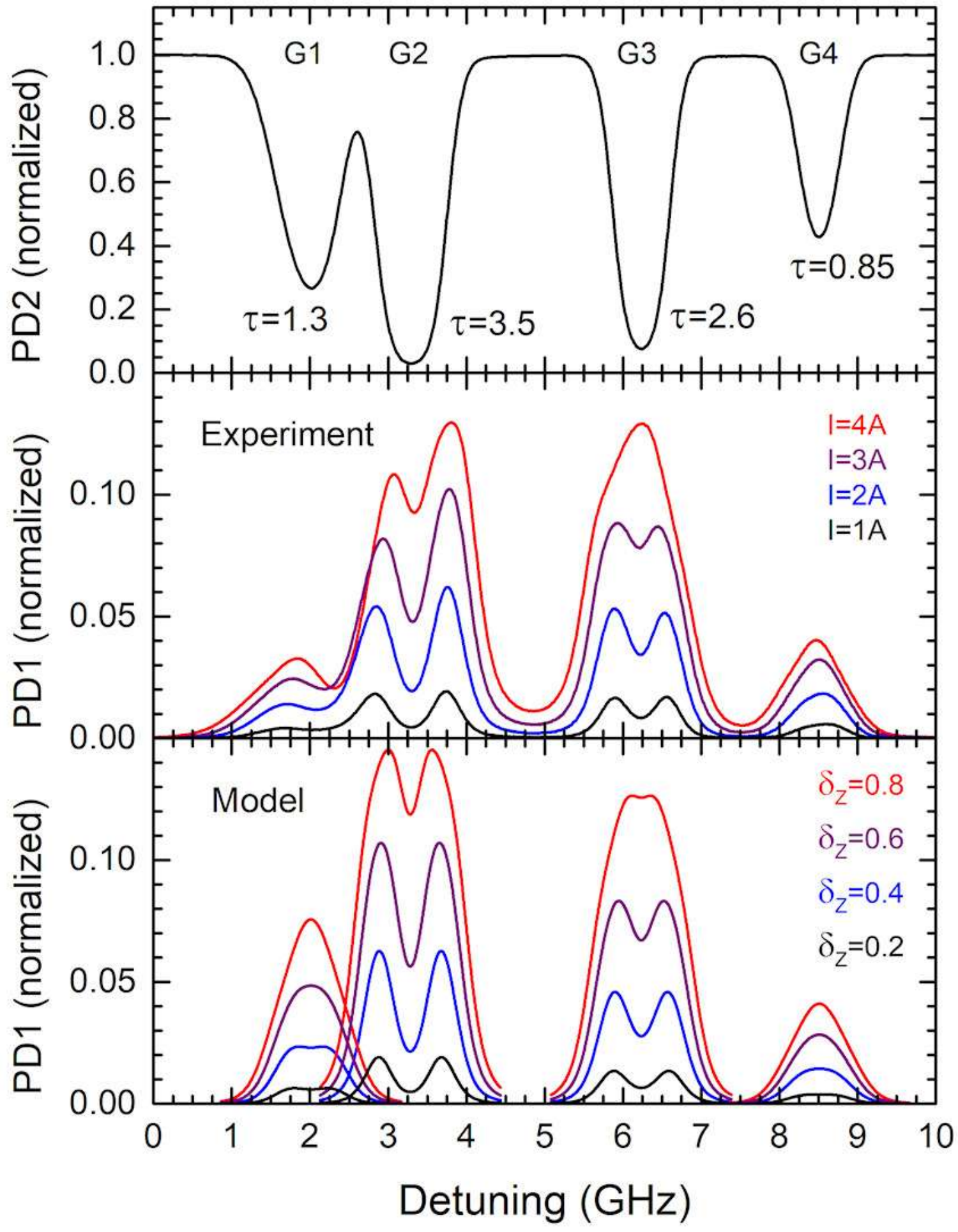

*Figure 37. This graph shows MOF data taken with a Rb cell temperature of 50C. The top plot displays the normalized PD2 spectrum, from which we measured the $\tau_0$ values shown. The middle plot gives the normalized PD1 signal at different values of the coil current (with B = 33.3 Gauss/A). The lower plot shows 16 separate MOF models, each assuming a single atomic transition with $\sigma$ = 330 MHz. The models calculated at different $\tau_0$ values were shifted in detuning to line up with the experimental data. Note that all model parameters were extracted from the measurements, so the model contained no adjustable parameters.*

## On-resonance transmission

Another variation of the MOF experiment is to focus on the signal $T(\delta = 0)$ at the center of a Doppler-broadened dip, for which data can be collected simply by using the cursor feature on the oscilloscope. Figure 38 shows some example data and modeling calculations, focusing on the G3 and G4 dips to avoid the overlap distortions in G1 and G2. Note that the theory curves again include no adjustable parameters, which was also the case in Figure 37. Overall the model reproduces the data reasonably well, although there are some small systematic differences that likely arise from the model's simplifying assumption of a single atomic transition (while the real G3 and G4 dips average over multiple transitions to different upper-state hyperfine levels).

In the limit of small $\delta_Z$, the model can be simplified to give the analytic expression

$$\begin{aligned} T(\delta = 0) &\approx \theta_F e^{-\tau_0} \\ &\approx \frac{\tau_0^2 \delta_Z^2}{\pi} e^{-\tau_0} \end{aligned} \tag{22}$$

where $\theta_F$ is called the *Faraday angle*

$$\theta_F \approx \frac{\tau_0}{\sqrt{\pi}} \delta_Z \tag{23}$$

*Figure 38. The data points on this plot show measurements of the normalized on-resonant transmission of the MOF, taken with a Rb cell temperature of 40C. The accompanying lines show model calculations at $\delta = 0$, where solid lines show the full theory and dotted lines give the Faraday-rotation limit, which applies for small $\tau_0$ and $\delta_Z$.*

which is described more generally in Appendix 2. As seen in Figure 38, the Faraday limit works well at low $\delta_Z$, but deviates from the full theory at higher $\delta_Z$.

## Faraday rotation

Continuing with this series, Figure 39 shows another interesting Zeeman spectroscopy experiment, this time for viewing the Faraday rotation signal directly [1996Baa]. Modeling the difference signal PD1–PD2 is similar to the MOF analysis, beginning with the complex susceptibility

$$\chi(\delta) = \frac{\tau_0}{\sqrt{\pi}} D_+(\delta) + i\frac{\tau_0}{2} e^{-\delta^2} \tag{24}$$

and

$$\chi^{\pm} = \chi(\delta \pm \delta_Z) \tag{25}$$

Going through the polarization analysis for this optical layout then yields

$$\begin{aligned} I_1(\delta) &= \frac{I_0}{8}\left|(1-i)e^{i\chi^+} + (1+i)e^{i\chi^-}\right|^2 \\ I_2(\delta) &= \frac{I_0}{8}\left|(1-i)e^{i\chi^+} - (1+i)e^{i\chi^-}\right|^2 \end{aligned} \tag{26}$$

and our final experimental product is the difference signal

$$I_{diff}(\delta) = I_1(\delta) - I_2(\delta) \tag{27}$$

As can be seen in Figure 40, the models for the G3 and G4 spectral features accurately represent the observations with no freely adjustable model parameters.

We can also do the analysis in the simple-Faraday-rotation limit (small $\delta_Z$ and $\tau_0$), in which the electric field after the Rb cell becomes

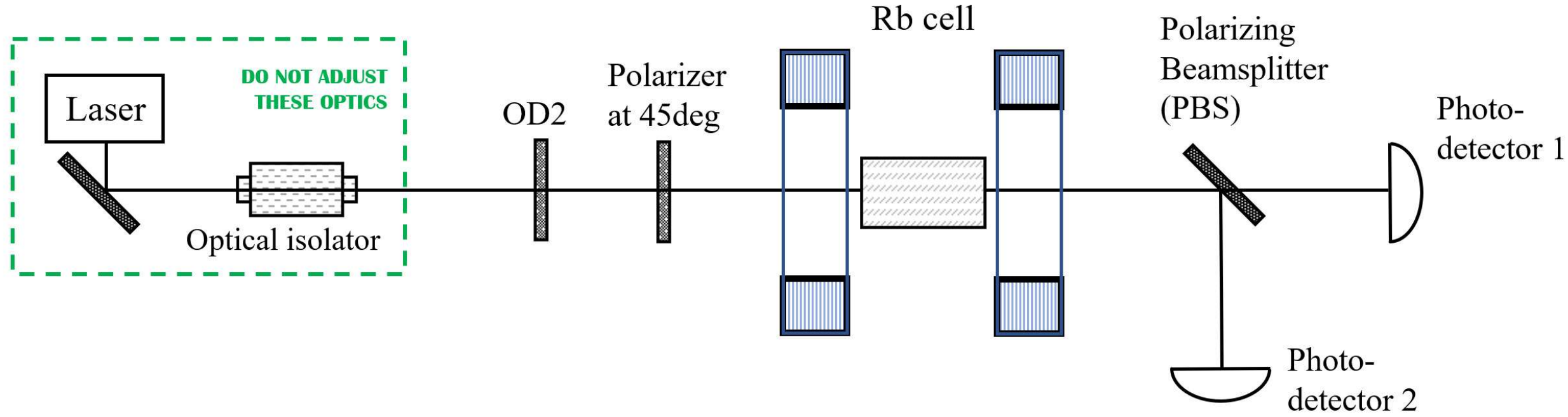

*Figure 39. This sketch shows an optical layout for directly measuring the Faraday rotation from a Rb atomic vapor [1996Baa]. The output from the experiment is the difference signal PD1-PD2.*

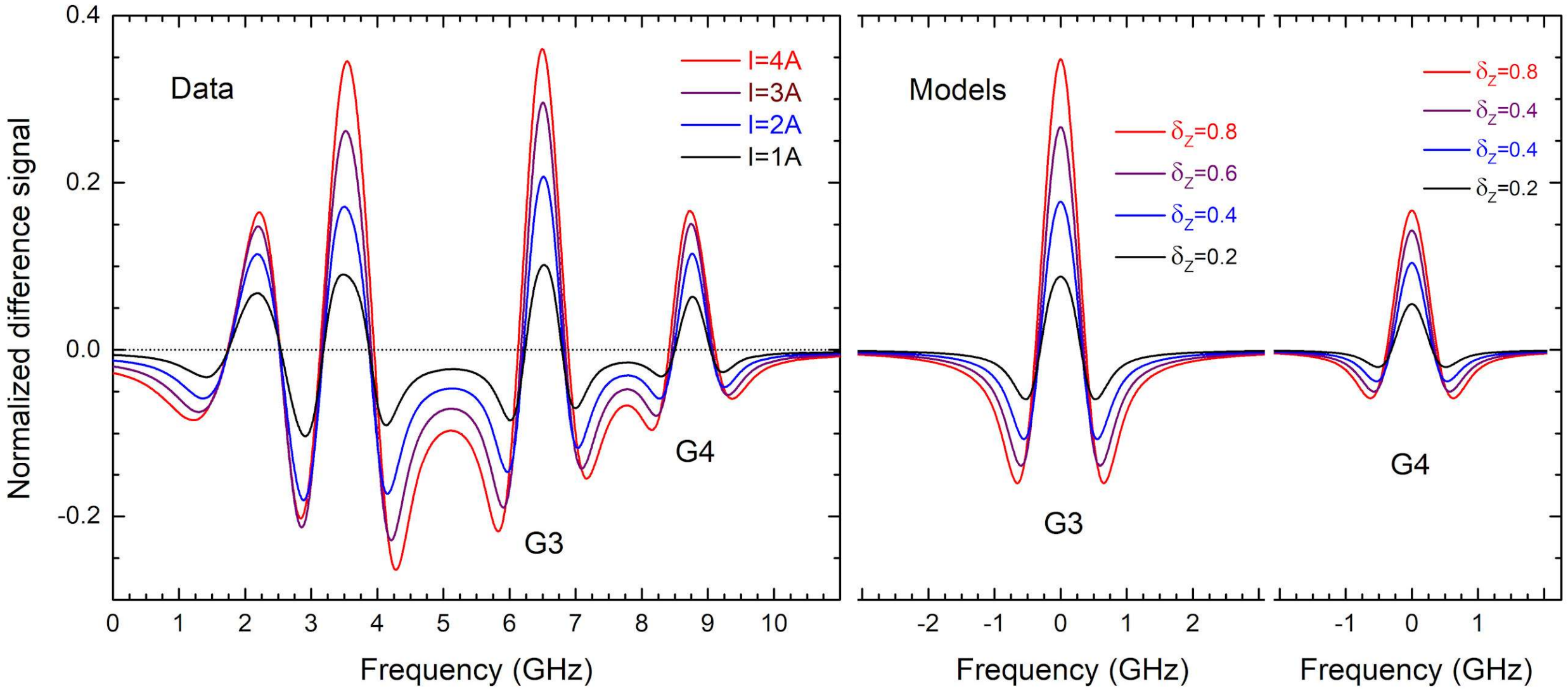

*Figure 40. The left plot shows normalized Faraday rotation data using the optical layout in Figure 39, for four different values of the coil current at a Rb cell temperature of 40C. Theoretical models for G3 and G4 are shown on the right, for $\delta_Z$ values that correspond to the experimental coil currents, using experimentally measured $\tau_0$ for G3 and G4 (1 and 0.33, respectively). The models contain no adjustable parameters.*

$$E(\delta) \approx E_0 \exp\left(-\frac{\tau_0}{2}e^{-\delta^2}\right)\begin{bmatrix}\sin\left(\frac{\pi}{4}+\theta_F\right)\\ \cos\left(\frac{\pi}{4}+\theta_F\right)\end{bmatrix} \approx E_0 \exp\left(-\frac{\tau_0}{2}e^{-\delta^2}\right)\frac{1}{\sqrt{2}}\begin{bmatrix}1+\theta_F\\ 1-\theta_F\end{bmatrix} \tag{28}$$

and the latter expression applies for small $\theta_F$. And this gives the measured intensities

$$I_1(\delta) \approx I_0 \exp\left(-\tau_0 e^{-\delta^2}\right)\left(\frac{1}{2}+\theta_F\right) \qquad I_2(\delta) = I_0 \exp\left(-\tau_0 e^{-\delta^2}\right)\left(\frac{1}{2}-\theta_F\right) \tag{29}$$

and the on-resonance difference signal is

$$I_1 - I_2 \approx 2I_0\theta_F e^{-\tau_0} \quad (\text{for } \delta = 0) \qquad T(\delta=0) \approx \frac{\Delta I}{I_0} \approx \frac{2\tau_0\delta_Z}{\sqrt{\pi}}e^{-\tau_0} \tag{30}$$

Here we see that $T(\delta = 0)$ is linearly proportional to $\delta_B$ and measurements of the difference signal in the lab could be done using the oscilloscope's math and cursor functions. Figure 41 shows that once again there is a reasonably good agreement between theory and measurements. In this plot we tweaked the theory with a 15% normalization adjustment to better match the data, but we would not expect perfect agreement between our single-transition toy model and the actual multi-transition Rb absorption features.

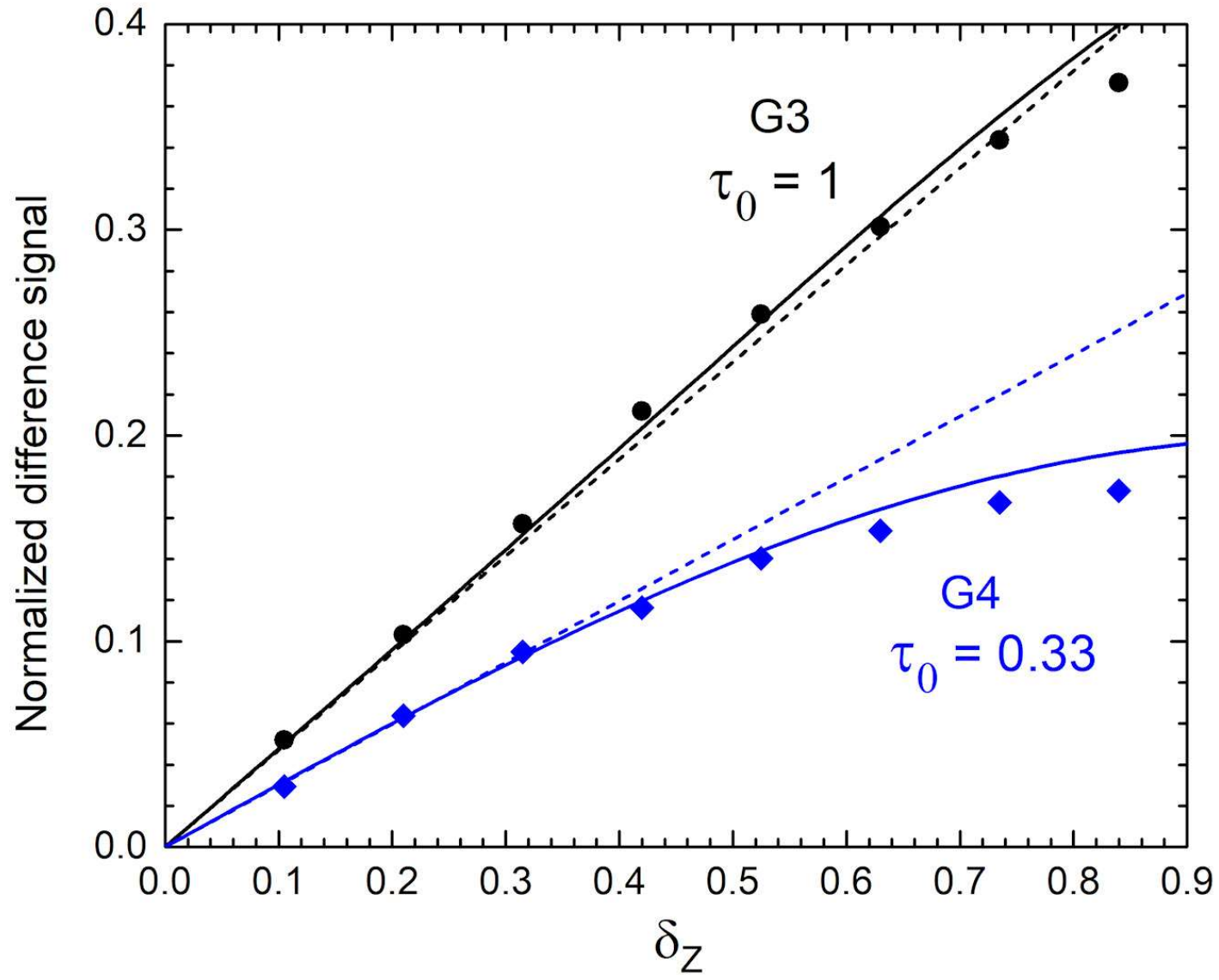

*Figure 41. The data points here show measurements of the on-resonance* $(\delta = 0)$ *Faraday rotation difference signals for the G3 and G4 spectral features. Solid lines show corresponding full-theory models, while dashed lines show the theory in the low-*$\delta_Z$ *limit. For this plot, both models (calculated with no adjustable parameters) were multiplied by 1.15 to better fit the data.*

## Summary and beyond

Our main goal in this paper was to produce a comprehensive guide to all the relatively easy experiments that can be performed with the Teachspin *Diode Laser Spectroscopy* instrument. No doubt we have overlooked some other examples and not fully accomplished this goal, but we feel that the experiments presented here are all suitable for the physics teaching lab. The theory can be challenging at times, and nontrivial alignment issues are present, but such is the nature of laser spectroscopy and AMO physics more generally. In our lab we have coupled this apparatus with Optical Pumping [2026Lib] and NMR [2025Lib] to create a full one-quarter AMO experimental "track".

If you would like to expand Rb spectroscopy even further, there are numerous more advanced experiments that have been proposed for the physics teaching lab. Some examples include 2-photon spectroscopy [2006Ols, 2009Jac], nonlinear magneto-optics [1999Bud], coherent population trapping [2009Bel], and even laser cooling and trapping of Rb atoms [1995Wie]. If you have sufficient time and budget, there is no shortage of fascinating AMO experiments you might incorporate into your undergraduate teaching lab.

## Acknowledgements

This work was supported in part by a generous donation from Beatrice and Sai-Wai Fu to the Physics Teaching Labs at Caltech. We also acknowledge support from Drs. Richard Karp and Vineer Bhansali, together with Caltech's long-standing support of outstanding laboratory instruction across many STEM fields.

Much of the theory work in this paper was done using Claude and ChatGPT. These relatively new AI tools are still rather crude in 2026, but they already show amazing facility at doing complex modeling calculations, and at explaining the underlying physics in considerable detail. It is not clear at present how students will use AI in the learning process, but clearly AI is already a remarkable resource for instructors!

*Contact:* For corrections, comments, or just to compare notes, please contact Kenneth G. Libbrecht, *kgl@caltech.edu*, or mail to: Mail-stop 264-33 Caltech, Pasadena, CA 91125.

## Appendix 1 – Polarization analysis

Many experiments in Atomic/Molecular/Optical (AMO) physics involve laser polarization, and the *Jones calculus* is a popular method for analyzing the effects of different polarization states. We begin by representing the polarization of a plane wave propagating in the $z$ direction by an electric-field vector (called the *Jones vector*)

$$\vec{E}(z,t) = \begin{bmatrix} \hat{E}_x \\ \hat{E}_y \end{bmatrix} = \begin{bmatrix} E_x e^{i\varphi_x} \\ E_y e^{i\varphi_y} \end{bmatrix} e^{i(\omega t - kz)} \tag{31}$$

where $E_x$ and $\varphi_x$ represent the amplitude and phase of the electric field at some point in space and time. Because our main concern is the light intensity hitting a photodetector (which records no phase information), we usually drop the $\exp[i(\omega t - kz)]$ propagation factor associated with traveling waves.

In laser optics, the most common polarization states are

$$|V\rangle = \begin{bmatrix} 1 \\ 0 \end{bmatrix}, \qquad |H\rangle = \begin{bmatrix} 0 \\ 1 \end{bmatrix}, \qquad |L\rangle = \frac{1}{\sqrt{2}} \begin{bmatrix} 1 \\ i \end{bmatrix}, \qquad |R\rangle = \frac{1}{\sqrt{2}} \begin{bmatrix} 1 \\ -i \end{bmatrix} \tag{32}$$

which describe light that is:
$|H\rangle$ = Linearly polarized in the horizontal direction (typically defined with respect to the tabletop)
$|V\rangle$ = Linearly polarized in the vertical direction
$|L\rangle$ = Left-circularly polarized (also called LCP or $\sigma^-$)
$|R\rangle$ = Right-circularly polarized (also called RCP or $\sigma^+$)
More generally, states with arbitrary $\hat{E}_x$ and $\hat{E}_y$ are often called "elliptically" polarized light.

Note that the polarization states $|V\rangle$ and $|H$ form an *orthogonal basis set*; any polarization state can be expressed as a linear combination of these two vectors. Similarly, $|L\rangle$ and $|R\rangle$ also form an orthogonal basis set. We often represent our laser in terms of one of these basis sets, depending on the nature of our experiments.

Students entering the physics teaching lab are almost always familiar with linear polarization (and linear polarizers), but many are quite unfamiliar with circular polarization states. We find it helpful to remind them that these are traveling waves where the electric-field vector executes a "corkscrew" pattern during propagation, as illustrated in Figure 42. Circularly polarized states are especially important in AMO (atomic/molecular/optical) physics because circularly polarized photons carry angular momentum. It is common to refer to the circular polarization states as $\sigma^+$ and $\sigma^-$, where $\sigma^+$ atomic transitions change the Zeeman quantum number $m$ by $\Delta m = +1$ and $\sigma^-$ transitions have $\Delta m = -1$.

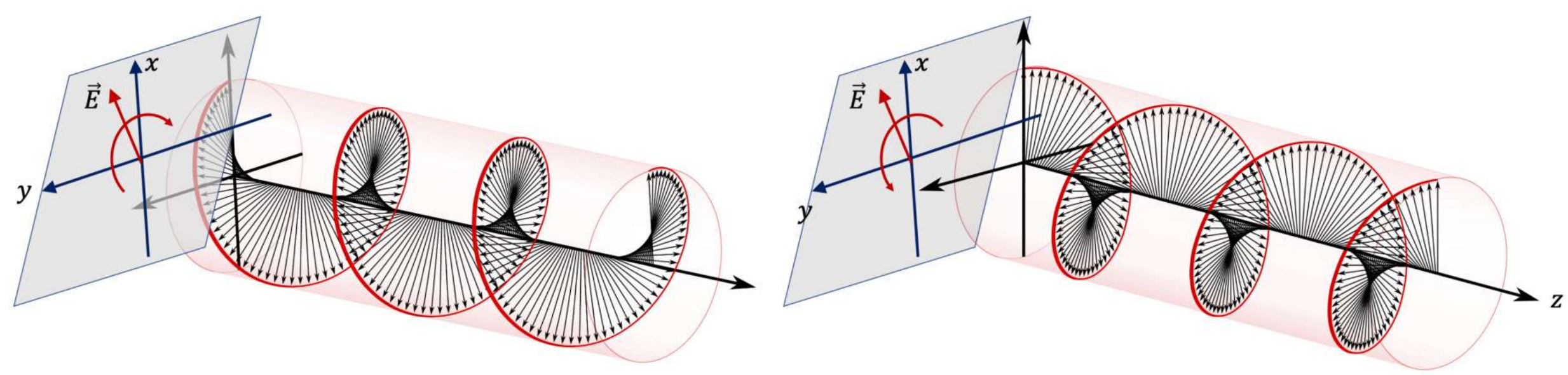

*Figure 42. In classical physics, the electric field vector in a circularly polarized laser beam follows a spiral pattern as it propagates, as shown in these sketches. There are only two choices for the "handedness" of the spiral relative to the propagation direction, and these are called right and left circularly polarized light. In quantum physics, these polarizations are especially important because circularly polarized photons carry angular momentum. Thus $\sigma^+$ photon can excite atomic transitions Zeeman quantum number $m$ by $\Delta m = +1$ while $\sigma^-$ photons excite $\Delta m = -1$ transitions. Image from https://www.physicsbootcamp.org/section-polarization-of-light.html.*

Many optical components that affect polarization can be represented by *Jones matrices* that act on Jones vectors. For example, horizonal and vertical polarizers are expressed as

$$M_V = \begin{bmatrix} 1 & 0 \\ 0 & 0 \end{bmatrix}, \qquad M_H = \begin{bmatrix} 0 & 0 \\ 0 & 1 \end{bmatrix}, \tag{33}$$

If you pass polarized light through a polarizer, you can verify that $M_H|H\rangle = |H\rangle$ (the polarizer does nothing) but $M_H|V\rangle = 0$ (a crossed polarizer extinguishes the light).

In a transparent birefringent material like crystalline quartz, light that is linearly polarized in one direction (relative to an axis of the crystal structure) travels faster than the orthogonal linear polarization. Such materials are used to make *retardation plates*, which can be described by the Jones matrix

$$M_{retarder} = \begin{bmatrix} 1 & 0 \\ 0 & e^{i\varphi} \end{bmatrix} \tag{34}$$

where the retardation angle $\varphi$ depends on the thickness of the optical element (normally a plate of quartz or some other crystalline material). The two most used retardation plates are the quarter-wave plate (QWP) and half-wave plate (HWP), which have the Jones matrices

$$M_{QWP} = \begin{bmatrix} 1 & 0 \\ 0 & i \end{bmatrix}, \qquad M_{HWP} = \begin{bmatrix} 1 & 0 \\ 0 & -1 \end{bmatrix} \tag{35}$$

Note that because $i = \exp(i\pi/2)$, the $i$ in the QWP matrix corresponds to a phase shift of 90 degrees, which is a quarter-wave. This matrix element becomes $-1$ for a phase shift of 180 degrees.

Like linear polarizers, retardation plates have a distinct optical axis, and one often rotates these optical elements about laser beam axis. For example, if we rotate a QWP by an angle $\theta$, this changes the Jones matrix to

$$M_{QWP,rotated} = R(-\theta)M_{QWP}R(\theta) \tag{36}$$

where $R$ is the 2D rotation matrix

$$R(\theta) = \begin{bmatrix} \cos(\theta) & \sin(\theta) \\ -\sin(\theta) & \cos(\theta) \end{bmatrix} \tag{37}$$

For $\theta = -45$ degrees, this gives

$$\begin{aligned} M_{QWP,rotated} &= \frac{1}{\sqrt{2}}\begin{bmatrix} 1 & 1 \\ -1 & 1 \end{bmatrix}\begin{bmatrix} 1 & 0 \\ 0 & i \end{bmatrix}\frac{1}{\sqrt{2}}\begin{bmatrix} 1 & -1 \\ 1 & 1 \end{bmatrix} \\ &= \frac{1}{2}\begin{bmatrix} 1+i & -1+i \\ -1+i & 1+i \end{bmatrix} \\ &= e^{i\pi/4}\frac{1}{\sqrt{2}}\begin{bmatrix} 1 & i \\ i & 1 \end{bmatrix} \end{aligned} \tag{38}$$

where we used $1 \pm i = \sqrt{2}e^{\pm i\pi/4}$. (For $\theta = -45$ degrees, replace each $i$ with $-i$ in the matrix.) And we can usually ignore the $e^{i\pi/4}$ factor because the overall phase of the light is irrelevant once it hits the photodetector.

Because beamsplitters have one input beam and two output beams, they cannot be represented by a single Jones matrix. For example, a polarizing beamsplitter (PBS) acts like $|H\rangle$ for the transmitted beam and $|V\rangle$ for the reflected beam. These are often called the P and S polarizations; P because the electric field "plunges" into the waveplate and S because the electric field "skips" off the surface of the waveplate.

## A circular analyzer

In Figure 32, our optical layout contains a QWP followed by a PBS, and this specific combination of optical elements is often called a *circular analyzer* because it separates the input beam into its two circular-polarization components. To see how this works, start with an input laser beam in some arbitrary polarization state. Because $|L\rangle$ and $|R\rangle$ are an orthogonal basis set, we can express this input laser beam as

$$\begin{aligned} \vec{E}_1 &= A_L|L\rangle + A_R|R\rangle \\ &= \frac{A_L}{\sqrt{2}}\begin{bmatrix} 1 \\ i \end{bmatrix} + \frac{A_R}{\sqrt{2}}\begin{bmatrix} 1 \\ -i \end{bmatrix} \end{aligned} \tag{39}$$

which has a light intensity $I = |\vec{E}_1|^2 = A_L^2 + A_R^2$.

Next the beam goes through the 45-degree QWP, yielding

$$\begin{aligned} \vec{E}_{out} &= \frac{1}{\sqrt{2}}\begin{bmatrix} 1 & i \\ i & 1 \end{bmatrix}\left\{\frac{A_L}{\sqrt{2}}\begin{bmatrix} 1 \\ i \end{bmatrix} + \frac{A_R}{\sqrt{2}}\begin{bmatrix} 1 \\ -i \end{bmatrix}\right\} \\ &= iA_L\begin{bmatrix} 0 \\ 1 \end{bmatrix} + A_R\begin{bmatrix} 1 \\ 0 \end{bmatrix} \end{aligned} \tag{40}$$

The last step is the polarizing beamsplitter that simply separates the linear polarizations, so the two output photodiodes record the signals

$$\begin{aligned} I_1 &= A_L^2 \\ I_2 &= A_R^2 \end{aligned} \tag{41}$$

and this is why we call this pair of optical elements (with these specific rotations) a circular analyzer.

## Appendix 2 – Atomic susceptibility

In the basic spectroscopy experiments described in this paper, we only needed to consider absorption of laser light in the Rb cell caused resonant photons exciting atomic transitions. In the Zeeman spectroscopy experiments, however, we also must include the frequency-dependent *index of refraction* of the vapor as well. The theory is typically expressed in terms of the frequency-dependent *complex susceptibility* of the medium, and this quantity for atomic transitions can be found in many textbooks. We will begin with a review of that theory and go a step further by including Doppler broadening.

We start by writing the complex index of refraction as

$$\tilde{n} = n + i\kappa \tag{42}$$

where $n$ is the normal (non-absorbing) index of refraction and $\kappa$ is called the *extinction coefficient*. For an EM wave propagating in the $z$ direction, the electric field can be written

$$\begin{aligned} E(z,t) &= E_0 e^{i(\tilde{n}kz-\omega t)} \\ &= E_0 e^{-\kappa kz} E_0 e^{i(nkz-\omega t)} \end{aligned} \tag{43}$$

where $k = \omega/c$. The intensity of the wave decreases as it propagates, giving

$$I(z) = I_0 e^{-\alpha z} \tag{44}$$

where $\alpha = 2\kappa k = 2\kappa\omega/c$ is called the *absorption coefficient*.

We then define the complex *susceptibility* $\tilde{\chi}$ using the relation

$$\begin{aligned} \tilde{n}^2 &= 1 + \tilde{\chi} \\ &= 1 + \chi' + i\chi'' \end{aligned} \tag{45}$$

where $\chi' = \mathrm{Re}[\tilde{\chi}]$ and $\chi'' = \mathrm{Im}[\tilde{\chi}]$.

In our Rb vapor cell, $\tilde{n}$ is very close to unity, so taking the limit of $\tilde{\chi} \ll 1$ gives

$$\chi'_{vapor} \approx 2(n-1) \tag{46}$$
$$\chi''_{vapor} \approx 2\kappa = \frac{\alpha c}{\omega}$$

If the laser passes though the Rb cell of length $L$, then the intensity will be diminished by a factor $e^{-\alpha L}$, so our earlier definition of the optical depth is just $\tau = \alpha L = 2\kappa\omega L/c = \chi''_{vapor}\omega L/c$. For a single Doppler-broadened optical transition, the optical depth has a Gaussian profile

$$\tau(\delta) = \tau_0 e^{-\delta^2} \tag{47}$$

where $\delta = (f - f_0)/\sigma$ and $\sigma$ is the Doppler-broadening parameter that arises from the Maxwell-Boltzmann velocity distribution of the atoms. Putting all this together, it is useful to define an effective $\chi''_{cell}$ for the whole Rb cell

$$\chi''_{cell}(\delta) = \frac{\tau_0}{2} e^{-\delta^2} \tag{48}$$

With this we write the electric field exiting the cell as

$$E = E_0 e^{i\tilde{\chi}_{cell}} \tag{49}$$
$$= E_0 e^{i\chi'_{cell}} e^{-\chi''_{cell}}$$

After all these definitions, the complex susceptibility $\tilde{\chi}_{cell}$ fits normal physics conventions and gives us the proper absorption profile from the gas cell, expressed in terms of the optical depth. All that remains is to determine $\chi'_{cell}$, which will tell us how the *phase* of the electric field is changed by Rb atomic interactions.

One way to derive $\chi'_{cell}$ stems from the *Kramers-Kronig relation*, which connects the real and imaginary parts of the susceptibility for a broad range of physical phenomena involving light interacting with dielectric materials (a subject that is beyond the scope of this paper). The known Gaussian absorption profile defines $\mathrm{Im}(\chi)$, and the K-K relation tells us that $\mathrm{Re}[\chi(\delta)]$ and $\mathrm{Im}[\chi(\delta)]$ are linked via a Hilbert transformation. Performing this mathematical transformation gives

$$\tilde{\chi}_{cell}(\delta) = i\frac{\tau_0}{2} e^{-\delta^2} + \frac{\tau_0}{\sqrt{\pi}} D_+(\delta) \tag{50}$$

where $D_+$ is called *Dawson's integral*

$$D_+(t) = e^{-t^2} \int_0^t e^{s^2} \, ds \tag{51}$$

This integral often comes up when dealing with Gaussian functions (so much so that it has its own Wikipedia page), and Dawson has his name attached to it because he found a convenient method for numerically evaluating $D_+(t)$ in [1897Daw]. In the modern era, of course, evaluating $D_+(t)$ is no different from evaluating $\cos(\theta)$ – it's just another function and your computer already knows how to calculate. Figure 43 shows a plot of $\mathrm{Re}[\tilde{\chi}_{cell}]$ and $\mathrm{Im}[\tilde{\chi}_{cell}]$, which you can think of as effectively the index of refraction (specifically $n-1$) and the absorption coefficient for one pass through the Rb vapor cell.

Once we have the susceptibility, the electric field exiting the Rb can be written as

$$\begin{aligned} E(\delta) &\approx E_0 e^{i\tilde{\chi}_{cell}} \\ &\approx E_0 \exp\left(-\frac{\tau_0}{2}e^{-\delta^2}\right)\exp\left(\frac{i\tau_0}{\sqrt{\pi}}D_+(\delta)\right) \end{aligned} \tag{52}$$

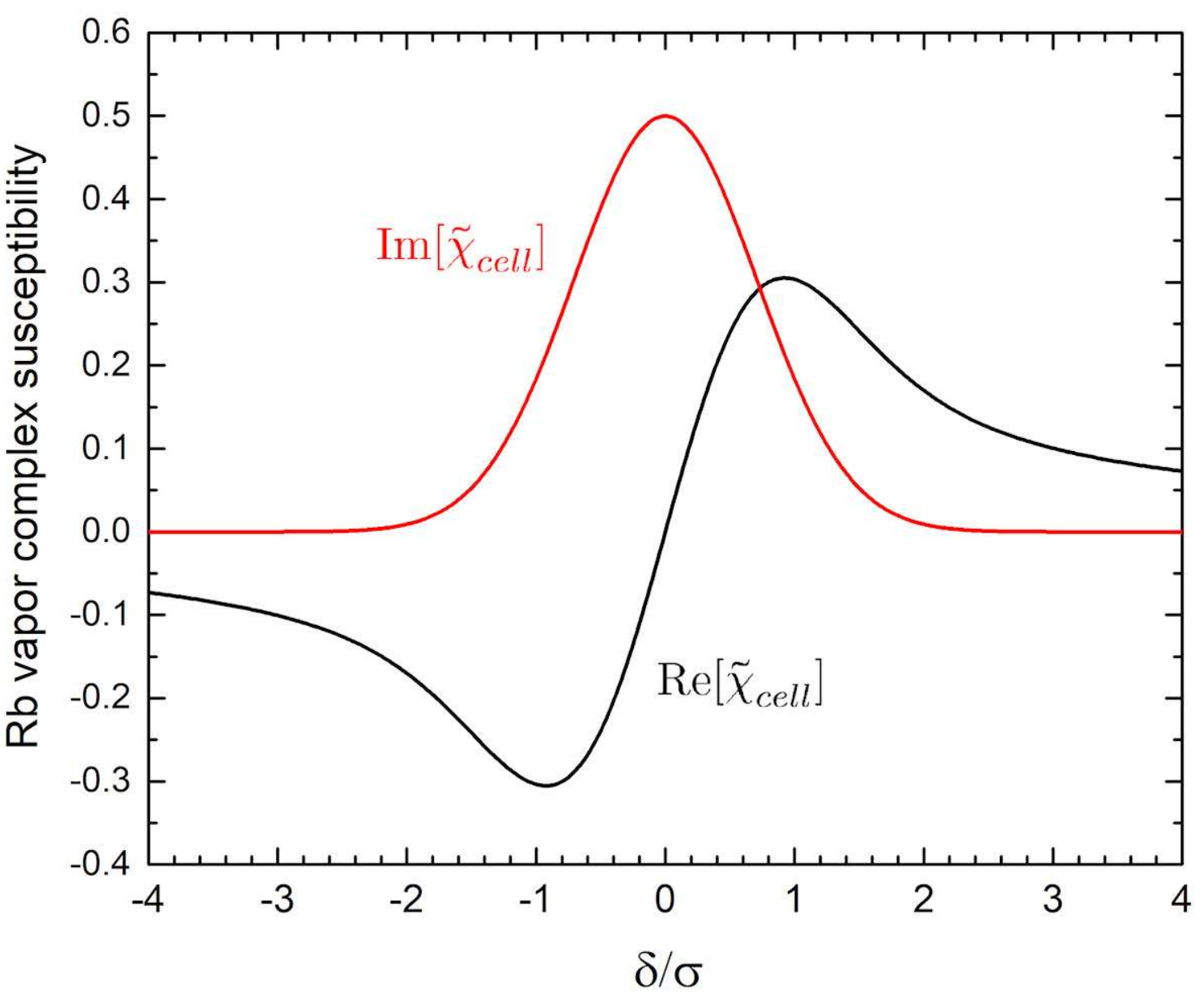

*Figure 43. This plot shows the complex susceptibility of the Rb vapor cell for $\tau_0 = 1$, defined in Equation (50).*

where $E_0$ is the electric field entering the cell and $\tau_0$ is the optical depth of the Rb cell on resonance. This equation encapsulates much of what we need to know about how electric fields propagate through our Rb cell, including both absorption and phase propagation, for the simplest case of a plane wave and polarization-independent absorption. If we use polarized laser light, then we must consider how the different polarizations pass through the Rb cell, which is described in the sections for each specific experiment.

In many textbooks, the complex susceptibility of an atomic vapor is presented in terms of the natural linewidth of an atomic transition. For our case, however, the Doppler width near room temperature ($\sigma \approx 310$ MHz) is much greater than the natural linewidth of the D2 transitions ($\Gamma/2\pi \approx 6$ MHz). Thus we work in the limit $\Gamma/\sigma \to 0$, and in this limit $\Gamma$ does not appear in our expression for $\tilde{\chi}_{cell}(\delta)$.

Note also that our complex susceptibility derives entirely from classical physics, even though its ultimate source is from quantum-mechanical atomic transitions. The Gaussian form of the absorption coefficient arises from the Maxwell-Boltzmann velocity distribution of vapor atoms, and the Kramers-Kronig is a feature of classical electromagnetism and its interaction with dielectric materials.

## Faraday rotation

If we send linearly polarized light into the Rb vapor cell and assume that $\chi''$ is small or independent of circular polarization (or both), then the above formalism gives the output electric field

$$\begin{aligned} E(\delta) &\approx E_0 \exp\left(-\frac{\tau_0}{2}\chi''\right)\frac{1}{2}\left\{\begin{bmatrix}1\\ i\end{bmatrix}\exp\left(\frac{i\tau_0}{2}\chi'_+\right)+\begin{bmatrix}1\\ -i\end{bmatrix}\exp\left(\frac{i\tau_0}{2}\chi'_-\right)\right\} \\ &= E_0 \exp\left(-\frac{\tau_0}{2}\chi''\right)\frac{1}{2}e^{i\overline{\varphi}}\left\{\begin{bmatrix}e^{i\varphi_+}+e^{i\varphi_-}\\ i(e^{i\varphi_+}-e^{i\varphi_-})\end{bmatrix}\right\} \\ &= E_0 \exp\left(-\frac{\tau_0}{2}\chi''\right)e^{i\overline{\varphi}}\begin{bmatrix}\cos(\theta_F)\\ \sin(\theta_F)\end{bmatrix} \end{aligned} \tag{53}$$

where we have defined the new variables $\varphi_\pm = (\tau_0/4)\chi'_\pm$, $\overline{\varphi} = (\varphi_+ + \varphi_-)/2$ and $\theta_F = \varphi_+ - \varphi_-$.

In this limit, we see that the output polarization vector is simply linearly polarized light with the polarization angle rotated by $\theta_F$ relative to where it started, and this phenomenon is called *Faraday rotation*. For the special case of our Doppler-broadened susceptibility, this becomes

$$E(\delta) \approx E_0 \exp\left(-\frac{\tau_0}{2}e^{-\delta^2}\right)\begin{bmatrix}\cos(\theta_F)\\ \sin(\theta_F)\end{bmatrix} \tag{54}$$

(after dropping the unimportant overall phase factor $e^{i\overline{\varphi}}$). On resonance, we can use the fact that $D_+(\delta_B) - D_+(\delta_B) \approx 2\delta_B$ for small $\delta_B$ to give the Faraday angle

$$\theta_F \approx \frac{\tau_0}{\sqrt{\pi}}\delta_B \tag{55}$$

Note that the concept of Faraday rotation – a simple rotation of the plane of polarization after passing through a vapor cell – is only strictly valid in two limits, either $\tau_0 \ll 1$ or $\delta_B \ll 1$, and we are not really working in either of these limits in our experiments. Nevertheless, the Faraday rotation is often roughly valid in the lab, and it is a very useful limiting case for our model calculations. In the MOF experiment, for example, the on-resonance transmission can be expressed as

$$I(\delta = 0) \approx I_0 \frac{\tau_0^2\delta_B^2}{\pi}e^{-\tau_0} = I_0\theta_F^2 e^{-\tau_0} \tag{56}$$

and this expression should match experimental measurements at small $\delta_B$.

As a historical aside, Michael Faraday discovered this rotation of the polarization angle in 1845 while passing light through borosilicate glass in the presence of a large magnetic field. The Faraday rotation is most often associated with solid materials, and the optical isolator used in all the above experiments is one application of the effect. In 1898 Damiano Macaluso and Orso Mario Corbino discovered that the Faraday rotation is much enhanced in atomic vapors near resonant transitions, as we have seen in the above experiments, and this phenomenon is often called the Macaluso-Corbino effect [1898Mac, 1996Baa]. Over time, the theory evolved into the more general concept of a complex atomic susceptibility function.